\documentclass[a4paper,11pt]{article}
\pdfoutput=1
\usepackage{jheppub}
\usepackage[T1]{fontenc}
\usepackage{tikz}
\usepackage[compat=1.1.0]{tikz-feynman}
\usetikzlibrary{decorations.markings}
\usetikzlibrary{positioning}
\usepackage{subcaption}
\usetikzlibrary{snakes}
\tikzset{
    fermion/.style={
        postaction={decorate},
        decoration={
            markings,
            mark=at position 0.5 with {\arrow{>}}
        }
    },
    antifermion/.style={
        postaction={decorate},
        decoration={
            markings,
            mark=at position 0.5 with {\arrow{<}}
        }
    }
}
\tikzset{
    fermionTB/.style={
        postaction={decorate},
        decoration={
            markings,
            mark=at position 0.5 with {\arrow{>}}
        }
    }
}
\tikzset{
    fermionBT/.style={
        postaction={decorate},
        decoration={
            markings,
            mark=at position 0.5 with {\arrow{<}}
        }
    }
}
\usepackage{amstext,amssymb}
\usepackage{amsmath}
\usepackage{graphicx}
\usepackage{xspace}
\usepackage{color}
\usepackage{units}
\usepackage{multirow}
\usepackage{slashed}
\usepackage{comment}
\usepackage{hyperref}
\usepackage{appendix}
\usepackage{url}

\tikzfeynmanset{warn luatex=false}

\title{\boldmath Radiative Dirac neutrino masses and dark matter in a $U(1)_{B-L}$ extended model
}

\author[a]{ Chayan Majumdar}
\author[b]{, Utkarsh Patel}
\author[c]{, Supriya Senapati}
\author[d,e]{, Sudhanwa Patra}
\affiliation[a]{Institute of Modern Physics, Chinese Academy of Sciences, Lanzhou, 730000, China}
\affiliation[b]{Saha Institute of Nuclear Physics, 1/AF Bidhan Nagar, Kolkata 700064, India}
\affiliation[c]{Department of Applied Physics and MIIT Key Laboratory of Semiconductor Microstructure and Quantum Sensing, Nanjing University of Science and Technology, Nanjing 210094, China}
\affiliation[d]{Department of Physics, Indian Institute of Technology Bhilai, Durg 491002, India}
\affiliation[e]{Institute of Physics, Sachivalya Marg, Bhubaneswar-751005, India}
\vspace*{0.2cm}
\emailAdd{chayanmajumdar@impcas.ac.cn}
\emailAdd{utkarsh.patel@saha.ac.in}
\emailAdd{ssenapati@njust.edu.cn}
\emailAdd{sudhanwa@iitbhilai.ac.in}
\abstract{
We study a \(U(1)_{B-L}\) extension of the Standard Model (SM) in which Dirac neutrino masses are generated radiatively at the one-loop level through the exchange of new beyond the SM fields. This framework establishes a direct connection between neutrino mass generation and the dark sector, with the stability of the dark matter ensured by a residual discrete $\mathcal{Z}_6$ symmetry arising from the spontaneous breaking of \(U(1)_{B-L}\). We investigate the resulting charged lepton flavor violating processes and dark matter phenomenology, saturating relic observations and direct-detection constraints, and analyze the collider signatures of the dark sector at the Large Hadron Collider, its proposed high luminosity extension and at a future muon collider. We have identified excellent prospects for observing the considered dark matter candidates in these colliders, even with lower integrated luminosities than the proposed one.}

\begin{document} 
\maketitle
\flushbottom
\section{Introduction}
\label{sec:intro}
 
The origin of neutrino masses and the nature of dark matter (DM) are two of the most compelling open problems in particle physics and cosmology. While neutrino oscillation experiments have established the existence of nonzero neutrino masses, their smallness cannot be accommodated within the Standard Model (SM) without extending its field content or symmetries. This inadequacy points unambiguously to physics beyond the SM (BSM) and raises the fundamental question of the nature of neutrinos, which may be either Majorana or Dirac fermions. While Majorana neutrino mass models have been extensively explored~\cite{Minkowski:1977sc, Gell-Mann:1979vob, Mohapatra:1979ia,Schechter:1980gr,Schechter:1981cv,Foot:1988aq,Zee:1980ai,Zee:1985id,Babu:1988ki,Babu:1988ig,Ma:2006km,Cai:2017jrq}, the Dirac neutrino scenario remains equally viable~\cite{Ma:2014qra,Ma:2015mjd,Borah:2017leo,Borboruah:2024lli, Borah:2025fkd, Borah:2020boy, Ma:2016mwh, Boudjema:2025okq} and has gained increasing attention, particularly in frameworks where neutrino masses are generated radiatively~\cite{Mohapatra:1987hh,Mohapatra:1987nx,Branco:1978bz,Gu:2007ug,Farzan:2012sa,Ma:2017kgb,Han:2018zcn,Bonilla:2018ynb,Calle:2018ovc,Jana:2019mez,Babu:2019mfe,Jana:2020joi,Paul:2024prs,Dasgupta:2019rmf}. Such mechanisms naturally explain the smallness of neutrino masses through loop suppression, avoiding the need for unnaturally tiny Yukawa couplings and/or large new physics scale. If neutrinos are Dirac particles, the particle spectrum must be extended to include right-handed neutrinos $\nu_R$, which are singlets under the SM gauge group. In the absence of additional symmetries, the renormalizable Yukawa interaction
\begin{equation}
\mathcal{L}_Y = - y^\nu_{ij}\, \overline{L_i}\, \widetilde{H}\, \nu_{Rj} + \text{h.c.}
\end{equation}
leads to tree-level neutrino masses after electroweak symmetry breaking, $ m^\nu_{ij} = \frac{y^\nu_{ij} v}{\sqrt{2}} \, $ where $L$ and $\widetilde{H} = i\sigma_2 H$ are the SM lepton doublet and dual of Higgs doublet, respectively and $v$ corresponds to \textit{vev} acquired by $H$. Reproducing the observed neutrino mass scale then requires Yukawa couplings $y^\nu \sim \mathcal{O}(10^{-11})$, which is theoretically disfavored~\cite{tHooft:1979rat}. This motivates the introduction of additional symmetries that forbid the tree-level contribution and allow neutrino masses to arise only at the loop level or through higher-dimensional operators. A well-motivated realization of this idea is provided by a gauged $U(1)_{B-L}$ extension of the SM, which is anomaly-free in the presence of three right-handed neutrinos. For suitable $B-L$ charge assignments~\cite{Montero:2007cd,Machado:2010ui,Machado:2013oza}, the renormalizable Yukawa interaction is forbidden, and the leading contribution to neutrino masses is generated via the dimension-five operator
\begin{equation}
\mathcal{L}_5 = -\frac{h_{ij}}{\Lambda}\, \overline{L_i}\, \widetilde{H}\, \nu_{Rj}\, \sigma + \text{h.c.}
\end{equation}
where $\sigma$ is a scalar singlet responsible for the spontaneous breaking of the $U(1)_{B-L}$ symmetry. After symmetry breaking, naturally suppressed Dirac neutrino masses are generated, with their smallness controlled by the new physics scale and, in radiative realizations, by loop factors. The associated scalar and gauge sectors, including a massive $Z'$ boson, give rise to rich phenomenological signatures testable at colliders, low-energy experiments, and cosmological observations.

Independently, a wide range of astrophysical and cosmological observations provide strong evidence for the existence of DM, for which particle interpretation remains one of the most well-motivated and extensively studied possibilities. Radiative mechanisms offer an attractive framework to address both neutrino mass generation and the explanation of viable DM candidate in a unified manner. In such scenarios, neutrino masses are generated at the loop level via the exchange of new particles, which can simultaneously serve as viable DM candidates, as illustrated by the scotogenic type of models~\cite{Ma:2006km}. The stability of DM and the suppression of tree-level neutrino mass terms are typically ensured by additional symmetries in BSM frameworks~\cite{Farzan:2012ev,Restrepo:2013aga,Jana:2019mgj,Jana:2019mez}. Owing to the loop-induced suppression, the associated new physics scale can naturally lie at the TeV scale without requiring unnaturally small couplings, rendering these models predictive and testable in collider experiments, low-energy observables, and cosmological probes. A systematic approach for the minimal models to generate Dirac neutrino masses radiatively by utilizing the d = 5 effective operator has been explored in Ref. \cite{Jana:2019mgj}.

In this work, we aim to simultaneously address three open problems : the origin of tiny neutrino masses, the stability of DM, and the presence of experimentally testable new physics within a minimal and theoretically consistent gauge extension of the SM. In particular, a gauged $U(1)_{B-L}$ symmetry is among the simplest anomaly-free extensions of the SM once three right-handed neutrinos are included, and it provides a natural setting for Dirac neutrino mass generation without introducing explicit lepton-number violation. Along with three right-handed neutrinos in this $U(1)_{B-L}$ gauge extension, we have added a pair of vector-like fermions \(\Psi_{L,R}\) and an extended scalar sector comprising of one \(SU(2)_L\) doublet \(\phi\) and three singlet scalars \(\sigma\), \(\eta_1\), and \(\eta_2\). These new particles running in the loop can facilitate the Dirac mass generation mechanism for neutrinos; the smallness of neutrino masses is ensured due to loop suppression. Furthermore, these new states can participate in charged lepton flavor-violating (cLFV) mechanisms, which can test the framework using current and/or future projections from relevant experiments. In this framework, the spontaneous breaking of the \(U(1)_{B-L}\) gauge symmetry leaves a residual unbroken discrete symmetry that can stabilize the DM candidate. In particular, the symmetry breaking pattern $ U(1)_{B-L} \to  \mathcal{Z}_6 $ emerges in our setup. Since the SM Higgs doublet \(H\) is neutral under \(U(1)_{B-L}\), the structure of the residual symmetry is entirely determined by the \(U(1)_{B-L}\) charge of the singlet scalar \(\sigma\). The resulting \(\mathcal{Z}_6\) symmetry acts as a residual dark symmetry and ensures the stability of the DM candidates. Therefore, a key feature of our construction is that the stability of DM is not imposed by hand through an \emph{ad hoc} discrete symmetry, but instead arises automatically from the spontaneous breaking pattern $U(1)_{B-L}\to \mathcal{Z}_6$, by virtue of assigned quantum numbers for the scalars in the spectrum. Depending on the mass spectrum of the new particles, the model can accommodate either a scalar or a fermionic dark matter candidate, and we explore both possibilities in our analysis. We further find that the dark sector exhibits rich collider phenomenology, with the dominant collider signatures arising from the electroweak production of \(\mathcal{Z}_6\)-odd scalar states, allowing either scalar or fermionic DM depending on the mass ordering, making the model predictive and experimentally testable. While radiative Dirac neutrino mass models and $U(1)$ extensions have been explored previously~\cite{Bonilla:2018ynb,Calle:2018ovc,Saad:2019bqf,Jana:2019mez,Chowdhury:2022jde,Babu:2024zoe}, the present work goes beyond earlier studies by providing a unified and systematic phenomenological analysis of this specific $\mathcal{Z}_6$-protected setup. In particular, we simultaneously incorporate constraints from cLFV, DM relic abundance and direct-detection searches, and collider signatures of the $\mathcal{Z}_6$-odd states. Moreover, we present a collider study not only for the current LHC run, even briefly discussing the prospects of proposed high luminosity LHC, HL-LHC, in particular scenario where current LHC run alone will not be sufficient enough to produce a discoverable signature as well as for a future muon collider, identifying regions of parameter space where the dark sector can be probed efficiently even with luminosities below standard benchmark expectations.

The paper is organized as follows: In section~\ref{sec:mod}, the framework, along with the entire scalar sector particle spectrum and the Dirac mass generation of neutrinos at the one-loop level, is discussed. New BSM contributions to cLFV decays within the framework and the most stringent current and future sensitivity projections on these cLFV observables from relevant experiments have been explored in section~\ref{sec:LFV}. We have discussed about both the fermionic and scalar DM candidate phenomenology in section~\ref{sec:DM}. The collider prospects of these DM candidates in the context of the current $pp$ and proposed lepton $\mu^+\mu^-$ collider have been presented in section~\ref{sec:collider}. In section~\ref{sec:conclusion}, we have presented the conclusion of the work and further outlooks. 

\section{The Model}
\label{sec:mod}
We consider an extension of the SM symmetry with a gauged $U(1)_{B-L}$ group, where $B$ and $L$ correspond to Baryon and Lepton numbers, respectively. The particle content of the model and their charge assignments under the considered gauge group $SU(3)_C \times SU(2)_L \times U(1)_Y \times U(1)_{B-L}$ have been presented in table~\ref{tab:model}. In addition to the SM fields, the model contains three generations of right-handed neutrinos $\nu_{iR}$ and a pair of vector-like fermions $\Psi_{iL(R)}$ and an extended scalar sector consisting of additional $SU(2)_L$ doublet $\phi$, three SM-singlet scalars $\sigma$, $\eta_1$, and $\eta_2$. The $U(1)_{B-L}$ charges for the right-handed neutrinos have been assigned in such a way that the model remains triangle-anomaly free~\cite{Montero:2007cd, Machado:2010ui, Machado:2013oza}. Another charge assignment $\lbrace -1, -1, -1 \rbrace$ exists in the literature to keep the framework anomaly-free. However, we will use the former one in our subsequent analysis, as this particular choice will not affect the paper's overall conclusion. 

\begin{table}[htb]
\centering
\begin{tabular}{|c|c|c|c|c|c|}
\hline
    & Particle & $SU(3)_C$ & $SU(2)_L$ & $U(1)_Y$ & $U(1)_{B-L}$ \\
    \hline \hline
    Quarks & $Q_L$ & 3 & 2 & $\frac{1}{6}$ & $\frac{1}{3}$ \\
    & $u_R$ & 3 & 1 & $\frac{2}{3}$ & $\frac{1}{3}$ \\
    & $d_R$ & 3 & 1 & $-\frac{1}{3}$ & $\frac{1}{3}$ \\
    \hline \hline
    Leptons & $\ell_L$ & 1 & 2 & $-\frac{1}{2}$ & $-1$ \\
    & $e_R$ & 1 & 1 & $-1$ & $-1$ \\
    & $\nu_{1R}$ & 1 & 1 & $0$ & $5$ \\
    & $\nu_{iR}~ (i=2,3)$ & 1 & 1 & $0$ & $-4$ \\
    & $\Psi_{iL(R)}$ & 1 & 1 & $0$ & $\frac{1}{2}$ \\
    \hline \hline
    Scalars & $H$ & 1 & 2 & $\frac{1}{2}$ & $0$ \\
    & $\sigma$ & 1 & 1 & $0$ & $3$ \\
    & $\eta_1$ & 1 & 1 & $0$ & $\frac{9}{2}$ \\
    & $\eta_2$ & 1 & 1 & $0$ & $\frac{3}{2}$ \\
    & $\phi$ & 1 & 2 & $\frac{1}{2}$ & $\frac{3}{2}$ \\
    \hline
\end{tabular}
\caption{Particle content and charge assignments of the model.}
\label{tab:model}
\end{table}

The most general renormalizable scalar potential consistent with the gauge symmetries is given by,
\begin{align}
V(H,\phi,\eta_1,\eta_2,\sigma) =\;&
-\mu_H^2 H^\dagger H + \mu_\phi^2 \phi^\dagger \phi
+ \mu_{\eta_1}^2 \eta_1^\ast \eta_1
+ \mu_{\eta_2}^2 \eta_2^\ast \eta_2
+ \mu_\sigma^2 \sigma^\ast \sigma \nonumber \\
& + \lambda_H (H^\dagger H)^2
+ \lambda_\phi (\phi^\dagger \phi)^2
+ \lambda_{\eta_1} (\eta_1^\ast \eta_1)^2
+ \lambda_{\eta_2} (\eta_2^\ast \eta_2)^2
+ \lambda_\sigma (\sigma^\ast \sigma)^2 \nonumber \\
&+ \frac{\lambda_{H\phi}}{2} [(H^\dagger H)(\phi^\dagger \phi) +(H^\dagger \phi)(\phi^\dagger H)]+
+ \lambda_{H\eta_1} (H^\dagger H)(\eta_1^\ast \eta_1)
 \nonumber \\
&+ \lambda_{H\eta_2} (H^\dagger H)(\eta_2^\ast \eta_2) + \lambda_{H\sigma} (H^\dagger H)(\sigma^\ast \sigma) + \lambda_{\phi\eta_1} (\phi^\dagger \phi)(\eta_1^\ast \eta_1) \nonumber \\
& + \lambda_{\phi\eta_2} (\phi^\dagger \phi)(\eta_2^\ast \eta_2) + \lambda_{\eta_1\eta_2}(\eta_1^\ast \eta_1)(\eta_2^\ast \eta_2)
 + \lambda_{\phi\sigma} (\phi^\dagger \phi)(\sigma^\ast \sigma) \nonumber \\
& + \lambda_{\eta_1\sigma} (\eta_1^\ast \eta_1)(\sigma^\ast \sigma) + \lambda_{\eta_2\sigma} (\eta_2^\ast \eta_2)(\sigma^\ast \sigma) + \lambda_{\eta_1\eta_2}^{\prime}[\eta_1^\ast \eta_2^3 + \text{h.c.}] \nonumber \\
& + \lambda_m [(H^\dagger \phi) \eta_2\sigma^\ast + \text{h.c.}] + \mu_1 [H^\dagger \phi \eta_2^\ast + \text{h.c.}]
+ \mu_2 [\eta_2 \sigma \eta_1^\ast + \text{h.c.}] \nonumber\\
& + \mu_3 [\sigma^\ast \eta_2^2 + \text{h.c.}] .
\end{align}
After spontaneous symmetry breaking, the scalar fields are expanded around their vacuum expectation values (\textit{vev}s) as
\begin{align}
\sigma &= \frac{1}{\sqrt{2}}(v_\sigma + \sigma_R^0 + i \sigma_I^0), \qquad
\eta_i = \frac{1}{\sqrt{2}}(\eta_{iR}^0 + i \eta_{iI}^0) \quad (i=1,2), \nonumber \\
\phi &= 
\begin{pmatrix}
\phi^+ \\ \frac{1}{\sqrt{2}}(\phi_R^0 + i\phi_I^0)
\end{pmatrix}, \qquad
H =
\begin{pmatrix}
h^+ \\ 
\frac{1}{\sqrt{2}}(v_H + h_R^0 + i h_I^0)
\end{pmatrix}.
\end{align}
Minimization of the scalar potential with respect to $v_H$ and $v_\sigma$ yields,
\begin{equation}
\mu_H^2 = \lambda_H v_H^2 + \frac{1}{2}\lambda_{H\sigma} v_\sigma^2,
\qquad
\mu_\sigma^2 = -\left(\frac{1}{2}\lambda_{H\sigma} v_H^2 + \lambda_\sigma v_\sigma^2 \right).
\end{equation}

In this framework, the spontaneous breaking of the $U(1)_{B-L}$ symmetry leaves a residual discrete $\mathcal{Z}_6$ symmetry via $v_\sigma$. Now, the residual symmetry becomes $\mathcal{Z}_6$ instead of $\mathcal{Z}_3$ due to the presence of half-integer $B-L$ charges of BSM scalars present in the particle content~\cite{Jana:2019mgj}. Now, defining one of the generators of this remnant symmetry as $\omega = \text{exp}(2\pi i/6)$~\footnote{The entire $\mathcal{Z}_6$ group can be written as $\lbrace 1, \omega, \omega^2, \omega^3 = -1, \omega^4, \omega^5 \rbrace$.}, each generic field $\chi$ in scalar sector with $B-L$ charge $\chi_{BL}$ will transform as $\chi \to \omega^{2\chi_{BL}(\text{mod}~6)}\chi$ under $\mathcal{Z}_6$. Field having half-integer $B-L$ charge has to be first rescaled by a factor of 2 to understand its transformation property under the remnant symmetry. Now, it is easy to show that within the BSM scalars $\lbrace \eta_1, \eta_2, \sigma, \phi \rbrace$ having non-zero $B-L$ charge, $\sigma$ transforms as even particle and others transforms as odd particles under remnant $\mathcal{Z}_6$. Moreover, SM Higgs doublet $H$ remains $Z_6$ even for its zero $B-L$ charge. As a result, the scalar mass matrices decompose into independent $\mathcal{Z}_6$-even and $\mathcal{Z}_6$-odd sectors. In the CP-even $\mathcal{Z}_6$-even basis $(h_R^0, \sigma_R^0)^T$, the mass-squared matrix is
\begin{equation}
\mathcal{M}^{\mathcal{Z}_6\text{-even}}_{\text{even}} =
\begin{pmatrix}
2\lambda_H v_H^2 & \lambda_{H\sigma} v_H v_\sigma \\
\lambda_{H\sigma} v_H v_\sigma & 2\lambda_\sigma v_\sigma^2
\end{pmatrix}.
\end{equation}
Diagonalization leads to two physical neutral scalars with masses squared,
\begin{equation}
M_{h,H}^2 =
\lambda_H v_H^2 + \lambda_\sigma v_\sigma^2
\mp \sqrt{(\lambda_H v_H^2 - \lambda_\sigma v_\sigma^2)^2
+ \lambda_{H\sigma}^2 v_H^2 v_\sigma^2},
\end{equation}
where the mass eigenstates are related to the gauge eigenstates by
\begin{align}
h &= h_R^0 \cos\theta - \sigma_R^0 \sin\theta, \\
H &= h_R^0 \sin\theta + \sigma_R^0 \cos\theta,
\end{align}
with mixing angle
\begin{equation}
\tan 2\theta = \frac{\lambda_{H\sigma} v_H v_\sigma}
{\lambda_H v_H^2 - \lambda_\sigma v_\sigma^2}.
\end{equation}
The lighter state $h$ is identified with the SM-like Higgs boson, with $v_H = 246$~GeV. The symmetry-breaking pattern of the model is
\begin{equation}
SU(2)_L \times U(1)_Y \times U(1)_{B-L}
\xrightarrow{\,v_\sigma\,}
SU(2)_L \times U(1)_Y
\xrightarrow{\,v_H\,}
U(1)_{\rm em}.
\end{equation}
Throughout this work, we will assume $v_\sigma \gg v_H$, and in the absence of new physics below the TeV scale, a particularly safe choice would be $v_\sigma \gtrsim \mathcal{O}(1)$ TeV. The CP-even $\mathcal{Z}_6$-odd scalar mass-squared matrix, in the basis
$(\phi_R^0, \eta_{1R}^0, \eta_{2R}^0)^T$, is given by
\begin{equation}
\mathcal{M}^{\mathcal{Z}_6\text{-odd}}_{\text{even}} =
\begin{pmatrix}
M_\phi^2 & 0 &\frac{\mu_1^\prime v_H}{\sqrt{2}} \\
0 & M_{\eta_1}^2 & \frac{\mu_2 v_\sigma}{\sqrt{2}} \\
\frac{\mu_1^\prime v_H}{\sqrt{2}}  & \frac{\mu_2 v_\sigma}{\sqrt{2}} & M_{\eta_2}^2
\end{pmatrix},
\label{Moddodd}
\end{equation}
where
\begin{align}
M_\phi^2 &= \frac{1}{2}\left(\lambda_{H\phi} v_H^2
+ \lambda_{\phi\sigma} v_\sigma^2 + 2\mu_\phi^2 \right), \nonumber \\
M_{\eta_1}^2 &= \frac{1}{2}\left(\lambda_{H\eta_1} v_H^2
+ \lambda_{\sigma\eta_1} v_\sigma^2 + 2\mu_{\eta_1}^2 \right), \nonumber \\
M_{\eta_2}^2 &= \frac{1}{2}\left(\lambda_{H\eta_2} v_H^2
+ \lambda_{\sigma\eta_2} v_\sigma^2 + 2\mu_{\eta_2}^2 \right), \nonumber \\
\mu_1^\prime &= \mu_1 + \frac{\lambda_mv_\sigma}{2\sqrt{2}}.
\label{eq:mDScalars}
\end{align}
This matrix can be diagonalized by a unitary transformation $U$, yielding three physical CP-even $\mathcal{Z}_6$-odd scalars $S_i$ $(i=1,2,3)$:
\begin{equation}
\label{eq:dScalarmass}
\begin{pmatrix}
S_1 \\ S_2 \\ S_3
\end{pmatrix}
= U
\begin{pmatrix}
\phi_R^0 \\ \eta_{1R}^0 \\ \eta_{2R}^0
\end{pmatrix}.
\end{equation}
The mixing matrix $U$ can be parameterized by three Euler angles $\theta_{ij}$, with
\begin{align}
\tan 2\theta_{13} &= \frac{\sqrt{2}\mu_1^\prime v_H}{M_\phi^2 - M_{\eta_2}^2}, \\
\tan 2\theta_{23} &= \frac{\sqrt{2}\mu_2v_\sigma}{M_{\eta_1}^2 - M_{\eta_2}^2}, \\
\tan 2\theta_{12} &=
\frac{2\mu_1^\prime\mu_2v_Hv_\sigma}{2(M_\phi^2 - M_{\eta_1}^2)(M_{\eta_1}^2 + M_{\eta_2}^2 - M_\phi^2)
+ \mu_1^{\prime 2}v_H^2 - \mu_2^2v_\sigma^2}.
\end{align}
The CP-odd $\mathcal{Z}_6$-even scalars provide the Goldstone bosons eaten by the neutral gauge bosons, while the CP-odd $\mathcal{Z}_6$-odd scalars possess a mass structure identical to Eq.~\eqref{Moddodd}. Finally, there are two charged scalars $h^\pm$ and $\phi^\pm$, among which $h^\pm$ acts as Goldstone boson for $W^\pm$ gauge bosons. Furthermore, these $h^\pm, \phi^\pm$ do not mix, by costruction, due to their different $B-L$ and $Z_6$ charges. Therefore, only physical singly charged scalar $\phi^\pm$ has mass,
\begin{equation}
\label{eq:charSca}
m_{\phi^\pm} = M_\phi
= \sqrt{\frac{1}{2}\left(\lambda_{H\phi} v_H^2
+ \lambda_{\phi\sigma} v_\sigma^2 + 2\mu_\phi^2\right)}.
\end{equation}

In this work, we focus primarily on dark matter phenomenology and its detectability prospects within current and proposed hadron and muon colliders; therefore, the relevant scalar spectrum of this framework is discussed in greater detail. The discussion of the gauge boson structure of the model will be omitted from the subsequent discussion, as it remains similar to any generic $U(1)_X$-extended framework, for instance,~\cite{Montero:2007cd,Machado:2013oza,Jana:2019mgj}. 

\subsection{Generation of Neutrino Masses}
\label{sec:Numass}

\begin{figure}[htb!]
\centering
\includegraphics[width=0.5\textwidth]{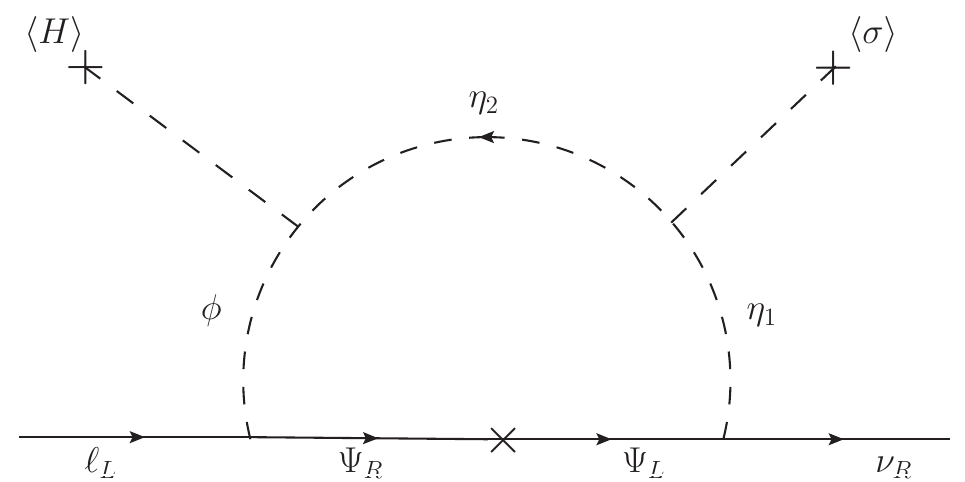}
\caption{One-loop Dirac mass generation for neutrinos where the BSM fields are running within the loop : vector-like isosinglet fermions $\Psi$, and the scalars from the extended scalar sector $\phi, \eta_1, \eta_2$, while the neutrino mass is generated via the \textit{vev}s of $H$ and $\sigma$. The insertion on the $\Psi$ line denotes the bare mass term $M_\Psi$ for the exotic fermion.}
\label{fig:numassmuon}
\end{figure}

The Yukawa sector responsible for quark and charged-lepton masses is given by
\begin{equation}
\mathcal{L}_{\rm Yuk} =
\sum_{\rm generations}
\left(
\overline{Q_L} Y^d H d_R
+ \overline{Q_L} Y^u \, \widetilde{H}^\dagger u_R
+ \overline{\ell_L} Y^\ell H e_R
\right),
\end{equation}
where we have omitted the generation and flavor indices of the fermions for simpler notations. After electroweak symmetry breaking, with
$\langle H \rangle = v_H/\sqrt{2}$,
the Dirac mass matrices for quarks and charged leptons are
\begin{equation}
M_u = Y^u v_H /\sqrt{2} , \qquad
M_d = Y^d v_H /\sqrt{2}, \qquad
M_e = Y^\ell v_H /\sqrt{2},
\end{equation}
where we have taken the Higgs \textit{vev} to be real, and we will work in a basis where the charged-lepton mass matrix is diagonal, without loss of generality.
The quark mass matrices can be diagonalized via bi-unitary transformations,
\begin{align}
M_u &= V_L^u \,
\text{diag}(m_u,m_c,m_t)\,
V_R^{u\dagger}, \\
M_d &= V_L^d \,
\text{diag}(m_d,m_s,m_b)\,
V_R^{d\dagger},
\end{align}
leading to the Cabibbo–Kobayashi–Maskawa (CKM) mixing matrix
\begin{equation}
V_{\rm CKM} = V_L^{u\dagger} V_L^d .
\end{equation}
In contrast to the charged fermions, neutrino masses are forbidden at tree level by the residual $\mathcal{Z}_6$ symmetry. Due to non-zero $U(1)_{B-L}$ charge assignment for right-handed neutrinos, and since we are mostly interested in the Dirac mass generation of neutrinos, the neutrino masses can be generated radiatively at the one-loop level. The relevant diagram is shown in figure~\ref{fig:numassmuon}. The loop involves the SM lepton doublet $\ell_L$, the SM Higgs doublet $H$, the inert scalar doublet $\phi$, the singlet scalars $\eta_{1,2}$, vector-like singlet fermions $\Psi_{L,R}$, and right-handed neutrinos $\nu_{iR}$ where $i = 2, 3$. The other right-handed neutrino $\nu_{1R}$ does not participate in this loop due to its different $U(1)_{B-L}$ charge assignment; consequently, this right-handed neutrino and its left-handed Dirac counterpart within the SM remain massless in the framework. As the current neutrino oscillation observation data support the existence of at least two mass-degenerate light neutrino states, this charge assignment goes well hand-in-hand with the current oscillation prediction. The relevant terms in the Lagrangian are,
\begin{align}
\mathcal{L} \supset\;&
M_\Psi \, \overline{\Psi_L} \Psi_R
+ y_1 \, \overline{\ell_L} \, \epsilon \, \phi^\ast \Psi_R
+ y_2 \, \sum_{i=2,3} \overline{\Psi_L} \eta_1 \nu_{iR} + \mu_1 (H^\dagger \phi \eta_2^\ast + \text{h.c.})
+ \mu_2 (\eta_2 \sigma \eta_1^\ast + \text{h.c.}) ,
\end{align}
where $\epsilon = i\sigma_2$ is the antisymmetric $SU(2)_L$ tensor. Here, $M_\Psi$ corresponds to bare mass term for the vector-like leptons and $y_i~(i=1,2)$ correspond to the relevant Yukawa couplings, we have also omitted the generational and flavor indices in this Lagrangian for sake of simplicity.

After spontaneous symmetry breaking, the interplay of the trilinear scalar couplings $\mu_{1,2}$ and the Yukawa interactions $y_{1,2}$ induces a Dirac mass for neutrinos at the one-loop level. Evaluating the loop integral using \texttt{Package-X}~\cite{Patel:2015tea}, the radiatively generated neutrino masses in the basis of physical $\mathcal{Z}_6$-odd scalar eigenstates $S_i$ $(i=1,2,3)$ can be expressed as
\begin{equation}
m_\nu^{\rm 1\text{-}loop}
=
\left|
\frac{y_1 y_2 \mu_1 \mu_2 v_H v_\sigma}{16\pi^2}
\, M_\Psi
\sum_{a,b,c=1}^3
U_{1a} U_{2c} U_{3b}
\,
\mathcal{I}(m_c, m_b, m_a, M_\Psi)
\right|.
\end{equation}
where the loop function $\mathcal{I}$ is given by
\begin{equation}
\mathcal{I}(m_1,m_2,m_3,m_4)
=
\sum_{i=1}^4
\frac{m_i^2 \ln \!\left(\frac{m_2^2}{m_i^2}\right)}
{\prod\limits_{j=1,\,j\neq i}^4 (m_i^2 - m_j^2)} ,
\end{equation}
with $\{m_i\} \equiv \{M_{S_1}, M_{S_2}, M_{S_3}, M_\Psi\}$. Throughout this analysis, the charged-lepton masses are neglected inside the loop, i.e.\ $m_\ell^2 \to 0$ considering $m_\ell \ll M_{S_i}, M_\Psi$. The condition $j \neq i$ avoids the unphysical singularities arising from degenerate mass terms in the denominator. This radiative mechanism naturally explains the smallness of neutrino masses through loop suppression, heavy mediator masses, and the residual $\mathcal{Z}_6$ symmetry.

To gain further insight, it is useful to extract a parametric scaling of the radiatively generated neutrino masses. For a generic spectrum without large hierarchies among the $\mathcal{Z}_6$-odd scalar and the vector-like fermion masses, the loop function scales as $\mathcal{I}(M_{S_i}, M_\Psi) \sim M_{\rm loop}^{-4}$ where $M_{\rm loop}$ denotes the characteristic mass scale of particles propagating inside the loop, typically set by $m_{S_i}$ or $M_\psi$. Under this approximation, the one-loop neutrino masses can be estimated as
\begin{equation}
m_\nu^{1-\text{loop}} \;\sim\;
\frac{y_1 y_2\,}{16\pi^2}\,
\frac{\mu_1 \mu_2 \, v_H v_\sigma}{M_{\rm loop}^4}\,
M_\Psi \,
\mathcal{U},
\label{eq:nu_scaling}
\end{equation}
where $\mathcal{U}$ represents the products of scalar mixing matrix elements,
$\mathcal{U} \equiv \sum_{a,b,c} U_{1a}U_{2b}U_{3c}$,
and is generically $\mathcal{O}(1)$ in the absence of special alignment. Numerically, taking representative values :
\begin{center}
$y_1,y_2 \sim 10^{-6},~~
\mu_1 \sim \mu_2 \sim \mathcal{O}(1~{\rm TeV}),~~
M_\Psi \sim M_{\rm loop} \sim \mathcal{O}(1~{\rm TeV}),~~ v_\sigma \sim \mathcal{O}(10~{\rm TeV}),$
\end{center}
one obtains $ m_\nu \sim \mathcal{O}(0.01~{\rm eV})$, which naturally lies in the range favored by current neutrino oscillation data. We emphasize that this estimate assumes no accidental cancellations among different contributions in Eq.~\eqref{eq:nu_scaling}. Such cancellations may occur for quasi-degenerate $\mathcal{Z}_6$-odd scalar masses or for specific alignments of the scalar mixing matrix, leading to further suppression of $m_\nu$. Analogous cancellation effects are well known in Ma-type radiative neutrino mass models; see, for example, Refs.~\cite{Ma:2006km,Kubo:2006yx}.

\section{Charged Lepton Flavor Violation (cLFV)}
\label{sec:LFV}

In this section, we analyze the charged lepton-flavor-violating processes $\mu \to e\gamma$, $\mu \to 3e$, and coherent $\mu$--$e$ conversion in nuclei within the present framework. For a comprehensive discussion of cLFV processes in radiative neutrino mass models, we refer the interested reader to Ref.~\cite{Toma:2013zsa}. Usually within the SM, cLFV amplitudes are strongly suppressed by the Glashow--Iliopoulos--Maiani (GIM) mechanism. In models containing electroweak-scale right-handed neutrinos, this suppression can be lifted due to mixing between left- and right-handed neutrinos, leading to sizable enhancements of cLFV rates~\cite{Ilakovac:1994kj, Deppisch:2004fa, Deppisch:2005zm, Ilakovac:2009jf, Alonso:2012ji, Dinh:2012bp, Ilakovac:2012sh, Abada:2012cq, Lee:2013htl}. In the present model, however, the residual $\mathcal{Z}_6$ symmetry forbids such mixing, and consequently the usual enhancement from $W$-$\nu$ loop diagrams is absent. Instead, cLFV transitions arise dominantly from one-loop diagrams involving the $\mathcal{Z}_6$-odd charged scalar $\phi^\pm$ and the SM-singlet vector-like fermion $\Psi_R$, which can still generate phenomenologically relevant effects.

\begin{figure}[htb!]
\centering
\includegraphics[width=0.27\textwidth]{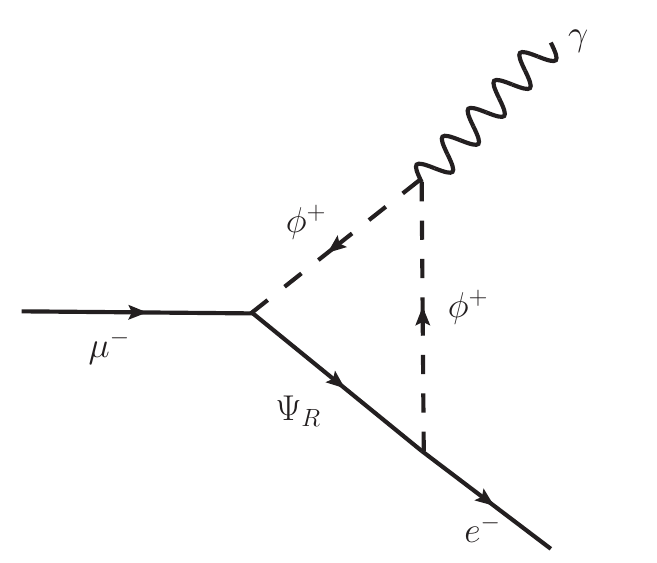}
\includegraphics[width=0.4\textwidth]{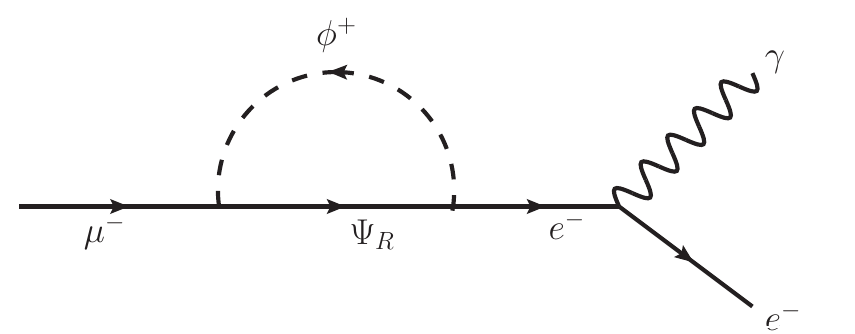}
\includegraphics[width=0.3\textwidth]{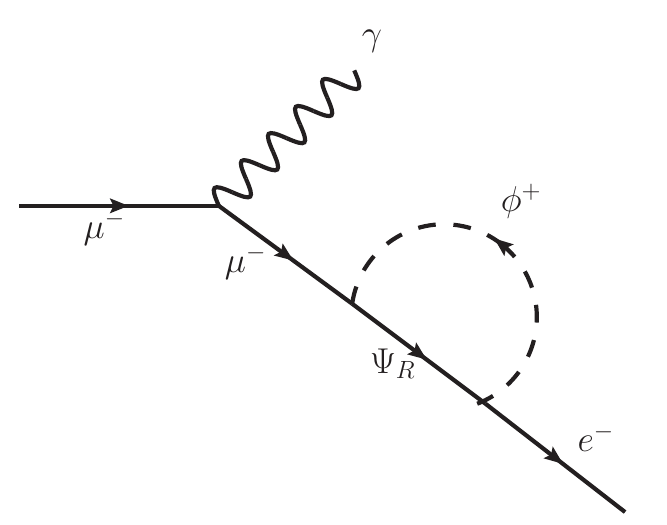}
\caption{New diagrams involving the vector-like isosinglet fermion $\Psi_R$ and the singly charged BSM scalar $\phi^+$ contributing to radiative muon decay $\mu \to e\gamma$ process.} 
\label{fig:mueg}
\end{figure}

The radiative decay $\mu \to e\gamma$ is induced at the one-loop level through the diagrams shown in figure~\ref{fig:mueg}.
The corresponding branching ratio is given by
\begin{equation}
\text{BR}(\mu \to e\gamma)
= \frac{3 (4\pi)^3 \alpha_{\rm em}}{4 G_F^2}\, |A_D|^2,
\end{equation}
where $G_F$ denotes the Fermi constant, $\alpha_{\rm em}=e^2/(4\pi)$ is the electromagnetic fine-structure constant, and $A_D$ is the dipole form factor arising from photon-penguin diagrams. In the limit $m_e \ll m_\mu \ll m_{\phi^+}, M_{\Psi}$, the dipole amplitude is
\begin{equation}
A_D = \frac{y_1^2}{2(4\pi)^2}\, \frac{1}{m_{\phi^+}^2}\, F(x),
\label{eq:AD}
\end{equation}
with loop factor $F(x) = \frac{1}{6(1-x)^4}
\left[ 1 - 6x + 3x^2 + 2x^3 - 6x^2 \ln x \right]$ and $x \equiv m_{\phi^+}^2/M_{\Psi}^2$.

\begin{figure}[htb!]
\centering
\includegraphics[width=0.33\textwidth]{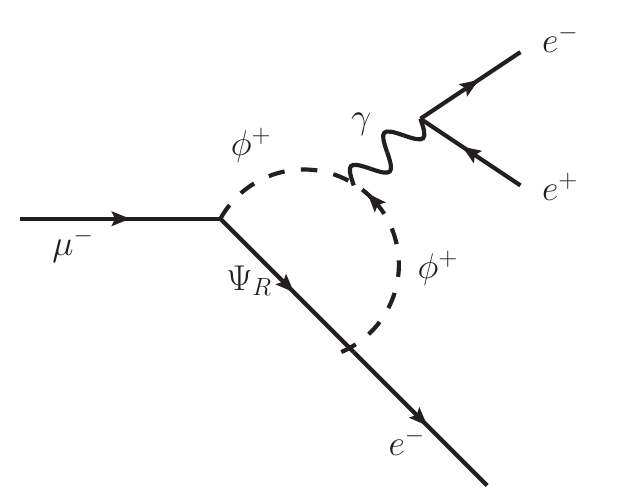}
\includegraphics[width=0.40\textwidth]{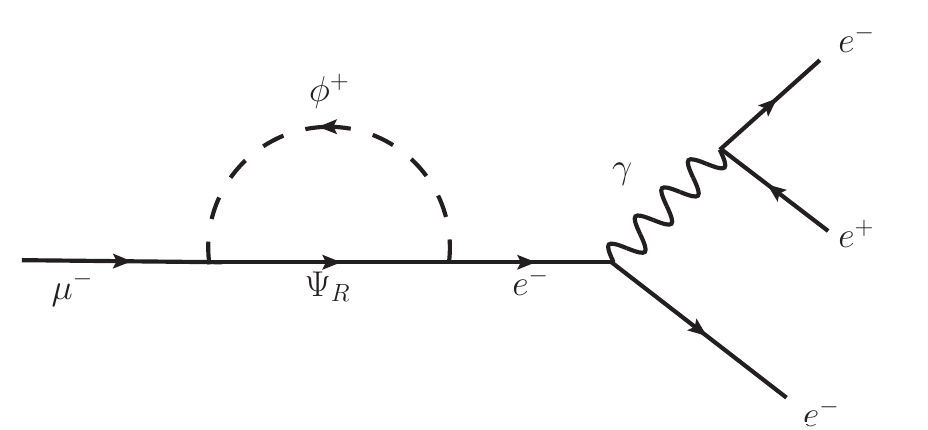}
\includegraphics[width=0.25\textwidth]{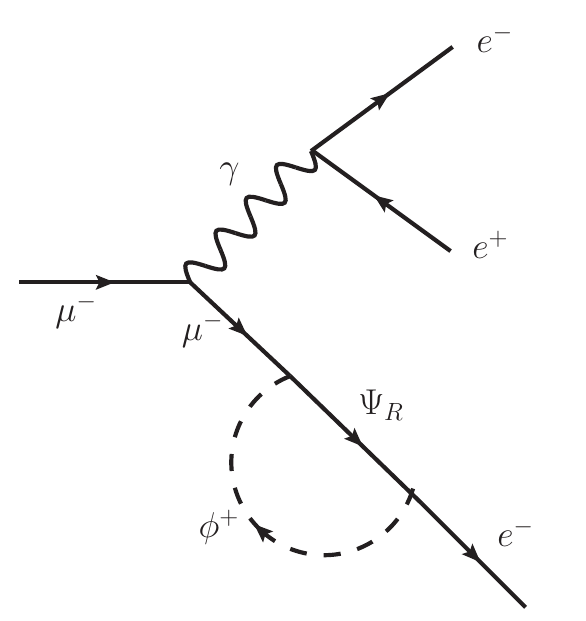}
\includegraphics[width=0.4\textwidth]{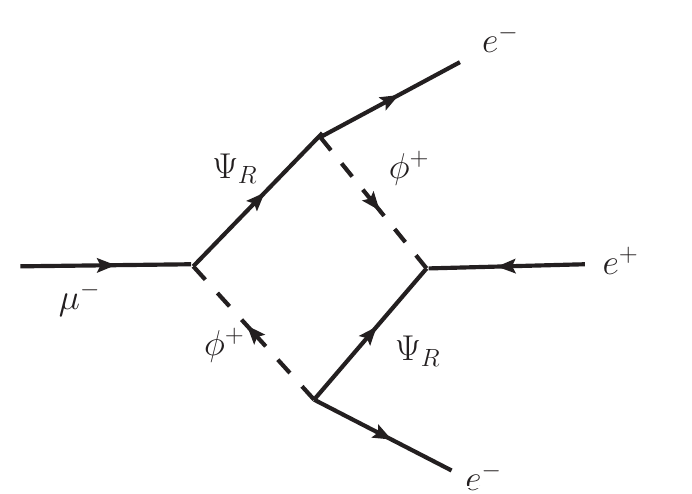}
\includegraphics[width=0.4\textwidth]{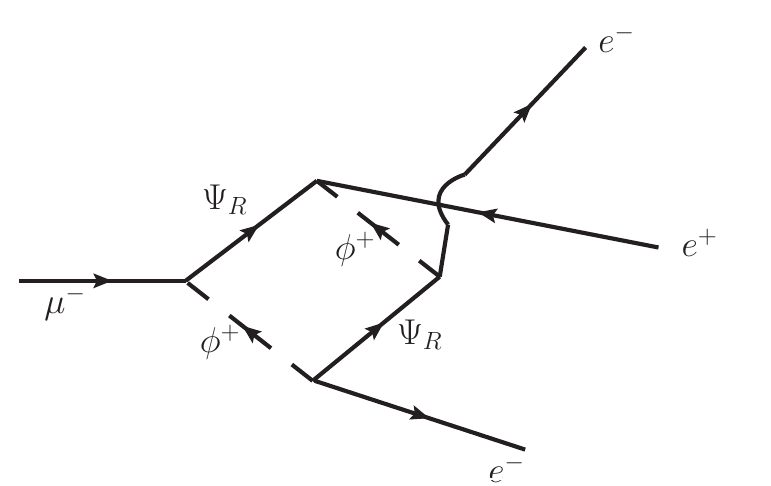}
\caption{New diagrams contributing to three-body decay of muon $\mu \rightarrow 3e$ process. In the upper panel, the Penguin diagrams are shown, while the lower panel corresponds to the box diagrams. These new contributions are arising due to the running of BSM particles $\Psi_R$ and $\phi^+$ running within the loop.} 
\label{fig:mu3e}
\end{figure}

Next, we consider the three-body decay $\mu \to 3e$. The relevant penguin and box diagrams are displayed in figure~\ref{fig:mu3e}.
In this model, contributions from $Z$-boson penguins are negligible due to their suppression by charged-lepton masses~\cite{Toma:2013zsa}, while Higgs-penguin diagrams are further suppressed by the small Yukawa couplings. Consequently, only photon-penguin and box-diagram contributions are retained. The branching ratio for $\mu \to 3e$ can be written as,
\begin{align}
\text{BR}(\mu \to 3e) =\;&
\frac{3 (4\pi)^3 \alpha_{\rm em}^2}{8 G_F^2}
\Bigg[
|A_{ND}|^2
+ |A_D|^2
\left(
\frac{16}{3}\ln\frac{m_\mu}{m_e}
- \frac{22}{3}
\right)
+ \frac{1}{6}|B|^2 \nonumber \\
&\hspace{1.5cm}
+ \left(
-2 A_{ND} A_D^\ast
+ \frac{1}{3} A_{ND} B^\ast
- \frac{2}{3} A_D B^\ast
+ \text{h.c.}
\right)
\Bigg].
\end{align}
Here, dipole form factor $A_D$ is defined in Eq.~\eqref{eq:AD}, while $A_{ND}$ arises from non-dipole photon-penguin diagrams and is given by
\begin{equation}
A_{ND} = \frac{y_1^2}{6(4\pi)^2}\, \frac{1}{m_{\phi^+}^2}\, G(x),
\label{eq:AND}
\end{equation}
with $G(x) = \frac{1}{6(1-x)^4}
\left[ 1 - 9x + 18x^2 - 11x^3 + 6x^3 \ln x \right]$ and $x = m_{\phi^+}^2/M_\Psi^2$. The quantity $B$ denotes the contribution from box diagrams,
\begin{equation}
e^2 B =
\frac{1}{(4\pi)^2 m_{\phi^+}^2}
\left[
\frac{1}{2} y_1^4 D_1(x,x)
+ y_1^4 x D_2(x,x)
\right],
\end{equation}
where the loop functions are
\begin{center}
$D_1(x,x) = \frac{1}{(1-x)^3}
\left( x^2 - 1 - 2x \ln x \right)$~~ and $D_2(x,x) = \frac{1}{(1-x)^3}
\left( 2(x-1) - (1+x)\ln x \right)$.
\end{center}

\begin{figure}[htb!]
\centering
\includegraphics[width=0.32\textwidth]{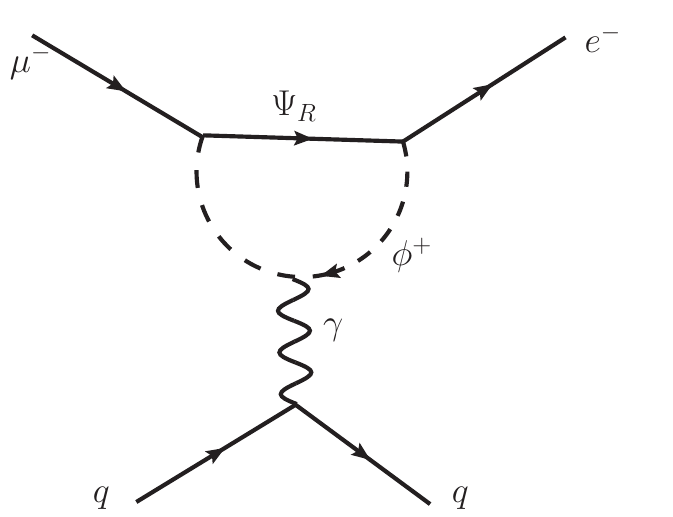}
\includegraphics[width=0.32\textwidth]{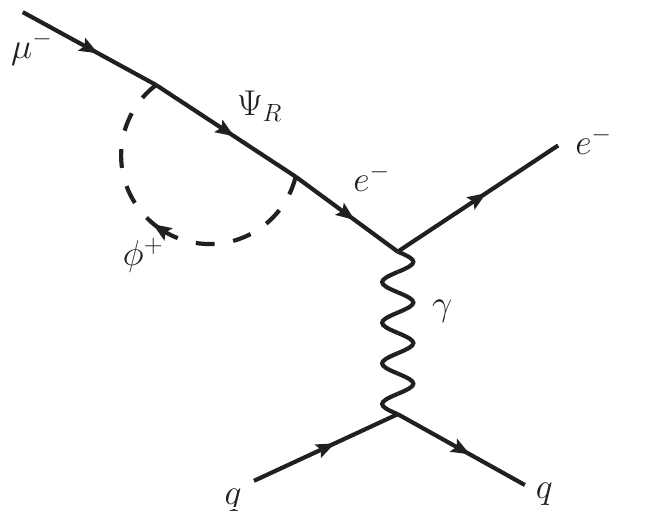}
\includegraphics[width=0.32\textwidth]{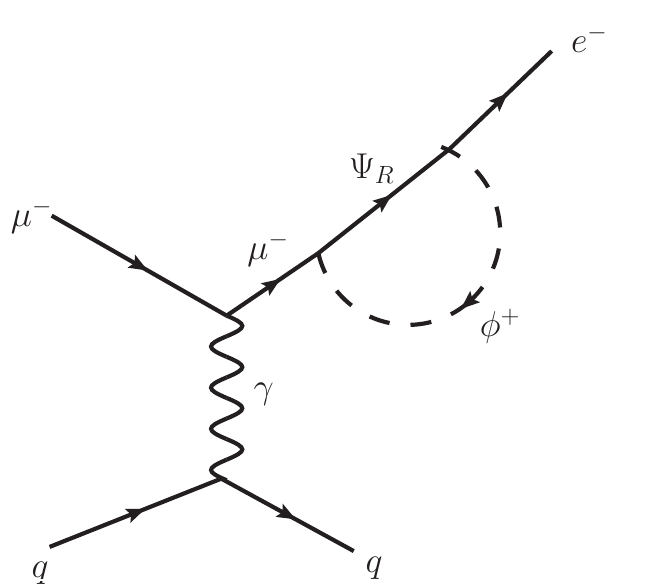}
\caption{New diagrams contributing to coherent $\mu$--$e$ conversion in nuclei are shown, involving quarks $q$ of the relevant generation and the incoming muon. The diagrams also depict the presence of the BSM particles $\Psi_R$ and $\phi^+$ running in the loop.}
\label{fig:mue}
\end{figure}

Finally, we now turn to coherent $\mu$--$e$ conversion in nuclei, which represents one of the most sensitive probes of cLFV in upcoming experiments. The relevant diagrams are shown in figure~\ref{fig:mue}.
In the present model, box-diagram contributions are absent because the residual $\mathcal{Z}_6$ symmetry forbids couplings between the charged scalar $\phi^\pm$ and quarks. As in the case of $\mu \to 3e$, $Z$-boson penguin contributions are suppressed by charged-lepton masses and are therefore neglected. The normalized $\mu$--$e$ conversion rate is given by~\cite{Kuno:1999jp, Arganda:2007jw}
\begin{align}
\text{CR}(\mu\text{--}e,\, \text{Nucleus}) =\;&
\frac{1}{\Gamma_{\rm capt}}
\frac{p_e E_e m_\mu^3 G_F^2 \alpha_{\rm em}^3 Z_{\rm eff}^4 F_p^2}
{8\pi^2 Z}
\Big[\Big| \textcolor{blue}{e A_D D}+ 
(Z+N)(g_{LV}^{(0)} + g_{LS}^{(0)}) \nonumber \\
& \hspace{3.5cm} + (Z-N)(g_{LV}^{(1)} + g_{LS}^{(1)})\Big|^2 + \left( L \leftrightarrow R \right)
\Big],
\end{align}
where $D$ denotes the nuclear overlap integral associated with the dipole operator contribution to coherent $\mu$--$e$ conversion, $\Gamma_{\rm capt}$ is the total muon capture rate, $p_e$ and $E_e$ are the momentum and energy of the emitted electron, $Z_{\rm eff}$ denotes the effective atomic charge~\cite{Chiang:1993xz}, and $F_p$ is the nuclear matrix element. The quantities $Z$ and $N$ represent the number of protons and neutrons in the nucleus, respectively. Numerical values of these nuclear parameters for experimentally relevant targets can be found in Refs.~\cite{Kitano:2002mt,Arganda:2007jw}.
The effective couplings $g_{XK}^{(0,1)}$ $(X=L,R;\, K=S,V)$ are defined as
\begin{align}
    & g_{XK}^{(0)} = \frac{1}{2} \sum_{q=u,d,s} \left( g_{XK(q)} G_K^{q,p} + g_{XK(q)} G_K^{q,n} \right),~ g_{XK}^{(1)} = \frac{1}{2} \sum_{q=u,d,s}
\left( g_{XK(q)} G_K^{q,p} - g_{XK(q)} G_K^{q,n} \right), \nonumber
\end{align}
where the nucleon matrix elements $G_K^{q,p(n)}$ are taken from
Refs.~\cite{Kuno:1999jp, Kosmas:2001mv, Arganda:2007jw}. 
In the present model, the effective quark-level couplings are~\cite{Vicente:2014wga}
\begin{align}
& g_{LV(q)} = g_{LV(q)}^\gamma
= \frac{\sqrt{2}}{G_F} e^2 Q_q A_{ND}, \nonumber \\
& g_{RV(q)} = g_{LV(q)}^\gamma \big|_{L \leftrightarrow R}, ~~ g_{LS(q)} \simeq 0,~~ g_{RS(q)} \simeq 0, \nonumber
\end{align}
where $Q_q$ is the electric charge of the quark $q$. The quantities
$A_D$ and $A_{ND}$ are defined in Eqs.~\eqref{eq:AD} and~\eqref{eq:AND}, respectively. A more complete EFT treatment of coherent $\mu$--$e$ conversion, including improved nuclear and hadronic inputs, has recently been presented in Ref. \cite{Haxton:2024lyc}. In the present work, we employ the standard phenomenological formalism of Ref. \cite{Kitano:2002mt}, which is sufficient for the level of accuracy required in our numerical analysis. Finally, the current experimental bounds and projected future sensitivities for the cLFV processes considered in this work are summarized in table~\ref{tab:LFV}.
\begin{table}[htb]
\centering
\begin{tabular}{|c|c|c|}
\hline
    LFV Process & Present Bound & Future Sensitivity \\
    \hline \hline
    $\mu \rightarrow e \gamma$ & $1.5 \times 10^{-13}$ \cite{MEGII:2025gzr} & $6 \times 10^{-14}$ \cite{MEGII:2021fah} \\
    \hline
    $\mu \rightarrow 3e $ & $1.0 \times 10^{-12}$ \cite{SINDRUM:1987nra} & $\sim 10^{-16}$ \cite{Mu3e:2020gyw} \\
    \hline
    $\mu^- \rm{Au} \rightarrow e^- \rm{Au}$ & $7 \times 10^{-13}$ \cite{SINDRUMII:2006dvw} & -\\
    \hline
    $\mu^- \rm{Al} \rightarrow e^- \rm{Al}$ & - & $\sim 10^{-17}$ \cite{COMET:2025sdw} \\
    \hline
\end{tabular}
\caption{The experimental limits of present bounds as well as future sensitivity for cLFV processes. For our numerical analysis, we have considered the gold nucleus for the coherent $\mu$--$e$ conversion process, most stringent bound so far; however, for a future sensitivity projection of $\mu$--$e$ conversion, we have considered the Aluminium isotope, as only its future projection is available in the literature.}
\label{tab:LFV}
\end{table}

In our numerical analysis, the Yukawa couplings $y_1$ and $y_2$ are treated as real parameters, corresponding to a single generation of active neutrinos, without loss of generality. While a fully general treatment would require promoting $y_1$ and $y_2$ to $3\times 3$ matrices, thereby introducing additional flavor structure, such an extension does not qualitatively alter the conclusions of this study and is therefore not pursued here. We perform a comprehensive parameter scan over the relevant model parameters. The ranges adopted for the scan are summarized as follows:
\begin{center}
    $\lambda_{ij} \in [-4\pi,4\pi]$ with $i\neq j, i = \lbrace H, \phi, \sigma \rbrace, j = \lbrace \phi, \sigma, \eta_1, \eta_2 \rbrace$ \\
    $\mu_i, \mu_1^\prime \in [10^{-4}, 500]$ GeV with $i = \lbrace \phi, \eta_1, \eta_2, 2 \rbrace$,\\
    $v_\sigma \in [1, 10^3]$ TeV,~~ $M_\Psi \in [100, 10^5]$ GeV,~~ $\lbrace y_{1,2} \rbrace \in [10^{-6}, 1]$.
\end{center}
All scanned points are required to satisfy theoretical consistency conditions and the current upper bound on the radiatively generated light neutrino masses, $m_\nu^{\rm 1\text{-}loop} \lesssim \mathcal{O}(10^{-2})~\text{eV}$.

\begin{figure}[htb!]
\centering
\includegraphics[width=0.48\textwidth]{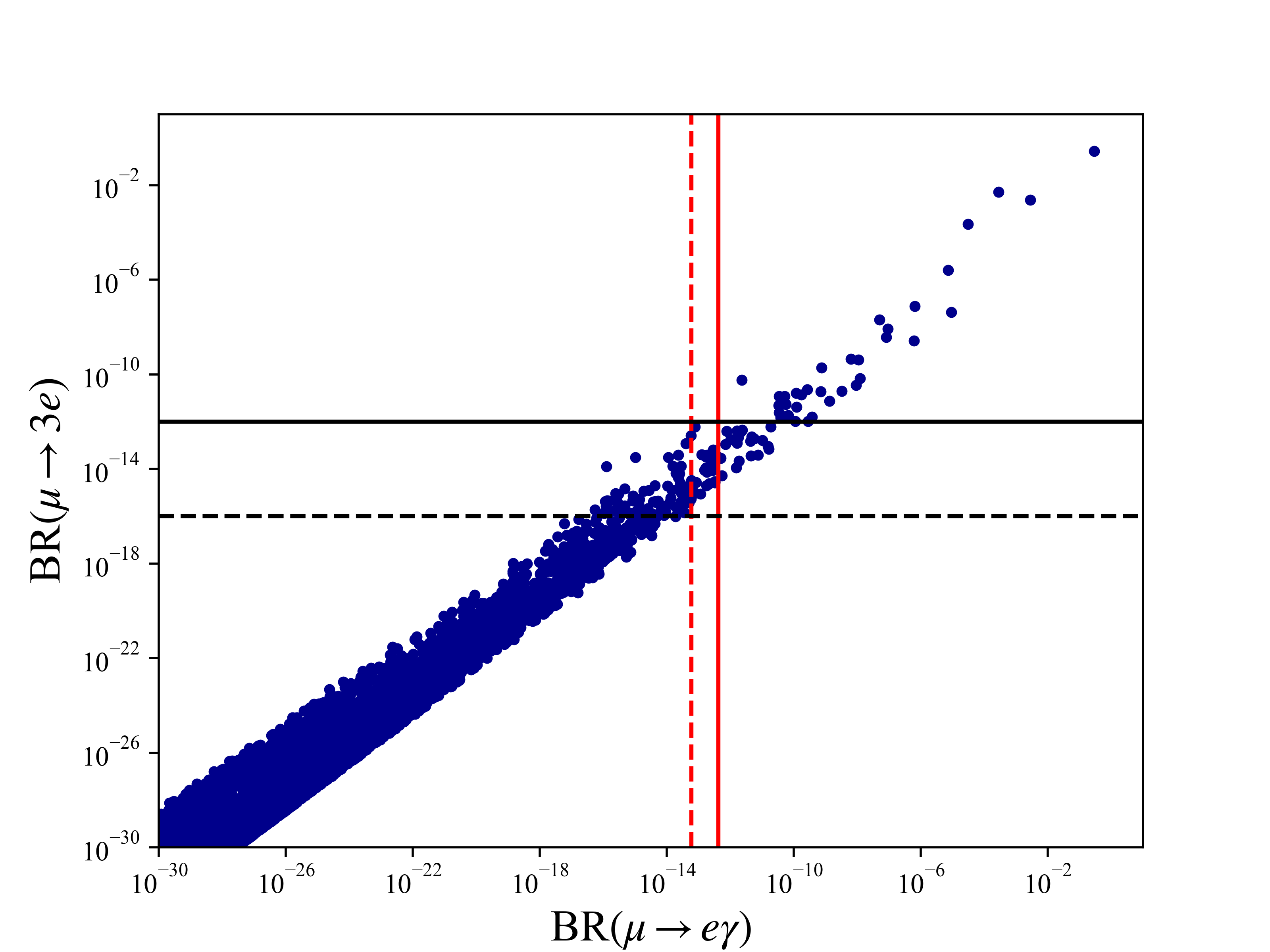}
\includegraphics[width=0.48\textwidth]{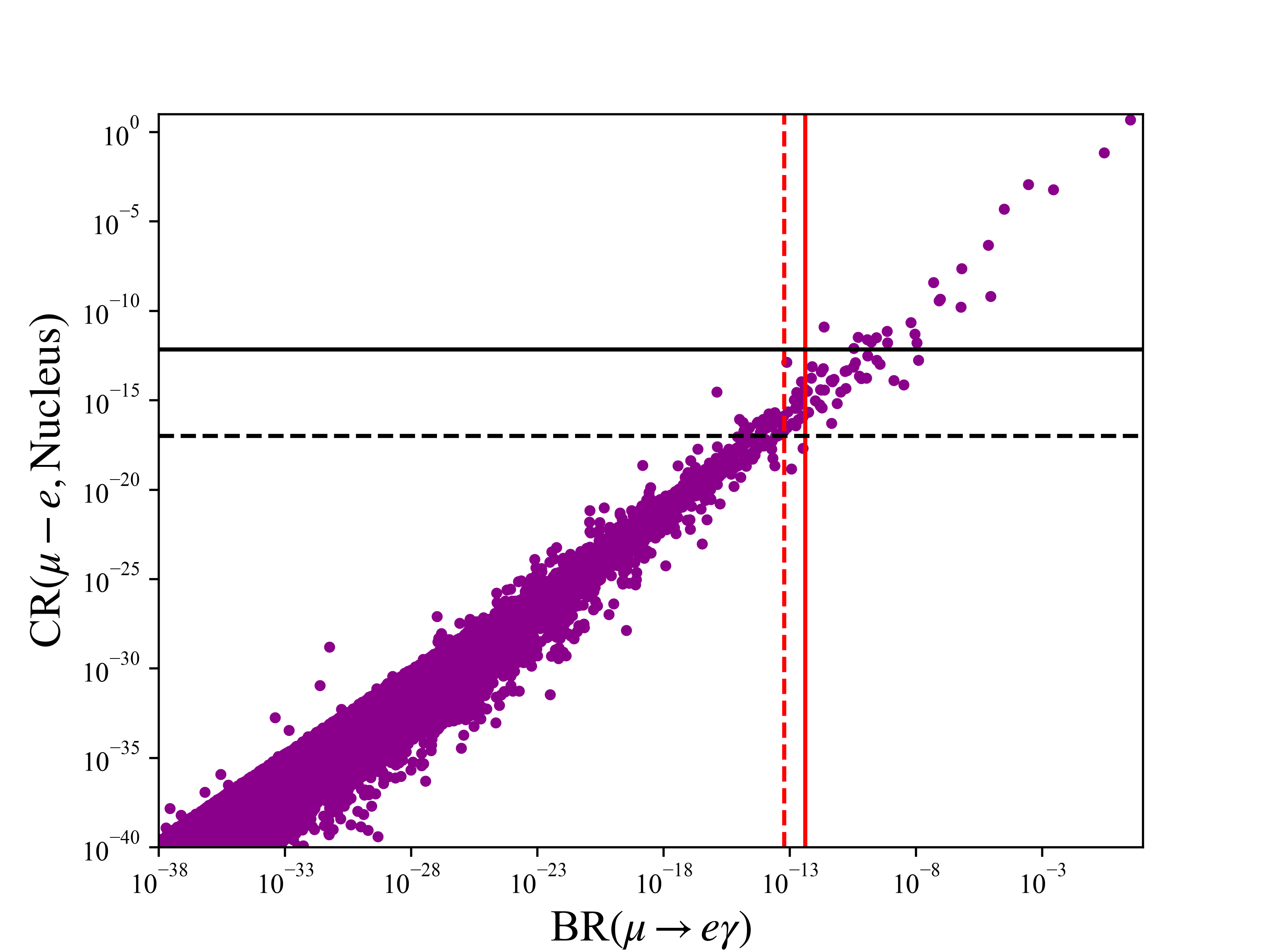}
\caption{Correlations among cLFV observables obtained from the parameter scan.
Left: $\text{BR}(\mu \to e\gamma)$ versus $\text{BR}(\mu \to 3e)$.
Right: $\text{BR}(\mu \to e\gamma)$ versus $\text{CR}(\mu\text{--}e,~\text{Nucleus})$ considering gold nucleus.
Solid (dashed) lines indicate current (projected) experimental sensitivities. In the right panel, there is no projected sensitivity available yet for gold isotope, so we have shown the future sensitivity for Aluminium isotope in this plane by black dashed line.}
\label{fig:LFV}
\end{figure}

The results of the scan are presented in figure~\ref{fig:LFV}. In the left panel, we display the allowed parameter space in the
$\text{BR}(\mu \to e\gamma)$ vs $\text{BR}(\mu \to 3e)$ plane in dark blue points, while the right panel shows the correlation between
$\text{BR}(\mu \to e\gamma)$ and the $\mu$--$e$ conversion rate in gold nucleus, $\text{CR}(\mu\text{--}e,\text{Au})$ using dark magenta points. All points shown satisfy the theoretical consistency conditions, such as perturbativity bounds on specific couplings and the neutrino mass bound quoted above. The red solid (dashed) line in both panels denotes the present (projected) experimental sensitivity on
$\text{BR}(\mu \to e\gamma)$, whereas the black solid (dashed) line indicate the corresponding current (projected) limit on $\text{BR}(\mu \to 3e)$ in the left panel while in the right panel we have denoted the current limit for gold ($^{192}_{\ 79}\mathrm{Au}$) nucleus on
$\text{CR}(\mu\text{--}e,\text{Au})$ by black solid line. As no future sensitivity has been reported in the literature for the gold isotope, we have shown here the projected sensitivity reach of the Aluminium isotope using a black dashed curve.

In the left panel, the nearly linear behaviour in the log-log representation suggests that $\mu \to 3e$ is strongly correlated with the radiative decay $\mu \to e \gamma$, as expected when dipole operators provide the dominant contribution. Most parameter points lie well below the present experimental limits, although a significant fraction enters the region accessible to upcoming experiments. The future sensitivity of $\mu \to 3e$ searches therefore probes a substantial portion of the parameter space already testable through improved $\mu \to e \gamma$ measurements. The right panel exhibits a similar correlation between $\mu \to e \gamma$ and $\mu$--$e$ conversion. The conversion rate spans a wider numerical range, reflecting additional dependence on nuclear matrix elements and non-dipole contributions.
These results highlight strong correlations among different cLFV observables arising from a common underlying mechanism for radiative neutrino mass generation. The interplay between the Yukawa couplings and the scalar mass spectrum simultaneously governs the magnitude of light neutrino masses and cLFV decay rates, leading to characteristic correlations in the corresponding observable planes. In particular, future improvements in the sensitivities to $\mu \to e\gamma$, $\mu \to 3e$, and $\mu$--$e$ conversion experiments are expected to probe a substantial region of the parameter space consistent with neutrino mass constraints, thereby providing a complementary and powerful test of the model.

\section{Dark Matter Phenomenology}
\label{sec:DM}
In this section, we investigate the dark matter phenomenology associated with our model. The presence of a residual $\mathcal{Z}_6$ symmetry after $U(1)_{B-L}$ breaking allows the framework to accommodate dark sector particles, whose stability is ensured by this discrete symmetry. Owing to the mass structure of the new particles, the model naturally accommodates the possibility of two distinct dark matter scenarios: one in which the dark matter candidate is a scalar field and another in which it is a Dirac fermion. We therefore perform a comprehensive study of both possibilities, treating each case independently to determine the corresponding viable parameter space consistent with the observed relic density. For the numerical analysis, we implement our model in the \texttt{SARAH}~\cite{Staub:2013tta} package tool, and perform evaluations with \texttt{SPheno}~\cite{Porod:2011nf} and \texttt{micrOMEGAs}~\cite{Belanger:2018ccd} and cross-verify our particle spectrum and their masses using \texttt{MATHEMATICA}~\cite{Mathematica}. The thermal relic abundance is computed and compared with the latest cosmological observations reported by the \texttt{Planck} collaboration~\cite{Planck:2018vyg} within $2\sigma$ uncertainty. Furthermore, we examine the constraints arising from direct detection experiments such as \texttt{XENONnT}~\cite{XENON:2025vwd}, \texttt{PandaX}~\cite{PandaX:2024qfu}, and \texttt{LZ}~\cite{LZ:2024zvo}, identifying the regions of parameter space that remain compatible with current experimental limits. We further impose constraints from electroweak precision observables, in particular the $\rho$ parameter~\cite{ParticleDataGroup:2024cfk}, which is sensitive to mass splittings within the electroweak scalar doublet $\phi$. Requiring consistency with experimental limits on $\Delta \rho$~\cite{Baak:2014ora} places an upper bound on the mass difference between the charged and neutral components of $\phi$, thereby restricting regions with large custodial symmetry breaking. In addition, bounds on charged Higgs bosons obtained at \texttt{LEP} within two-Higgs-doublet model frameworks~\cite{LEPChargedScalar} can be reinterpreted in our scenario, since the charged component $\phi^{\pm}$ shares the same electroweak gauge quantum numbers and hence exhibits similar production mechanisms. These searches impose a robust lower bound $m_{\phi^{\pm}} \gtrsim 80$--$100~\mathrm{GeV}$, with the precise limit depending on the decay modes. Following a conservative approach, we ensure that $m_{\phi^{\pm}} > 100~\mathrm{GeV}$ for our analysis, and that the mass splitting between the charged and neutral components satisfies $\Delta m (\phi^{+},\phi^0) \lesssim 20~\mathrm{GeV}$. While the LHC can provide stronger but model-dependent constraints on $m_{\phi^{\pm}}$, we do not consider these bounds in the present analysis, as a comprehensive and dedicated collider analysis is beyond the scope of this work and is left for future investigation. These constraints are explicitly shown in our parameter scans, and only points that satisfy electroweak precision tests and collider limits are retained for further scrutiny, further reducing the allowed parameter space.

We organize our DM analysis by treating the fermionic and scalar dark matter scenarios independently. For each possibility, we first outline the relevant input parameters, such as the couplings and mass hierarchies that determine the corresponding annihilation and co-annihilation behaviors. We then examine the dominant channels that govern the freeze-out processes and establish the resulting relic abundance. Finally, we discuss the obtained plots and numerical results, ensuring that the relevant input parameter choices, interaction channels, and phenomenological outcomes are discussed in a self-contained manner.
\subsection{Fermionic DM}
\label{subsec:fDM}
The dark sector of the model contains two generations of iso-singlet vector-like fermions $\Psi_{i}, \text{ with } i=1,2$ and three scalars $\eta_1, \eta_2, \phi$ which combine to form three scalar mass eigenstates $S_1, S_2, S_3$ as shown in Eq.~\eqref{eq:dScalarmass}. The lightest among these states becomes a viable DM candidate. In this subsection, we study the DM phenomenology for the scenario when fermion $\Psi_1$ is the DM candidate (i.e., when $M_{\Psi_1}< M_{\Psi_{2}}~\&~M_{S_i}$), with $i=1,2,3$.
\subsubsection{Dominant Channels}
\label{subsubsec:dc1}
For $\Psi_1$ as DM, the Feynman diagrams at tree level for all possible annihilation channels are depicted in figures~\ref{fig:DM1FD1}-\ref{fig:DM1FD4}, and for co-annihilations with $\Psi_{2}$ in figures~\ref{fig:DM1FD5},~\ref{fig:DM1FD6}, and for co-annihilations with $S_{1,2,3}$ in figures~\ref{fig:DM1FD7},~\ref{fig:DM1FD8}. The strength of these interactions (DM annihilations and co-annihilations) relative to the Hubble expansion rate in the early universe plays an important role in deciding the DM freeze-out temperature and its relic abundance. Also, it is worth noting that the $\Psi_1$ can annihilate to SM states via both $Z, Z'$ vector bosons and also via scalar interactions mediated via the new exotic scalars. Thus, the model parameters most relevant to $\Psi_1$ phenomenology are listed below: 
\begin{equation}
\, M_{\Psi_1},~~M_{Z'},~~g_{B-L},~~v_\sigma,~~\Delta M(\Psi_1,\Psi_2),~~\Delta M(\Psi_1,S_1),
\label{eq:parameters1}
\end{equation}
where $M_{\Psi_1}$ is the mass of fermion DM, $M_{Z'}$ is the $Z'$ boson mass, $g_{B-L}$ is the dark sector $U(1)_{B-L}$ coupling strength, $v_\sigma$ is the associated \textit{vev} of the scalar $\sigma$ for $U(1)_{B-L}$ breaking, $\Delta M(\Psi_1,\Psi_2)$ is the mass splitting of $\Psi_1$ with $\Psi_2$, and $\Delta M(\Psi_1,S_1)$ is the mass splitting between $\Psi_1$ and the lightest mass state~$(S_1)$ from the dark scalar sector. For the purpose of this subsection, the couplings in Eq.~\eqref{eq:mDScalars} are taken in such a way that the exotic doublet scalar $\phi$ completely represents $S_1$ in the mass basis (i.e. achieved by setting $A=B=0$, and following $\mu_\phi^2\ll\mu_{\eta_1}^2,\mu_{\eta_2}^2$ in Eq.~\eqref{eq:mDScalars}).

\subsubsection{Numerical results}
\label{subsubsec:nr1}
\label{subsubsec:domfDM}
\begin{figure}[htbp]
    \centering
    \includegraphics[width=0.8\textwidth]{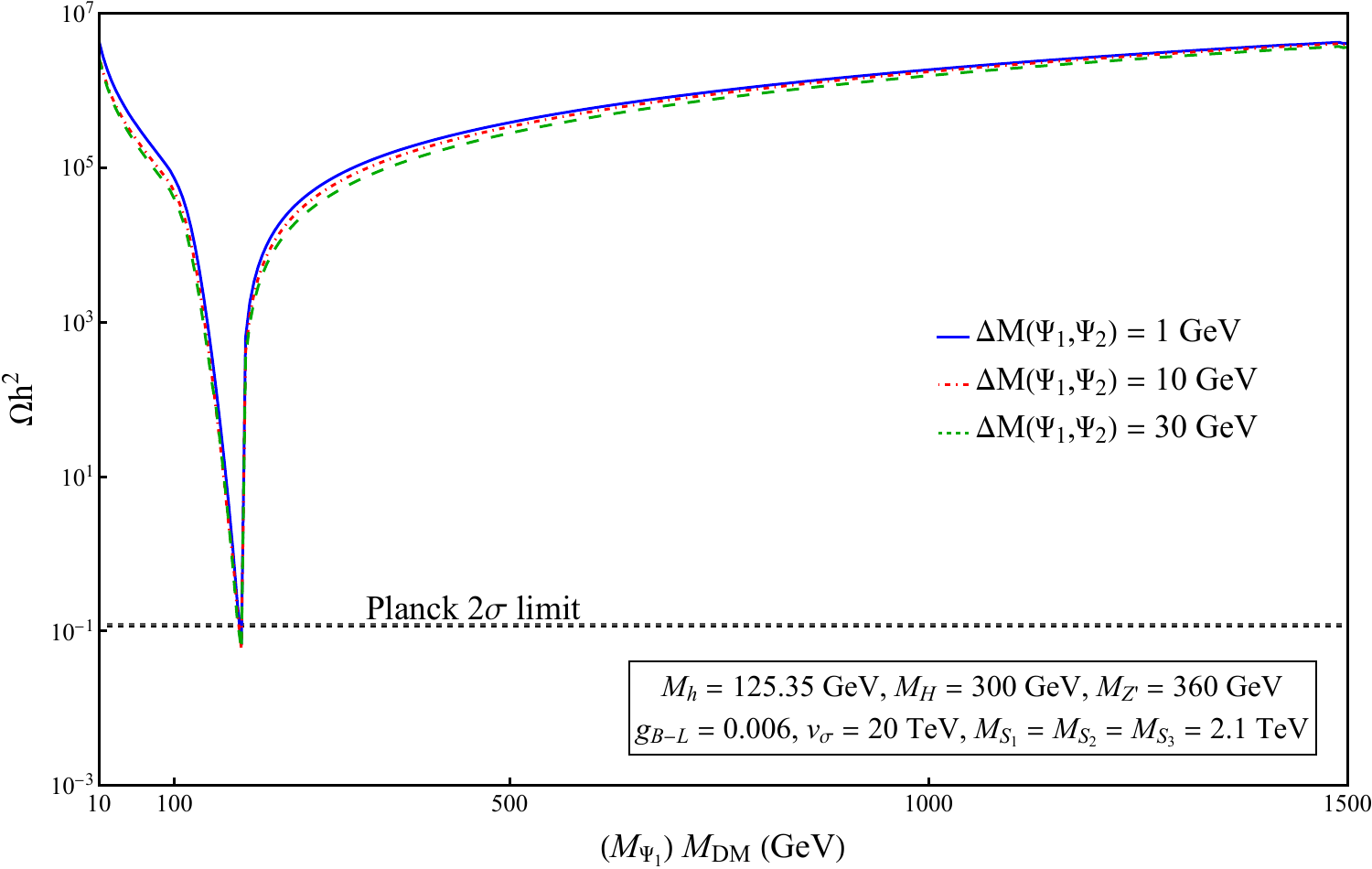}
    \caption{Fermionic dark matter relic density ($\Omega h^{2}$) as a function of the dark matter mass ($M_{\Psi_1}$). The different curves correspond to variations in the mass splitting $\Delta M(\Psi_{1},\Psi_{2})$: $1~\rm GeV$ (solid blue), $10~\rm GeV$ (dot-dashed red), and $30~\rm GeV$ (dashed green). The fixed parameters are $M_{h}=125.35~\rm GeV$, $M_{H}=300~\rm GeV$, $M_{Z^{\prime}}=360~\rm GeV$, $g_{B-L}=0.006$, $v_{\sigma}=20~\rm TeV$, and $M_{S_1}=M_{S_2}=M_{S_3}=2.1~\rm TeV$. The horizontal band denotes the \texttt{Planck} $2\sigma$ limit on the observed dark matter relic abundance.}
    \label{fig:relic_p1}
\end{figure}
In this subsection, we present our numerical scans of the fermionic DM relic density and the corresponding direct detection analysis. For the purpose of these scans, the masses of the dark sector scalars, $S_{1,2,3}$, are taken to be higher than the mass of $\Psi_1$. Firstly, in figure~\ref{fig:relic_p1}, we plot the DM relic density~$(\Omega h^2)$ as a function of DM mass for three different values of mass splitting between $\Psi_1$ and $\Psi_2$, with relatively smaller $\Delta M(\Psi_1,\Psi_2)=1~\rm GeV$, intermediate $\Delta M(\Psi_1,\Psi_2)=10~\rm GeV$ and larger $\Delta M(\Psi_1,\Psi_2)=30~\rm GeV$, as shown by the solid blue, dot-dashed red, and dashed green curves, respectively. We have kept other model parameters fixed as : $M_{h}=125.35~\rm GeV$ (SM-like Higgs boson), $M_{H}=300~\rm GeV$, $M_{Z^{\prime}}=320~\rm GeV$, $g_{B-L}=0.006$, $v_{\sigma}=20~\rm TeV$, and $M_{S_1}=M_{S_{2}}=M_{S_{3}}=2~\rm TeV$. The dotted black horizontal lines indicate the \texttt{Planck} $2\sigma$ bound~$(0.118\le\Omega h^2\le 0.122)$~\cite{Planck:2018vyg} for dark matter relic density. From the plot it is evident that, contrary to the usual expectation where increasing the mass splitting between the dark matter particle and its heavier partners suppresses co-annihilation and thus enhances the relic abundance, an opposite behavior is observed in this case. This behavior can be explained using the following reasoning: in the case of the lightest state $\Psi_{1}$ of our two-generation exotic fermion species, although the heavier state $\Psi_{2}$ share the same quantum numbers, its larger mass suppress its annihilation rate through reduced available kinematic phase space or weaker effective coupling after mass diagonalization. Thus, co-annihilations involving the heavier state dilute the overall annihilation efficiency during freeze-out. When $\Psi_{2}$ contributes significantly to the thermal number density but annihilates less efficiently than $\Psi_{1}$, its presence increases the effective degrees of freedom in the plasma without providing a proportionate enhancement to the total annihilation cross section. As a result, the effective thermally averaged cross section $\langle \sigma_{\text{eff}} v \rangle$ becomes smaller, and the obtained relic abundance increases with decreasing value of $\Delta M(\Psi_1,\Psi_2)$. Increasing the mass splitting then suppresses the population of the heavier state, thereby removing the dilution and \emph{reducing} the relic density, leading to a behavior opposite to the standard co-annihilation trend. Additionally, the annihilation cross section of $\Psi_2$ is similar to the DM annihilation cross section, but $\Psi_2$ being heavier, it cannot contribute sufficiently in freeze-out. This effect is typically modest in magnitude and, in realistic parameter regions, usually induces variations in the relic abundance at the level of a few orders of magnitude or less, as is evident from the plot here. Apart from this, a resonance funnel around DM mass of $M_{Z'}/2\sim160~\rm GeV$ is observed in the plot where the obtained DM relic for all three plot-lines is within the \texttt{Planck} $2\sigma$ requirements.
\begin{figure}[htbp]
    \centering
    \includegraphics[width=0.8\textwidth]{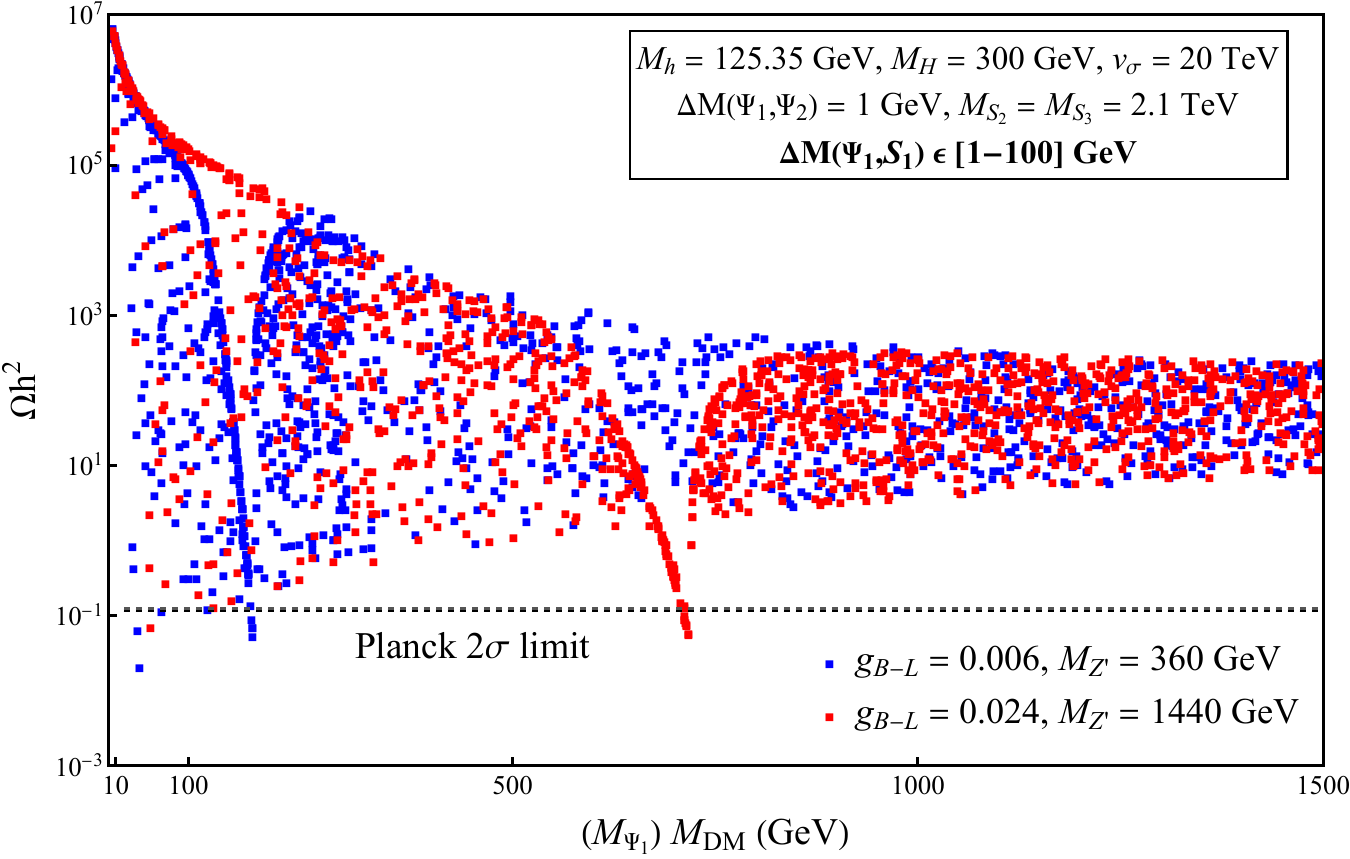}
    \caption{Fermionic dark matter relic density ($\Omega h^{2}$) as a function of the dark matter mass ($M_{\Psi_1}$). The blue (red) band of plot points corresponds to $g_{B-L}=0.006$ and $M_{Z^{\prime}}=360~\rm GeV$ ($g_{B-L}=0.024$ and $M_{Z^{\prime}}=1440~\rm GeV$). The width of each band arises from varying $\Delta M(\Psi_{1},S_{1})$ in the range $[1,100]~\rm GeV$. The other parameters are fixed at $M_{h}=125.35~\rm GeV$, $M_{H}=300~\rm GeV$, $v_{\sigma}=20~\rm TeV$, $\Delta M(\Psi_{1},\Psi_{2})=1~\rm GeV$, and $M_{S_{2}}=M_{S_{3}}=2.1~\rm TeV$. The horizontal band denotes the \texttt{Planck} $2\sigma$ limit on the observed dark matter relic abundance.}
    \label{fig:relic_p3}
\end{figure}

In figure~\ref{fig:relic_p3}, we plot the DM relic~$(\Omega h^2)$ as a function of fermionic DM mass for two different set of input parameters: $g_{B-L}=0.006$, $M_{Z'}=360~\rm GeV$ as shown by blue squares, and $g_{B-L}=0.024$, $M_{Z'}=1440~\rm GeV$ as represented by red squares. Both these parameter choices satisfy the constraint from LEP measurements on heavy neutral boson mass, $M_{Z^\prime}/g_{B-L} \gtrsim 3.59$ TeV~\cite{Electroweak:2003ram}. For both scenarios, we have kept the other model parameters fixed as: $M_{h}=125.35~\rm GeV$, $M_{H}=300~\rm GeV$, $v_{\sigma}=20~\rm TeV$, $\Delta M(\Psi_{1},\Psi_{2})=1~\rm GeV$, and $M_{S_2}=M_{S_{3}}=2.1~\rm TeV$. The black dotted lines indicate the \texttt{Planck} $2\sigma$ bound. The mass splitting $\Delta M(\Psi_1, S_1)$ is varied within a range of [1, 100] GeV in both these scenarios. This mass splitting drives the strength of co-annihilations of DM with $S_1$, and thus a band of final relic abundance is obtained. The final relic abundance decreases with decreasing mass splitting
$\Delta M(\Psi_1,S_1)$. In particular, when the splitting is reduced to ${\cal O}(1\,\mathrm{GeV})$, the resulting relic density is suppressed by several orders of magnitude compared to the case with a large splitting, $\Delta M(\Psi_1, S_1)\sim 100\,\mathrm{GeV}$. This dependence can be understood by the following argument: for larger mass splittings, the number density of $S_1$ is significantly Boltzmann suppressed during DM freeze-out, leading to a decrease in the effective co-annihilation cross-section involving $\Psi_1 S_1$ initial states, but such a Boltzmann suppression is not significant for smaller mass splittings. Consequently, varying $\Delta M(\Psi_1, S_1)$ within the range $\in (1,100)\,\mathrm{GeV}$ gives rise to band-like regions (blue \& red) of the obtained relic abundance, clearly reflecting the dependence of co-annihilation efficiency on the mass splitting. The plot also shows that the dependence of the final DM relic on these co-annihilations is stronger for lower dark matter masses. This behavior can be understood from the thermal population of the $S_1$ state at the time of freeze-out. The relative number density of the co-annihilating partner is governed by a Boltzmann suppression factor, $\exp(-\Delta M(\Psi_1, S_1)/T_f)$, where $ T_f\simeq M_{\Psi_1}/x_f$ is the freeze-out temperature with $ x_f\sim 20$–$30$. For lighter dark matter, the ratio $\Delta M(\Psi_1, S_1)/T_f$ is larger for a given mass splitting, implying a weaker Boltzmann suppression and hence a larger thermal abundance of the heavier state. As a result, co-annihilation processes involving the next-to-lightest state contribute more efficiently to the effective annihilation cross section, leading to a stronger depletion of the relic density. In contrast, for heavier dark matter masses, the co-annihilating partner becomes exponentially suppressed at freeze-out, rendering co-annihilation effects subdominant and leaving the relic abundance primarily determined by dark matter self-annihilation. Additionally, the resonance funnels around $M_{Z'}/2$ are evident for both blue and red plot regions. 
\begin{figure}[htbp]
    \centering
    \includegraphics[width=0.47\textwidth]{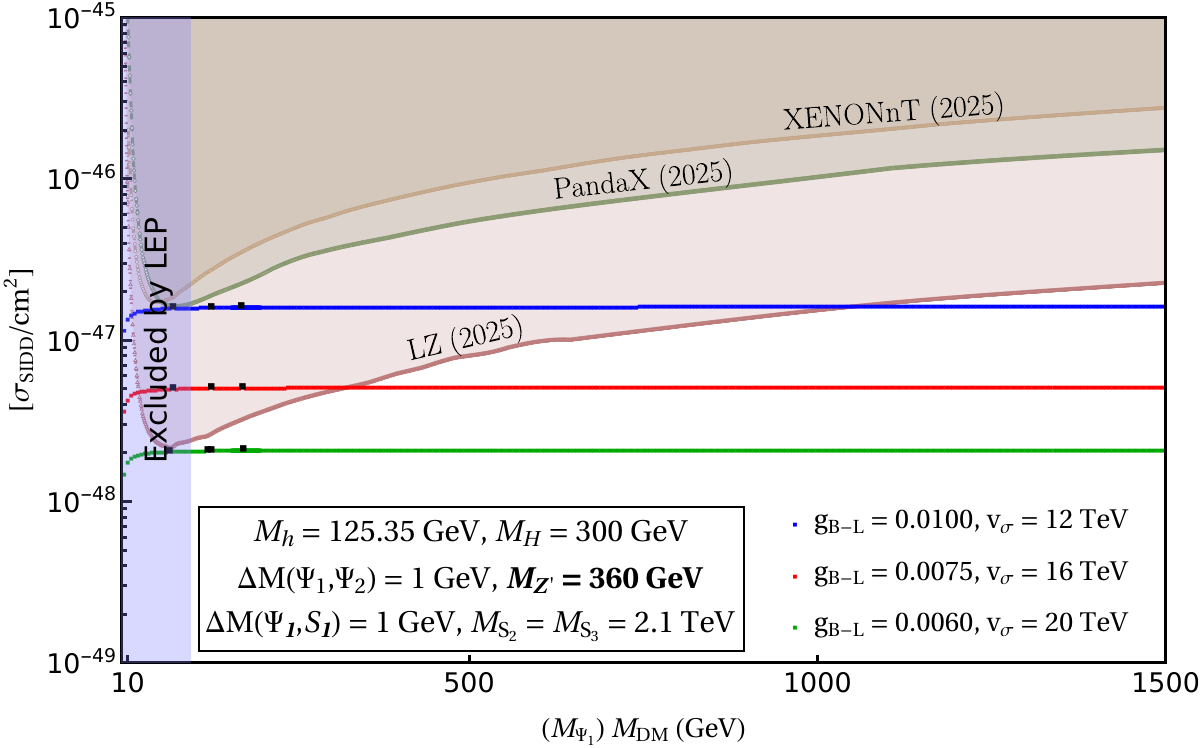}\qquad
    \includegraphics[width=0.47\textwidth]{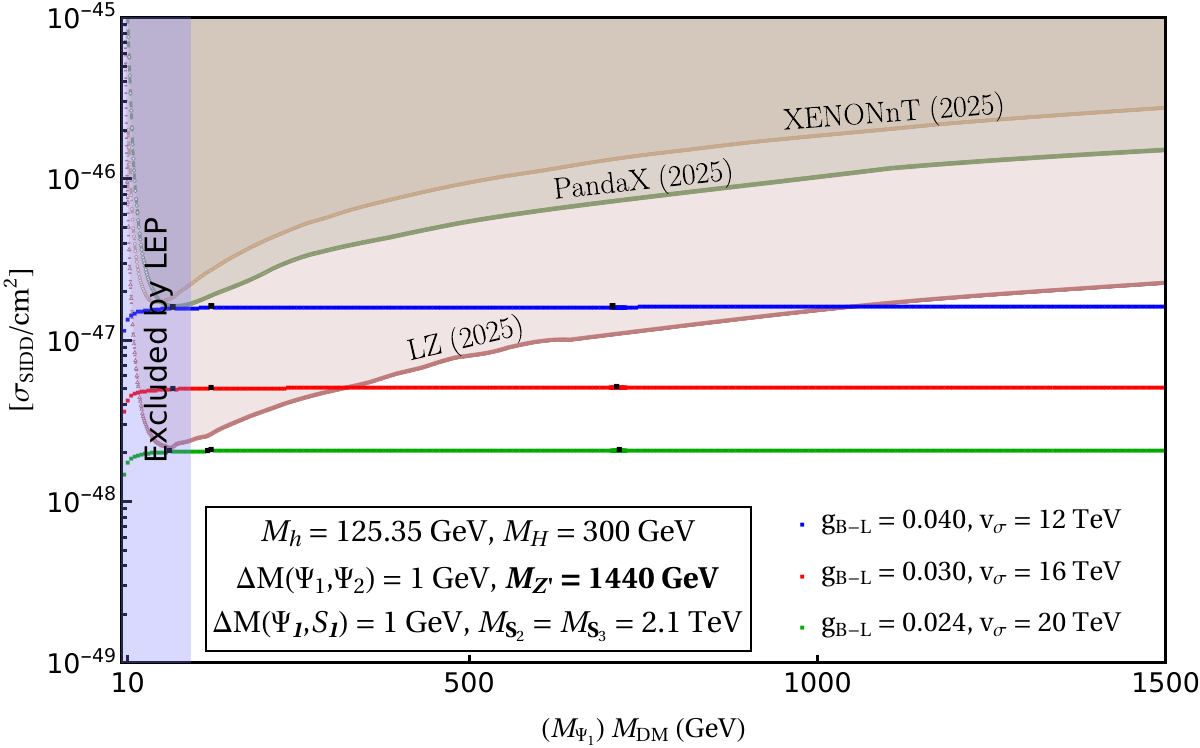}
    \caption{Spin-independent direct detection (SIDD) cross-section ($\sigma_{\rm SIDD}$) as a function of the fermionic dark matter mass ($M_{\Psi_1}$). The left (right) panel corresponds to $M_{Z^{\prime}} = 360~\mathrm{GeV}$ ($M_{Z^{\prime}} = 1440~\mathrm{GeV}$). In the left panel, the benchmark choices are $g_{B-L} = 0.01$, $v_{\sigma} = 12~\mathrm{TeV}$ (blue), $g_{B-L} = 0.0075$, $v_{\sigma} = 16~\mathrm{TeV}$ (red), and $g_{B-L} = 0.006$, $v_{\sigma} = 20~\mathrm{TeV}$ (green). The corresponding benchmarks in the right panel are $g_{B-L} = 0.04$, $v_{\sigma} = 12~\mathrm{TeV}$ (blue), $g_{B-L} = 0.03$, $v_{\sigma} = 16~\mathrm{TeV}$ (red), and $g_{B-L} = 0.024$, $v_{\sigma} = 20~\mathrm{TeV}$ (green). The remaining parameters are fixed to $M_{h} = 125.35~\mathrm{GeV}$, $M_{H} = 300~\mathrm{GeV}$, $\Delta M(\Psi_{1}, \Psi_{2}) = 1~\mathrm{GeV}$, $\Delta M(\Psi_{1}, S_{1}) = 1~\mathrm{GeV}$, and $M_{S_2} = M_{S_3} = 2.1~\mathrm{TeV}$. The plots also include the latest exclusion limits from \texttt{XENONnT} (2025), \texttt{PandaX} (2025), and \texttt{LZ} (2025), along with a conservative exclusion band (shown in blue) derived from reinterpreted \texttt{LEP} (2013) bounds on $\phi^{+}$. The black points indicate regions of parameter space consistent with the observed dark matter relic abundance within the $2\sigma$ range reported by \texttt{Planck}.}
    \label{fig:DD1}
\end{figure}

In figure~\ref{fig:DD1}, we show the spin-independent direct detection (SIDD) cross section, $\sigma_{\rm SIDD}$, as a function of the fermionic dark matter mass $M_{\Psi_1}$ for three representative benchmark choices of the $U(1)_{B-L}$ gauge coupling~$(g_{B-L})$ and symmetry-breaking scale~$(v_\sigma)$. The left panel corresponds to $M_{Z'}=360~$ GeV, where the blue line represents $g_{B-L}=0.01$ with $v_{\sigma}=12~\rm TeV$, the red corresponds to $g_{B-L}=0.0075$ with $v_{\sigma}=16~$TeV, and the green line denotes $g_{B-L}=0.006$ with $v_{\sigma}=20~$TeV, while the right panel corresponds to $M_{Z'}=1440~$ GeV, where the blue line represents $g_{B-L}=0.04$ with $v_{\sigma}=12~\rm TeV$, the red corresponds to $g_{B-L}=0.03$ with $v_{\sigma}=16~$TeV, and the green line denotes $g_{B-L}=0.024$ with $v_{\sigma}=20~$TeV. The remaining input parameters for both panels are fixed to $M_h=125.35~\rm GeV$, $M_H=300~\rm GeV$, $\Delta M(\Psi_1,\Psi_2)=1~\rm GeV$, $\Delta M(\Psi_1,S_1)=1~\rm GeV$, and $M_{S_2}=M_{S_3}=2.1~\rm TeV$. The obtained value of SIDD cross section is dominantly governed by $Z'$-mediated interactions and follows the scaling $\sigma_{\rm SIDD} \propto g_{B-L}^4/M_{Z'}^4 \simeq 1/v_{\sigma}^4$~\cite{Taramati:2024kkn}, leading to a clear separation among the benchmark curves. Furthermore, for the considered parameter space, $\sigma_{\rm SIDD}$ exhibits an almost flat dependence on the dark matter mass, reflecting its dominant sensitivity to the mediator mass and gauge coupling rather than to $M_{\Psi_1}$. Comparing with the latest bounds from \texttt{PandaX} and \texttt{LZ}, we find that a sizable region of parameter space remains allowed from the experimental results for the cases when $v_\sigma$ is set at 16 and 20 TeV. As an illustrative example, for the red benchmark line (in left panel) corresponding to $g_{B-L}=0.0075$ and $v_{\sigma}=16~\mathrm{TeV}$, the predicted SIDD cross section remains consistent with the current exclusion limits from \texttt{LZ (2025)} for $M_{\Psi_1} \gtrsim 300~\mathrm{GeV}$. The left portion of the plots shows a blue band depicting the parameter space up to $M_{\Psi_1}=101 \text{ GeV}$ ruled out by the reinterpreted \texttt{LEP} bounds on $\phi^+$ mass as discussed in section~\ref{sec:DM}. Furthermore, the black points indicate the points where the relic density simultaneously satisfies the $2\sigma$ constraint from \texttt{Planck} results, clustering around the $Z'$-resonance condition $M_{\Psi_1}\simeq M_{Z'}/2$. A similar trend is observed for the green benchmark with $g_{B-L}=0.006$ and $v_{\sigma}=20~\rm TeV$, where the SIDD cross section is fully allowed by the latest experiments and also remains compatible with relic density constraints near the resonant regime. The right panel displays the corresponding results for a heavier mediator mass, $M_{Z'}=1440~\rm GeV$, and thus we see relic allowed black points around $M_{\Psi_1}\simeq M_{Z'}/2\simeq720\text{ GeV}$. Notably, the SIDD cross section remains largely unchanged compared to the left panel, demonstrating its weak dependence on $M_{Z'}$ once the scaling with $v_{\sigma}$ is fixed. In contrast, the relic density shows a strong sensitivity to the $Z'$ mass, resulting in a visible shift of the black points toward higher dark matter masses, consistent with the resonance condition $M_{\Psi_1}\simeq M_{Z'}/2$. Other plot behaviors remain similar to those discussed for the left panel. All these benchmark choices saturate the latest \texttt{LEP} and LHC bounds on the mass of an exotic neutral gauge boson and the corresponding gauge couplings~\cite{Electroweak:2003ram,ATLAS:2018qto}. These plots highlight the complementary roles of direct detection and relic density constraints in probing different aspects of the fermionic dark matter parameter space.

\subsection{Scalar DM : scenario I}
\label{subsec:sDM}
In the dark scalar sector, the model consists of three scalar particles, $\eta_1$, $\eta_2$, and $\phi$, which mix to form the mass eigenstates $S_1$, $S_2$, and $S_3$, as shown in Eq.~\eqref{eq:dScalarmass}. The parameters that control the masses of $S_i$ (with $i=1,2,3$) are given in Eq.~\eqref{eq:mDScalars}. By carefully choosing the values of these parameters, the model allows for the possibility of a singlet scalar dark matter coming from $\eta_1$, and of a doublet scalar dark matter coming from $\phi$. In our analysis below, we study both possibilities separately. First, in this subsection, we study the case in which the mass eigenstate $S_1$ receives its dominant contribution from the gauge eigenstate $\eta_1$. We also choose these values such that $S_1$ is the lightest among all the scalar states and thus becomes a viable dark matter candidate. Additionally, we ensure that $M_{S_1} < M_{\Psi_i}$, with $i=1,2$. 
\subsubsection{Dominant Channels}
\label{subsubsec:dc2}
As we discussed in the subsection~\ref{subsec:fDM} for fermionic DM, similarly for $S_1$ as DM, the Feynman diagrams at tree level for all possible annihilation channels are depicted in figures~\ref{fig:DM2FD1}-\ref{fig:DM2FD4}, and for co-annihilations with $S_{2,3}$ in figures~\ref{fig:DM2FD5},~\ref{fig:DM2FD6}, and for co-annihilations with $\Psi_{1,2}$ in figures~\ref{fig:DM2FD7},~\ref{fig:DM2FD8}. $S_1$ can annihilate to SM states via both $Z$ and $Z'$ vector bosons, and also via scalar interactions mediated via the SM Higgs and the new exotic scalars. Thus, the most relevant model parameters with $S_1$ phenomenology are listed below: 
\begin{equation}
\, M_{S_1},~~M_{Z'},~~g_{B-L},~~v_\sigma,~~\Delta M(S_1,S_2),~~\Delta M(S_1,S_3),~~\mu_1,~~\mu_2,~~\mu_3,~~\lambda_m,
\label{eq:parameters2}
\end{equation}
where $M_{S_1}$ is the mass of scalar DM, $M_{Z'}$ is the $Z'$ mass, $g_{B-L}$ is the dark sector coupling strength, $v_\sigma$ is the associated \textit{vev} with $U(1)_{B-L}$ breaking, $\Delta M(S_1, S_2), \Delta M(S_1, S_3)$ are the mass splitting of $S_1$ with $S_2$ and $S_3$, respectively, and $\mu_1, \mu_2$, $\mu_3$, $\lambda_m$ are couplings associated with trilinear couplings between different scalars, respectively.
\subsubsection{Numerical Results}
\label{subsubsec:nr2}
\begin{figure}[htbp]
    \centering
    \includegraphics[width=0.8\textwidth]{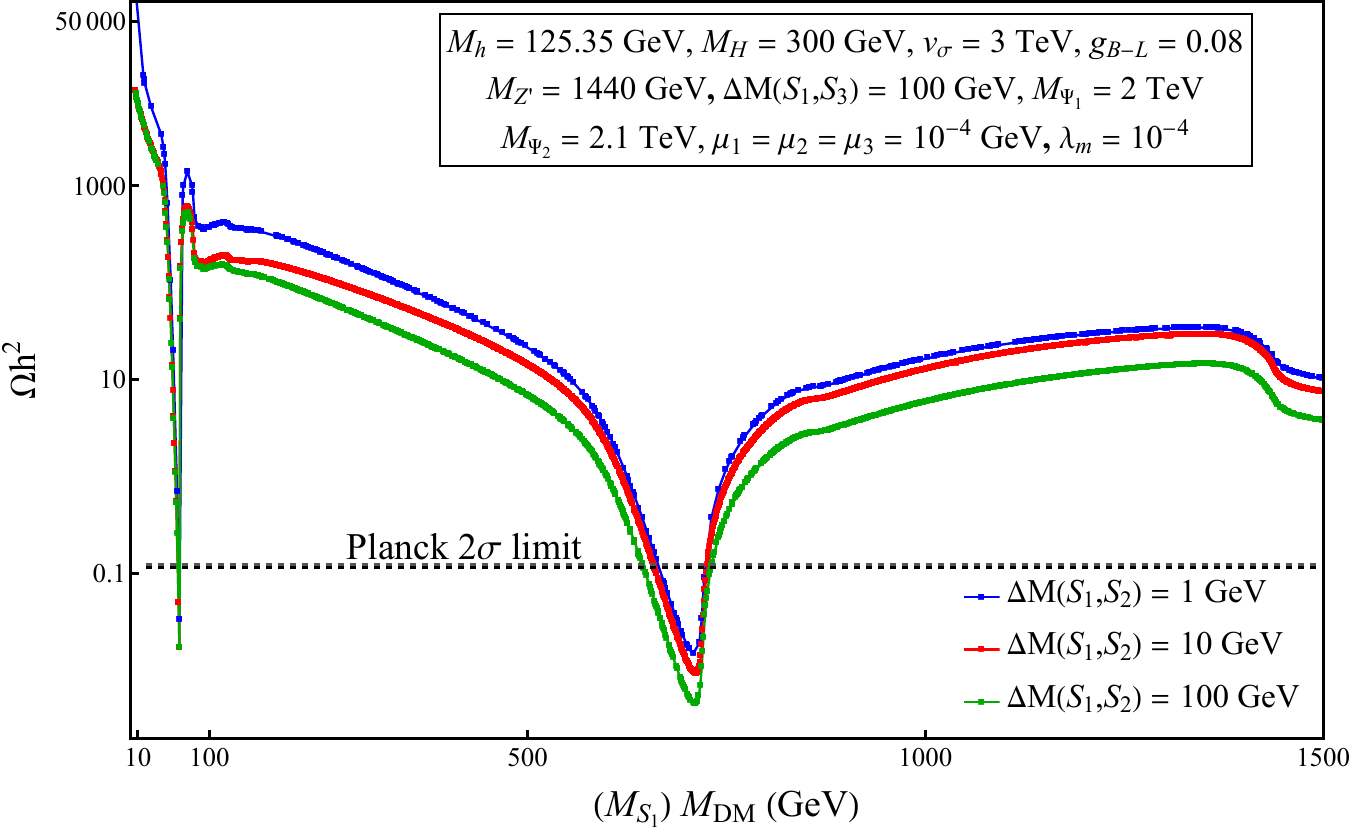}
   \caption{Scalar dark matter relic density ($\Omega h^{2}$) as a function of the dark matter mass ($M_{\mathrm{DM}}$) \textbf{with $\mathbf{S_1}$ state majorly coming from $\mathbf{\eta_1}$}, for three benchmark values of the scalar mass splitting $\Delta M(S_{1},S_{2})$: $1~\mathrm{GeV}$ (blue), $10~\mathrm{GeV}$ (red), and $100~\mathrm{GeV}$ (green). The model parameters are fixed to $M_{h} = 125.35~\mathrm{GeV}$, $M_{H} = 300~\mathrm{GeV}$, $v_{\sigma} = 3~\mathrm{TeV}$, $g_{B-L} = 0.08$, $M_{Z'} = 1440~\mathrm{GeV}$, and $\Delta M(S_{1},S_{3}) = 100~\mathrm{GeV}$. The fermion masses are taken as $M_{\Psi_{1}} = 2~\mathrm{TeV}$, $M_{\Psi_{2}} = 2.1~\mathrm{TeV}$. The scalar interaction parameters are fixed to $\mu_{1} = \mu_{2} = \mu_{3} = 10^{-4}~\mathrm{GeV}$ and $\lambda_{m} = 10^{-4}$. The horizontal line represents the $2\sigma$ relic density bound from \texttt{Planck}.}
    \label{fig:relic_sDM_p1}
\end{figure}
In this subsection, we discuss the numerical results obtained for the case in which the lightest scalar state $S_1$, mainly coming from the $\eta_1$ particle, constitutes dark matter. In figure~\ref{fig:relic_sDM_p1}, we show the scalar DM relic density $(\Omega h^2)$ as a function of the DM mass for three benchmark values of the scalar mass splitting $\Delta M(S_1, S_2)=1~\mathrm{GeV}$ (blue), $10~\mathrm{GeV}$ (red), and $100~\mathrm{GeV}$ (green). The fixed parameters for this scan are chosen as the SM-like and heavy scalar masses $M_{\mathrm{h}}=125.35~\mathrm{GeV}$ and $M_{H}=300~\mathrm{GeV}$, the $B\!-\!L$ sector parameters $v_{\sigma}=3~\mathrm{TeV}$, $g_{B-L}=0.08$, and $M_{Z'}=1440~\mathrm{GeV}$, the other scalar mass splitting $\Delta M(S_1, S_3)=100~\mathrm{GeV}$, and the iso-singlet vector-like fermions possess a hierarchical mass pattern, $M_{\Psi_{1}}=2~\mathrm{TeV}$, and $M_{\Psi_{2}}=2.1~\mathrm{TeV}$, with $\mu_1=\mu_{2}=10^{-4}$ GeV. The fermionic states $\Psi_i$ are taken to be heavier than $S_1$ and therefore do not participate in the freeze-out dynamics. As discussed earlier for the fermionic DM case, one might expect that increasing the mass splitting between the DM particle and its heavier partners suppresses co-annihilation effects and leads to an enhanced relic abundance. However, a qualitatively similar but opposite behavior is observed here as well. In the scalar sector, the heavier state $S_2$ (coming mostly from $\eta_2$ field) shares similar quantum numbers with the DM state $S_1$ (coming mostly from $\eta_1$ field) but exhibits reduced annihilation efficiency due to its larger mass and suppressed effective coupling after scalar mixing. Consequently, when this heavier scalar remains thermally populated during freeze-out, its co-annihilations dilute the overall annihilation efficiency for $S_1$. For smaller mass splittings, the heavier scalar states contribute appreciably to the effective number density while annihilating less efficiently than $S_1$, leading to a reduction in the effective thermally averaged cross section $\langle \sigma_{\rm eff} v \rangle$ and hence a larger relic abundance. As the mass splitting $\Delta M(S_1, S_2)$ increases, the thermal population of the heavier scalars becomes Boltzmann suppressed, thereby reducing their diluting effect and resulting in a decrease of the relic density. This explains the observed trend in figure~\ref{fig:relic_sDM_p1}, which is opposite to the standard expectation from co-annihilation-dominated scenarios. The overall variation in the relic abundance induced by varying the scalar mass splitting remains moderate, spanning a few orders of magnitude across the scanned parameter space, as evident in the figure. Finally, a resonance funnel around $M_{\rm DM} \simeq M_{Z'}/2 \sim 720~\mathrm{GeV}$ is clearly visible, where the relic density for all three benchmark mass splittings falls within the \texttt{Planck} $2\sigma$ allowed range.
\begin{figure}[htbp]
    \centering \includegraphics[width=0.8\textwidth]{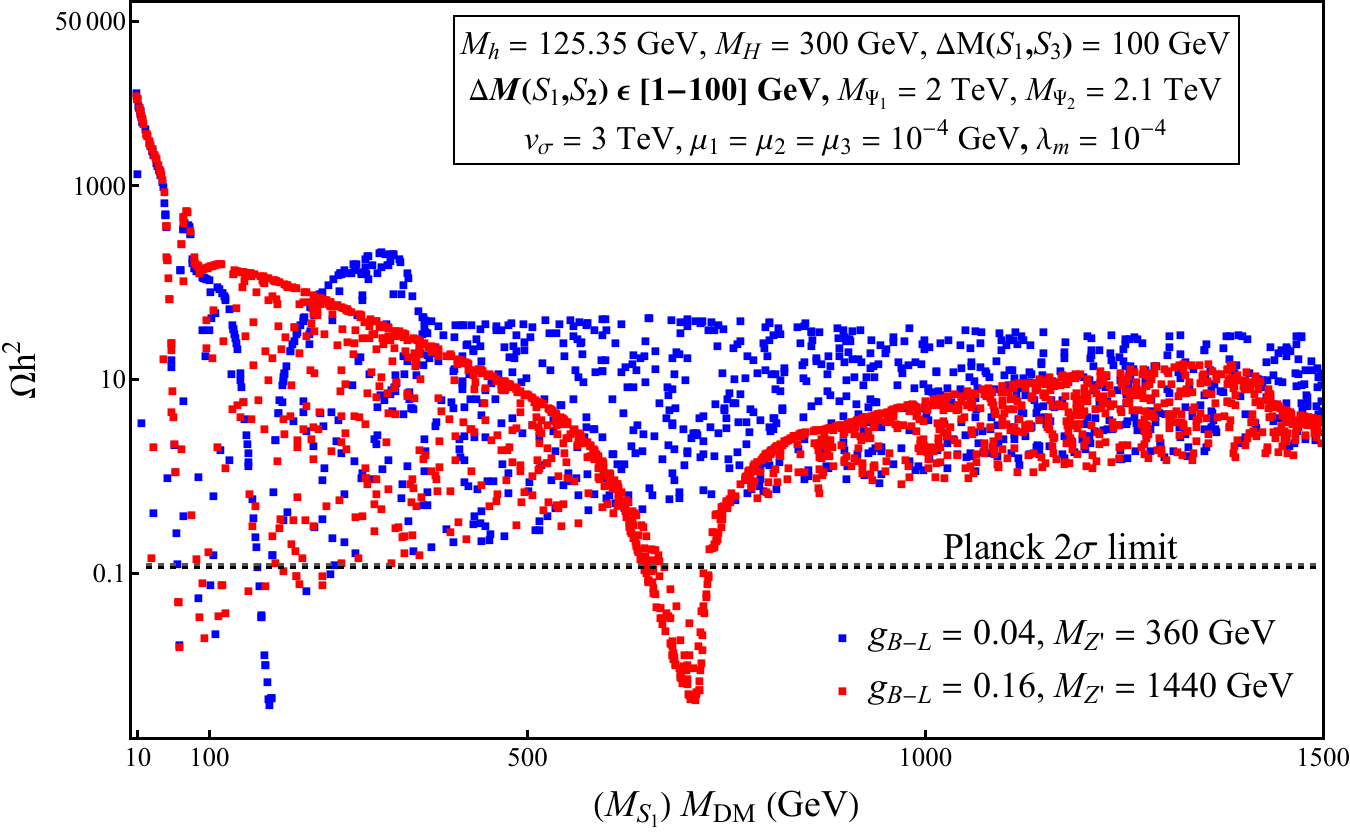}
    \caption{Scalar Dark Matter relic density ($\Omega h^{2}$) as a function of the Dark Matter mass ($M_{DM} = M_{S_1}$)\textbf{ with $\mathbf{S_1}$ state majorly coming from $\mathbf{\eta_1}$}. The blue squares correspond to $g_{B-L}=0.04$ and $M_{Z^{\prime}}=360~\mathrm{GeV}$, while the red squares correspond to $g_{B-L}=0.16$ and $M_{Z^{\prime}}=1440~\mathrm{GeV}$. The width of the bands arises from varying $\Delta M(S_{1},S_2)$ in the range $[1-100]~\mathrm{GeV}$. Other fixed parameters are $M_{\text{h}}=125.35~\mathrm{GeV}$, $M_{H}=300~\mathrm{GeV}$, $v_{\sigma}=3~\mathrm{TeV}$, $\Delta M(S_{1},S_{3})=100~\mathrm{GeV}$, $M_{\Psi_{1}}=2~\mathrm{TeV}$, $M_{\Psi_{2}}=2.1~\mathrm{TeV}$, $\mu_{1}=\mu_{2}=\mu_{3}=10^{-4}$ GeV and $\lambda_{m} = 10^{-4}$. The horizontal line indicates the \texttt{Planck} $2\sigma$ bound on the relic density.}
    \label{fig:relic_sDM_p3}
\end{figure}

In figure~\ref{fig:relic_sDM_p3}, we present the scalar DM relic density $(\Omega h^2)$ as a function of the dark matter mass for two different choices of the $B\!-\!L$ gauge coupling and the $Z'$ mass. The blue squares correspond to $g_{B-L}=0.04$ with $M_{Z'}=360~\mathrm{GeV}$, while the red squares represent the scenario : $g_{B-L}=0.16$ with $M_{Z'}=1440~\mathrm{GeV}$. The horizontal dashed black line indicates the \texttt{Planck} $2\sigma$ bound on the relic density. The width of each band arises from varying the scalar mass splitting $\Delta M(S_1, S_2)$ in the range $1$--$100~\mathrm{GeV}$, while keeping the remaining parameters fixed as: $M_{\text{h}}=125.35~\mathrm{GeV}$, $M_{H}=300~\mathrm{GeV}$, $v_{\sigma}=3~\mathrm{TeV}$, $\Delta M(S_{1}, S_{3})=100~\mathrm{GeV}$, $M_{\Psi_{1}}=2~\mathrm{TeV}$, $M_{\Psi_{2}}=2.1~\mathrm{TeV}$, $\mu_{1}=\mu_{2}=\mu_3=10^{-4}$ GeV and $\lambda_m = 10^{-4}$. In analogy to the fermionic DM case discussed earlier, the mass splitting $\Delta M(S_1, S_2)$ controls the strength of co-annihilation processes involving the next-to-lightest scalar state, $S_2$ (which has its major contributions from the doublet field, $\phi$, \footnote{
The notation $S_{1,2,3}$ denotes a mass-ordered basis rather than fixed field labels. The indexing is determined by the mass hierarchy of the physical states, which in turn is governed by the corresponding mass terms and coupling strengths in the model. Accordingly, \texttt{micrOMEGAs} assigns the index $1$ to the lightest state, $2$ to the next-to-lightest, and so on. In the present analysis, the state labeled as $S_2$ corresponds to the $\phi$ field, whereas in the earlier discussion related to figure~\ref{fig:relic_sDM_p1}, the same index was associated with the $\eta_2$ field. Therefore, the numerical labels $S_i$ are not fixed identifiers of specific fields but may change depending on the underlying mass ordering of the spectrum.}). As a result, scanning over this mass splitting leads to a band of final relic abundance for each benchmark choice of $(g_{B-L}, M_{Z'})$. The relic density decreases with decreasing $\Delta M(S_1, S_2)$ due to the enhancement of co-annihilation cross sections involving $\phi$, which increases the effective thermally averaged annihilation cross section and leads to a more efficient depletion of the DM abundance during freeze-out. It is also evident from the figure that the impact of scalar co-annihilations is more pronounced for lighter dark matter masses, as reflected by the broader bands towards the lower DM mass region, while the bands become comparatively narrower for heavier DM masses. This behavior can be understood from the Boltzmann suppression of the co-annihilating partner at freeze-out, governed by the factor $\exp[-\Delta M(S_1, S_2)/T_f]$, where $T_f \simeq M_{\rm DM}/x_f$ with $x_f \sim 20$--$30$. For lighter DM, the larger ratio $\Delta M(S_1, S_2)/T_f$ implies a larger thermal population of the heavier scalar, thereby enhancing co-annihilation effects. In contrast, for heavier DM masses, the co-annihilation channel becomes increasingly suppressed, rendering it subdominant and leaving the relic abundance primarily controlled by DM self-annihilation. Finally, resonance funnels around $M_{\rm DM} \simeq M_{Z'}/2$ are clearly visible for both benchmark choices, where the relic density falls within the \texttt{Planck} $2\sigma$ allowed range.
\begin{figure}[htbp]
    \centering
    \includegraphics[width=0.47\textwidth]{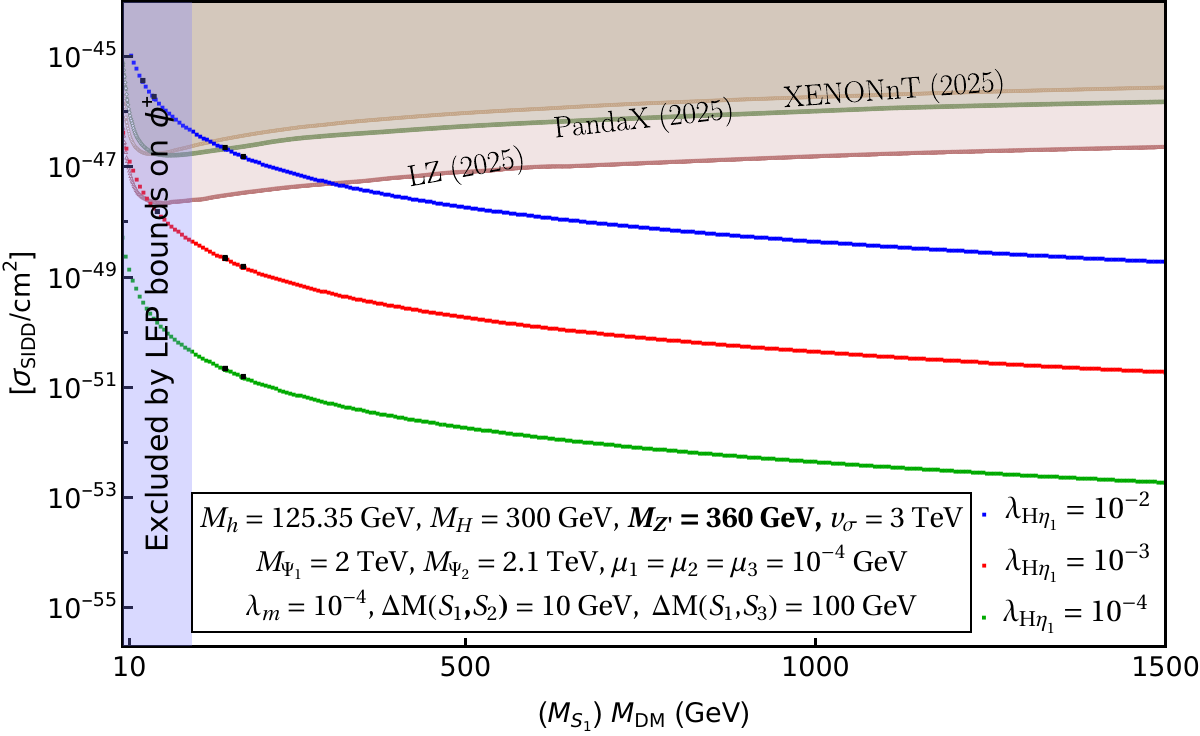}\qquad
    \includegraphics[width=0.47\textwidth]{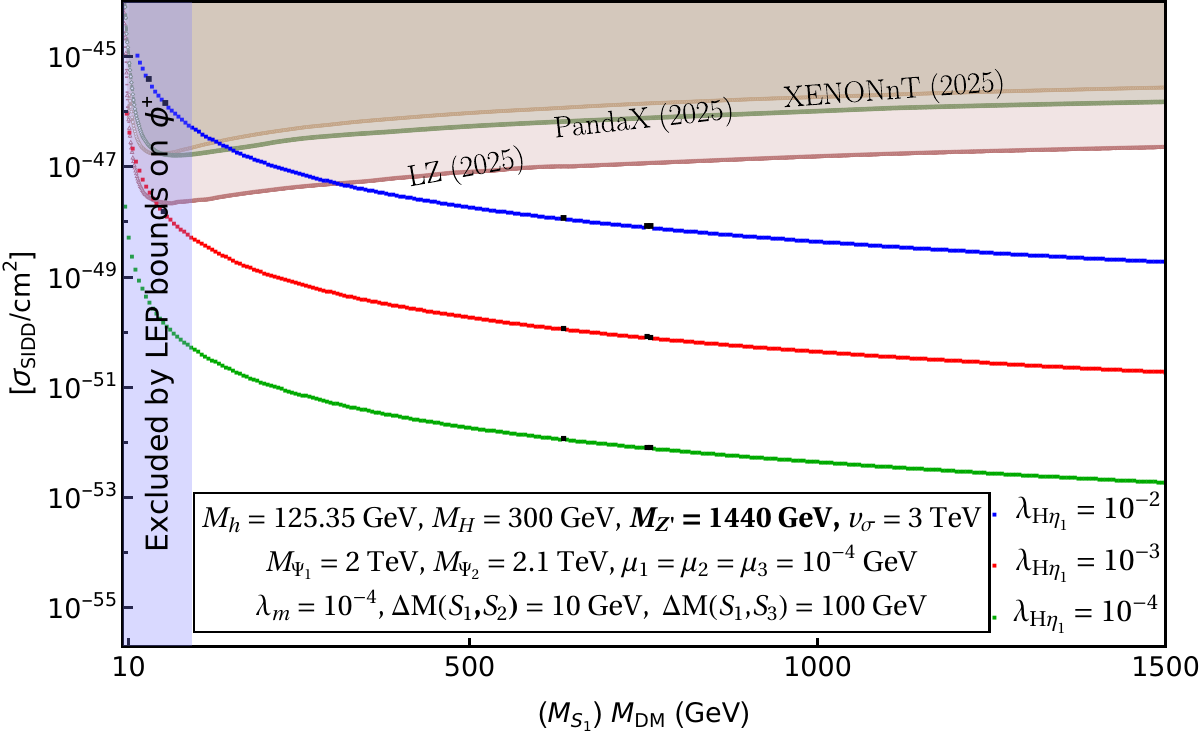}
    \caption{Spin-Independent Direct Detection (SIDD) cross-section ($\sigma_{\rm SIDD}$) as a function of the Scalar Dark Matter mass ($M_{\rm DM}$),\textbf{ with $\mathbf{S_1}$ state majorly coming from $\mathbf{\eta_1}$}, for two different values of $M_{Z'}$ with $M_{Z'}=360~\rm GeV$ in the left panel and $M_{Z'}=1440~\rm GeV$ in the right panel. The solid lines in both panels correspond to different choices of the Higgs portal coupling $\lambda_{H\eta_{1}}$: $\lambda_{H\eta_{1}}=10^{-2}$ (blue), $10^{-3}$ (red), and $10^{-4}$ (green). The fixed parameters are $M_{\text{h}}=125.35~\mathrm{GeV}$, $M_{H}=300~\mathrm{GeV}$, $v_{\sigma}=3~\mathrm{TeV}$, $M_{\Psi_{1}}=2~\mathrm{TeV}$, $M_{\Psi_{2}}=2.1~\mathrm{TeV}$, $\Delta M(S_{1},S_2)=10~\mathrm{GeV}$, $\Delta M(S_{1},S_{3})=100~\mathrm{GeV}$, $\mu_{1}=\mu_{2}=\mu_{3}=10^{-4}$ GeV and $\lambda_{m} = 10^{-4}$. The plots also include the latest exclusion limits from \texttt{XENONnT} (2025), \texttt{PandaX} (2025), and \texttt{LZ} (2025), along with a conservative exclusion band (shown in blue) derived from reinterpreted \texttt{LEP} (2013) bounds on $\phi^{+}$. The black points denote the regions of parameter space where the obtained relic is within $2\sigma$ range of \texttt{Planck} results.}
    \label{fig:DD2}
\end{figure}

In figure~\ref{fig:DD2}, we present the SIDD cross section, $\sigma_{\rm SIDD}$, as a function of the scalar dark matter mass $M_{\rm DM} \equiv M_{S_1}$ for two representative values of the mediator mass $M_{Z'}$. The left panel corresponds to $M_{Z'}=360~\mathrm{GeV}$, while the right panel shows the results for $M_{Z'}=1440~\mathrm{GeV}$. In both panels, the solid curves correspond to three benchmark choices of the Higgs portal coupling, $\lambda_{H\eta_1}=10^{-2}$ (blue), $10^{-3}$ (red), and $10^{-4}$ (green), with the remaining parameters are fixed as: $M_{\text{h}}=125.35~\mathrm{GeV}$, $M_{H}=300~\mathrm{GeV}$, $v_{\sigma}=3~\mathrm{TeV}$, $M_{\Psi_{1}}=2~\mathrm{TeV}$, $M_{\Psi_{2}}=2.1~\mathrm{TeV}$, $\Delta M(S_{1}, S_2)=10~\mathrm{GeV}$, $\Delta M(S_{1}, S_{3})=100~\mathrm{GeV}$, $\mu_{1}=\mu_{2}=\mu_3=10^{-4}$ GeV and $\lambda_m = 10^{-4}$. The shaded regions denote the current exclusion limits from \texttt{XENONnT} (2025), \texttt{PandaX} (2025), and \texttt{LZ} (2025). The left portion of the plots shows a blue band depicting the parameter space up to $M_{S_1}=110 \text{ GeV}$ ruled out by the reinterpreted \texttt{LEP} bounds on $\phi^+$ mass. The black points highlight regions of parameter space where the scalar relic density simultaneously satisfies the $2\sigma$ constraint from \texttt{Planck}. In contrast to the fermionic DM case, where the SIDD cross section is dominantly governed by $Z'$-mediated interactions, the scalar DM–nucleon scattering here is primarily controlled by Higgs-mediated processes through the portal coupling $\lambda_{H\eta_1}$. As a result, the predicted SIDD cross section exhibits a strong and monotonic dependence on $\lambda_{H\eta_1}$, with larger values of the portal coupling enhancing the effective scalar–nucleon interaction strength and leading to a clear separation among the benchmark curves. For a fixed value of $\lambda_{H\eta_1}$, the SIDD cross section is observed to decrease gradually with increasing dark matter mass. This behavior can be understood from the structure of the scalar mass, where the physical mass of $S_1$ receives contributions both from the Higgs portal term and from the bare mass parameter of the $\eta_1$ field~(refer Eq.~\eqref{eq:mDScalars}). As the dark matter mass is increased while keeping $\lambda_{H\eta_1}$ fixed, the bare mass term becomes increasingly dominant, effectively reducing the relative contribution of the Higgs portal interaction to the DM mass eigenstate. Consequently, the Higgs-mediated coupling to nucleons becomes less efficient, leading to a suppression of the SIDD cross section at higher dark matter masses. Comparing the two panels, we find that the SIDD cross section remains essentially unchanged for a given choice of $\lambda_{H\eta_1}$, demonstrating its weak sensitivity to the $Z'$ mass. This is expected since the dominant contribution arises from Higgs exchange rather than from $Z'$-mediated interactions. On the other hand, the relic density constraint shows a pronounced dependence on $M_{Z'}$, resulting in a visible shift of the black points toward higher dark matter masses in the right panel, consistent with the resonance condition $M_{\rm DM} \simeq M_{Z'}/2$. These results highlight the complementary roles of Higgs-portal direct detection and $Z'$-driven relic density constraints in probing the scalar dark matter parameter space.
\subsection{Scalar DM : scenario II}
\label{subsec:sDM2}
As discussed in sub-section~\ref{subsec:sDM}, the model consists of three scalar particles, $\eta_1$, $\eta_2$, and $\phi$, which mix to form the mass eigenstates $S_1$, $S_2$, and $S_3$, as shown in Eq.~\eqref{eq:dScalarmass}. In sub-section~\ref{subsec:sDM}, we studied the case of $\eta_1$ as $S_1$. Here, we study an alternate scenario where the mass eigenstate $S_1$ receives its dominant contribution from the gauge eigenstate $\phi$. We also choose the relevant parameter values such that $S_1$ is the lightest among all the scalar states and thus becomes a viable dark matter candidate. Additionally, we ensure that $M_{S_1} < M_{\Psi_i}$, with $i=1,2$. 
\subsubsection{Dominant Channels}
\label{subsubsec:dc3}
The most relevant model parameters with $S_1$ phenomenology are listed below: 
\begin{equation}
\, M_{S_1},~~M_{Z'},~~g_{B-L},~~v_\sigma,~~\Delta M(S_1,S_2),~~\Delta M(S_1,\Psi_1),~~\mu_1,~~\mu_2,~~\mu_3,~~\lambda_m,
\label{eq:parameters3}
\end{equation}
where $M_{S_1}$ is the mass of scalar DM, $M_{Z'}$ is the $Z'$ mass, $g_{B-L}$ is the dark sector coupling strength, $v_\sigma$ is the associated \textit{vev} with $U(1)_{B-L}$ breaking, $\Delta M(S_1, S_2), \Delta M(S_1, \Psi_1)$ are the mass splitting of $S_1$ with $S_2$ and $\Psi_1$, respectively, and $\mu_1, \mu_2$, $\mu_3$, $\lambda_m$ are couplings associated with trilinear couplings between different scalars, respectively.
\subsubsection{Numerical Results}
\label{subsubsec:nr3}
\begin{figure}[htbp]
    \centering
    \includegraphics[width=0.8\textwidth]{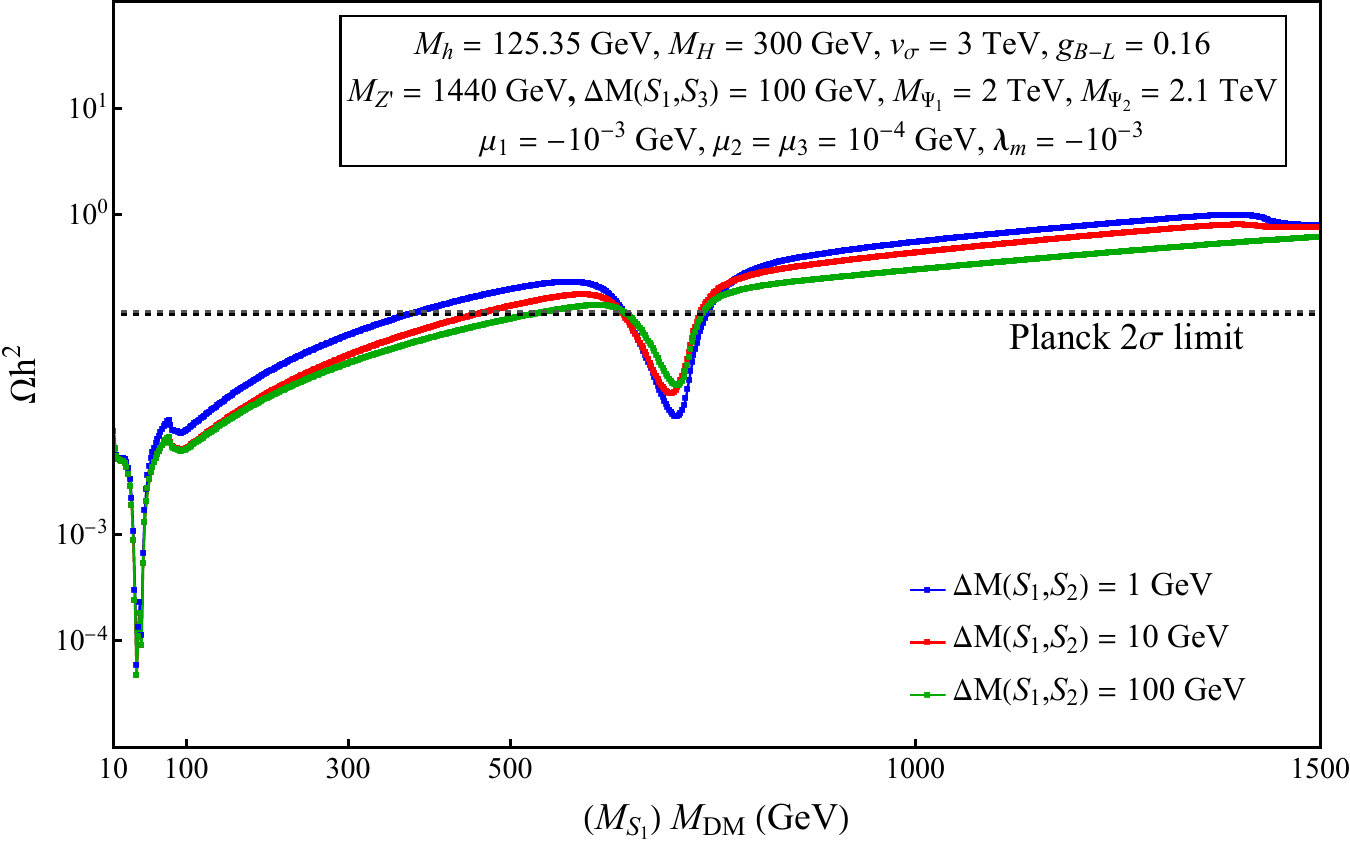}
    \caption{Scalar Dark Matter relic density ($\Omega h^{2}$) as a function of the Dark Matter mass ($M_{DM} = M_{S_1}$) \textbf{with $\mathbf{S_1}$ state majorly coming from $\phi$}, for three benchmark values of the scalar mass splitting $\Delta M(S_{1},S_{2})$: $1~\text{GeV}$ (blue), $10~\text{GeV}$ (red), and $100~\text{GeV}$ (green). The fixed parameters are $M_{\text{h}} = 125.35~\text{GeV}$, $M_{H} = 300~\text{GeV}$, $v_{\sigma} = 3~\text{TeV}$, $g_{B-L} = 0.16$, $M_{Z'} = 1440~\text{GeV}$, $\Delta M(S_{1},S_3) = 100~\text{GeV}$, and fermion masses $M_{\Psi_{1}} = 2~\text{TeV}$, $M_{\Psi_{2}} = 2.1~\text{TeV}$, with parameters $\mu_2 = \mu_{3} = 10^{-4}$ GeV, $\mu_1  = -10^{-3}$ GeV and $\lambda_m = -10^{-3}$. The horizontal line indicates the \texttt{Planck} $2\sigma$ limit.}
    \label{fig:relic_sDM_phi_p1}
\end{figure}
In this subsection, we discuss the numerical results obtained for the case in which the lightest scalar state $S_1$, originating from the neutral component of the scalar doublet $\phi$, constitutes dark matter. In figure~\ref{fig:relic_sDM_phi_p1}, we plot the scalar DM relic density $(\Omega h^2)$ as a function of the DM mass for three benchmark values of the scalar mass splitting $\Delta M(S_1, S_2)=1~\mathrm{GeV}$ (blue), $10~\mathrm{GeV}$ (red), and $100~\mathrm{GeV}$ (green). The fixed parameters for this scan are chosen as: $M_{\mathrm{h}}=125.35~\mathrm{GeV}$ and $M_{H}=300~\mathrm{GeV}$, the $B\!-\!L$ sector parameters $v_{\sigma}=3~\mathrm{TeV}$, $g_{B-L}=0.16$, and $M_{Z'}=1440~\mathrm{GeV}$, the other scalar mass splitting $\Delta M(S_1, S_3)=100~\mathrm{GeV}$, and the vector-like fermion $\Psi$ possess a hierarchy as, $M_{\Psi_{1}}=2~\mathrm{TeV}$, $M_{\Psi_{2}}=2.1~\mathrm{TeV}$, with $\mu_{2}=\mu_{3}=10^{-4}$ GeV, $\mu_{1}=-10^{-3}$ GeV and $\lambda_{m}=10^{-3}$~\footnote{From Eqs.~\eqref{eq:mDScalars} and \eqref{eq:charSca}, it can be seen that the mass splitting between the neutral and charged components of the scalar doublet $\phi$ depends sensitively on the parameters $\mu_1$ and $\lambda_m$. In particular, appropriate choices of these couplings allow one to control the splitting such that it remains within the limits imposed by electroweak precision tests, as discussed in section~\ref{sec:DM}. For the parameter space considered here, we ensure that the mass difference between the charged and neutral components remains sufficiently small~$\Delta m (\phi^{+},\phi^0) \lesssim 20~\mathrm{GeV}$ to satisfy these bounds, while simultaneously choosing sign of $\mu_1$ and $\lambda_m$ to be negative so that the neutral component is lighter than the charged one. This hierarchy ensures that the state $S_1$, originating from $\phi^0$, remains the lightest state in the exotic scalar sector and hence a viable dark matter candidate.}. For this choice of parameters, the scalar mixing is arranged such that the state $S_2$ predominantly originates from $\eta_1$, while $\eta_2$ contributes mainly to the heavier state $S_3$.

Compared to the singlet scalar DM scenario shown in figure~\ref{fig:relic_sDM_p1}, the relic abundance in this case is generically suppressed over a wide range of DM masses. This behavior can be attributed to the doublet nature of $S_1$, which allows for additional annihilation channels mediated by electroweak gauge interactions. As a result, the thermally averaged annihilation cross section is significantly enhanced, particularly in the low DM mass region, leading to a reduced relic density. This effect is more pronounced at smaller DM masses where gauge-mediated annihilation processes dominate. In contrast to the expectation from conventional co-annihilation scenarios, the dependence of the relic density on the mass splitting $\Delta M(S_1, S_2)$ remains relatively mild. This indicates that co-annihilation effects involving the heavier scalar state $S_2$ do not play a dominant role in determining the freeze-out dynamics for the chosen parameter space. Instead, the relic density is primarily governed by the efficient annihilation channels available to the doublet-like DM state. Finally, resonance funnels around $M_{\rm DM} \simeq M_{h}/2 \sim 62.5~\mathrm{GeV}$ and $M_{\rm DM} \simeq M_{Z'}/2 \sim 720~\mathrm{GeV}$ are clearly visible, where the annihilation cross section is resonantly enhanced via the Higgs- and $Z'$-mediated channels, respectively. In this region, the relic density for all three benchmark mass splittings falls within the \texttt{Planck} $2\sigma$ allowed range, as indicated in figure~\ref{fig:relic_sDM_phi_p1}.
\begin{figure}[htbp]
    \centering
    \includegraphics[width=0.8\textwidth]{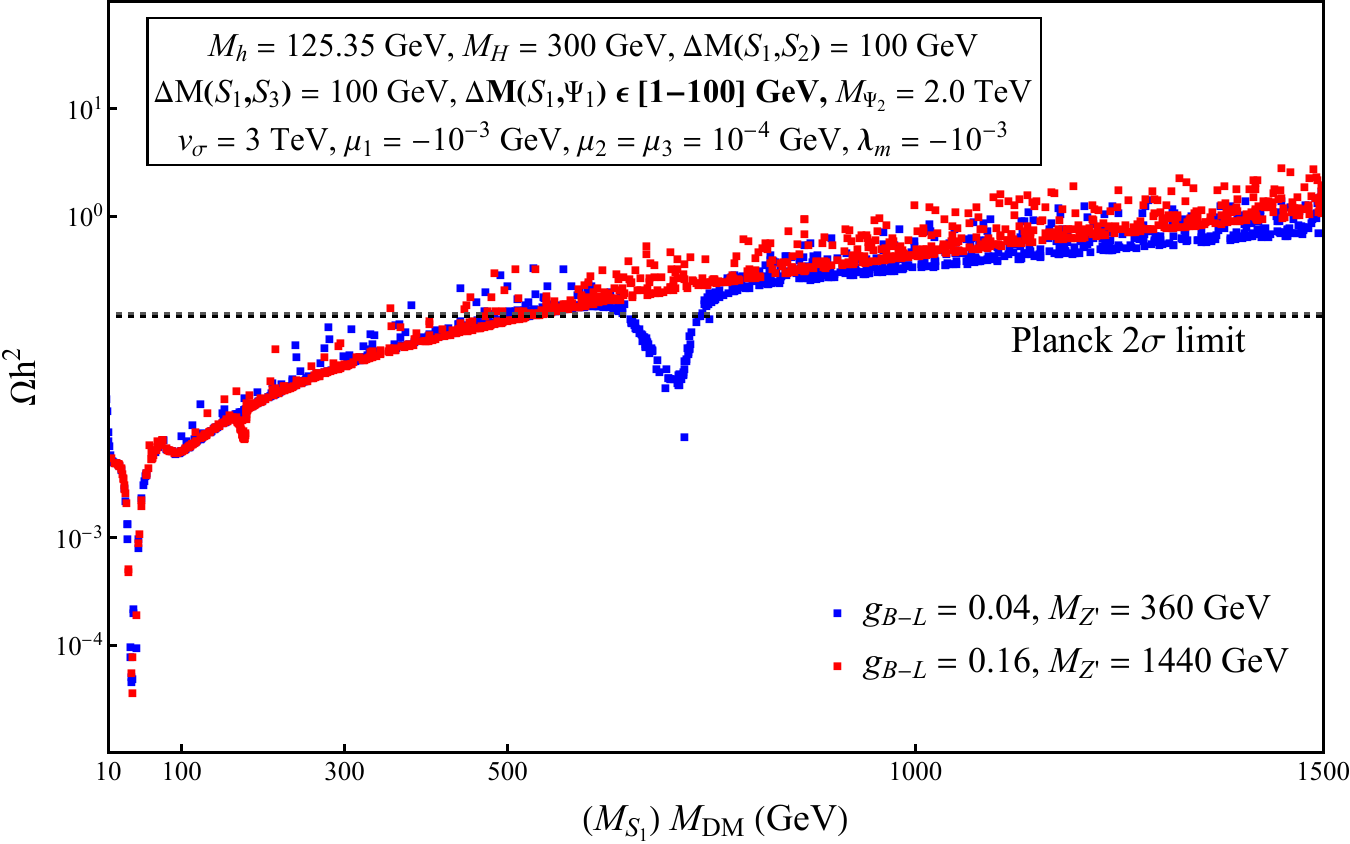}
    \caption{Scalar Dark Matter relic density ($\Omega h^{2}$) as a function of the Dark Matter mass ($M_{DM} = M_{S_1}$) \textbf{with $\mathbf{S_1}$ state majorly coming from $\phi$}. The blue squares correspond to $g_{B-L}=0.04$ and $M_{Z^{\prime}}=360~\mathrm{GeV}$, while the red squares correspond to $g_{B-L}=0.16$ and $M_{Z^{\prime}}=1440~\mathrm{GeV}$. The width of the bands arises from varying $\Delta M(S_{1},\Psi_1)$ in the range $[1-100]~\mathrm{GeV}$. Other fixed parameters are $M_{\text{h}}=125.35~\mathrm{GeV}$, $M_{H}=300~\mathrm{GeV}$, $v_{\sigma}=3~\mathrm{TeV}$, $\Delta M(S_{1},S_{3})=100~\mathrm{GeV}$, $M_{\Psi_{2}}=2.0~\mathrm{TeV}$, with parameters $\mu_2 = \mu_{3} = 10^{-4}$ GeV, $\mu_1 = -10^{-3}$ GeV and $\lambda_m = -10^{-3}$. The horizontal line indicates the \texttt{Planck} $2\sigma$ bound on the relic density.}
    \label{fig:relic_sDM_phi_p2}
\end{figure}

Next, in figure~\ref{fig:relic_sDM_phi_p2}, we show the scalar DM relic density $(\Omega h^2)$ as a function of the DM mass, focusing on the effects of co-annihilation of $S_1$ with the vector-like fermion $\Psi_1$. In this case, the mass splitting between the DM and the co-annihilating partner is varied within the range $\Delta M(S_1,\Psi_1)\in [1,100]~\mathrm{GeV}$, while keeping $\Delta M(S_1,S_2) = \Delta M(S_1,S_3)=100~\mathrm{GeV}$ fixed. The other parameters are chosen as in figure~\ref{fig:relic_sDM_phi_p1}. The two curves correspond to different choices of the $B\!-\!L$ gauge coupling and mediator mass, namely $g_{B-L}=0.04$, $M_{Z'}=360~\mathrm{GeV}$ (blue), and $g_{B-L}=0.16$, $M_{Z'}=1440~\mathrm{GeV}$ (red). From the figure, it is evident that the overall relic abundance remains suppressed over the entire DM mass range, consistent with the behavior observed in figure~\ref{fig:relic_sDM_phi_p1}. This can again be attributed to the doublet nature of $S_1$, which allows for efficient annihilation through electroweak gauge interactions. Although small mass splittings $\Delta M(S_1,\Psi_1)$ would, in principle, enhance co-annihilation effects, the already efficient annihilation of $S_1$ suppresses any significant modification to the effective thermally averaged cross section. As a result, even in the presence of a co-annihilating partner~$(\Psi_1)$, the relic density is primarily governed by the dominant annihilation channels of the DM itself. Consequently, despite varying the mass splitting $\Delta M(S_1,\Psi_1)$ over a wide range from $1$ to $100~\mathrm{GeV}$, only a relatively thin band in the relic density is obtained, indicating a very mild dependence on co-annihilation effects, in contrast to the stronger co-annihilation dependence observed for other DM candidates shown in figures~\ref{fig:relic_p3} and \ref{fig:relic_sDM_p3}. The resonance features remain visible in this plot as well. A Higgs-mediated resonance appears around $M_{\rm DM} \simeq M_{h}/2 \sim 62.5~\mathrm{GeV}$, while an additional resonance occurs at $M_{\rm DM} \simeq M_{Z'}/2$, whose position depends on the chosen $Z'$ mass. For the lighter mediator case ($M_{Z'}=360~\mathrm{GeV}$), the latter appears at lower DM masses, whereas for the heavier mediator ($M_{Z'}=1440~\mathrm{GeV}$), it is correspondingly shifted to higher masses, as expected. Again, the horizontal band corresponds to the \texttt{Planck} $2\sigma$ limit.
\begin{figure}[htbp]
    \centering
    \includegraphics[width=0.47\textwidth]{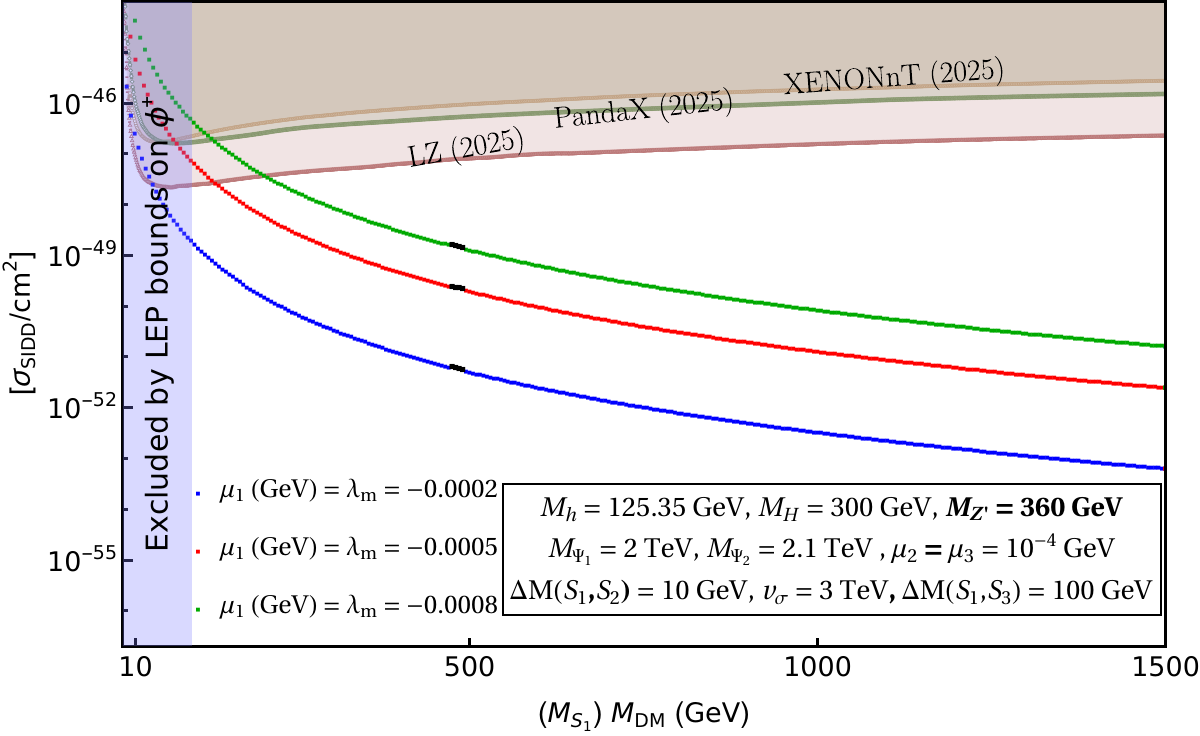}\qquad
    \includegraphics[width=0.47\textwidth]{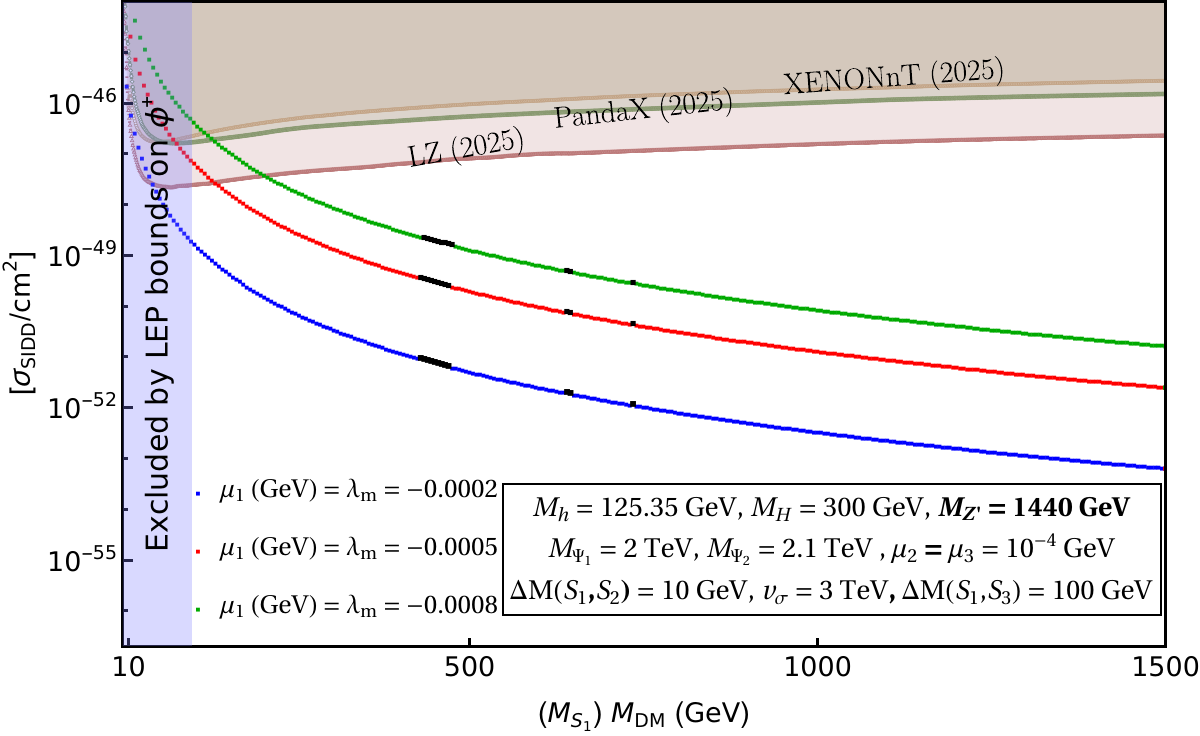}
    \caption{Spin-Independent Direct Detection (SIDD) cross-section ($\sigma_{\rm SIDD}$) as a function of the Scalar Dark Matter mass ($M_{\rm DM}$), \textbf{with $\mathbf{S_1}$ state majorly coming from $\phi$}, for two different values of $M_{Z'}$ with $M_{Z'}=360~\rm GeV$ in the left panel and $M_{Z'}=1440~\rm GeV$ in the right panel. In both panels, the benchmark curves correspond to: $\mu_1 = -0.0002 \text{ GeV}$ (blue), $ -0.0005 \text{ GeV}$ (red), and $ -0.0008 \text{ GeV}$ (green) with $\mu_1 = \lambda_m$ where $\lambda_m$ is the dimensionless quartic coupling. The remaining parameters are fixed at $M_{h}=125.35~\rm GeV$, $M_{H}=300~\rm GeV$, $M_{\Psi_{1}}=2~\rm TeV$, $M_{\Psi_{2}}=2.1~\rm TeV$, $\mu_2 = \mu_3 = 10^{-4} \text{ GeV}$, $\Delta M(S_{1},S_{2})=10~\rm GeV$, $\Delta M(S_{1},S_3)=100~\rm GeV$, and $v_{\sigma}=3~\rm TeV$. The plots also include the latest exclusion limits from \texttt{XENONnT} (2025), \texttt{PandaX} (2025), and \texttt{LZ} (2025), along with a conservative exclusion band (shown in blue) derived from reinterpreted \texttt{LEP} (2013) bounds on $\phi^{+}$. The black points denote the regions of parameter space consistent with the observed dark matter relic abundance within the $2\sigma$ range by \texttt{Planck}.}
    \label{fig:DD3}
\end{figure}

In figure~\ref{fig:DD3}, we present the SIDD cross section, $\sigma_{\rm SIDD}$, as a function of the scalar dark matter mass $M_{\rm DM} \equiv M_{S_1}$, for the case where the DM state $S_1$ predominantly originates from the neutral component $\phi_R^0$ of the scalar doublet. The left panel corresponds to $M_{Z'}=360~\mathrm{GeV}$, while the right panel shows the results for $M_{Z'}=1440~\mathrm{GeV}$. In both panels, the benchmark curves correspond to three representative choices of the parameters $\mu_1=-0.0002~\mathrm{GeV}$ (blue), $-0.0005~\mathrm{GeV}$ (red), and $-0.0008~\mathrm{GeV}$ (green) with $\mu_1 = \lambda_m$ where $\lambda_m$ being dimensionless coupling, keeping the remaining parameters fixed as specified in the figure caption. The shaded regions denote the current exclusion limits from \texttt{XENONnT} (2025), \texttt{PandaX} (2025), and \texttt{LZ} (2025), while the black points indicate regions consistent with the observed relic abundance within the $2\sigma$ range from \texttt{Planck}. The left portion of the plots show a blue band depicting the parameter space up to $M_{S_1}=100 \text{ GeV}$ ruled out by the reinterpreted \texttt{LEP} bounds on $\phi^+$ mass. In this scenario, the scalar DM--nucleon scattering is predominantly mediated by the Higgs boson, with the effective coupling controlled by the mixing structure of the scalar sector. As can be seen from Eqs.~\eqref{Moddodd}) and \eqref{eq:mDScalars}, the off-diagonal term proportional to $\mu_1' v_H$ induces mixing between the doublet scalar $\phi$ and the singlet states, where $\mu_1'=\mu_1 + \frac{\lambda_m v_\sigma}{2\sqrt{2}}$. Consequently, varying $\mu_1$ and $\lambda_m$ directly affects the doublet admixture in the physical DM state $S_1$, and hence its coupling to the SM Higgs. This behavior is clearly reflected in the figure, where larger magnitudes of $\mu_1$ and $\lambda_m$ enhance the Higgs portal interaction, leading to an increase in the predicted SIDD cross section and a clear separation among the benchmark curves. For a fixed choice of $\mu_1$ and $\lambda_m$, the SIDD cross section exhibits a mild decrease with increasing dark matter mass. This can be understood from the structure of the scalar mass matrix, where increasing $M_{\rm DM}$ reduces the relative contribution of the Higgs-induced mixing to the physical state, thereby suppressing the effective coupling to nucleons. As a result, the scattering cross section decreases gradually with increasing DM mass. Comparing the left and right panels, we observe that the SIDD cross section remains essentially unchanged for a given choice of $\mu_1$ and $\lambda_m$, indicating its weak sensitivity to the $Z'$ mass. This is expected since the dominant contribution to direct detection arises from Higgs-mediated interactions rather than from $Z'$ exchange. On the other hand, the distribution of the black points shows a clear shift toward higher dark matter masses in the right panel, consistent with the relic density constraint being influenced by the $Z'$-mediated resonance condition $M_{\rm DM} \simeq M_{Z'}/2$. Overall, these results highlight that while the relic abundance is sensitive to the $Z'$ sector, the direct detection prospects are primarily governed by the Higgs portal interaction controlled by the mixing parameter $\mu_1'$, leading to a strong dependence on $\mu_1$ and $\lambda_m$ but only a negligible dependence on $M_{Z'}$. In our DM numerical analysis, all parameter points satisfying the relic density and recent most direct detection bounds shown in the figures simultaneously satisfy the neutrino mass, the current cLFV constraints, and lepton mixing.

\section{Collider implications}
\label{sec:collider}
The study of final-state missing energy alongside a prominent visible SM signature in the context of $pp$ colliders, given their high-energy prospects, attracts the community's interest. Moreover, with existing LHC searches, future high-luminosity lepton colliders, specifically $\mu^+\mu^-$ colliders~\cite{MuCoL:2025quu}, provide an excellent testing ground for the dark sector predicted in this model. Owing to the clean experimental environment and the precisely known initial state, lepton colliders are particularly sensitive to charged electroweak states and missing-energy signatures, allowing a systematic probe of both fermionic and scalar DM realizations.

In both cases, the dominant collider signals arise from the electroweak production of $\mathcal{Z}_6$-odd inert scalars, which carry SM gauge charges and therefore couple directly to the $Z$, $\gamma$, and $W^\pm$ gauge bosons. At lepton colliders, the dominant production of the $\mathcal{Z}_6$-odd scalar sector proceeds through $s$-channel exchange of neutral gauge bosons. Since the $\mu^+\mu^-$ initial state is electrically neutral, pair production of inert scalars occurs primarily via $Z^\ast/\gamma^\ast$ mediation, leading to $\phi^+\phi^-$ and $\phi_i^0 \phi_j^0$ final states. Production channels involving $W^\pm$ exchange require $t$-channel neutrino mediation and are therefore sub-leading, lacking the resonant enhancement of the $s$-channel processes. Consequently, for center-of-mass energies relevant to the proposed muon colliders, the $Z/\gamma$-mediated channels dominate the sensitivity. In this section, we have presented the possible collider signatures of our considered DM particles and their detection prospects in context of LHC ($\sqrt{s} = 13$ TeV @ $\mathcal{L} = 140$ fb$^{-1}$) and $\mu^+\mu^-$ collider~\cite{MuCoL:2025quu} ($\sqrt{s} = 10$ TeV @ $\mathcal{L} = 10$ ab$^{-1}$). The leading production channels are therefore
\begin{equation}
pp / \mu^+ \mu^- \to \phi^+ \phi^-,
\qquad
pp / \mu^+ \mu^- \to \phi_i^0 \phi_j^0 \gamma,
\end{equation}
where $\phi^\pm$ and $\phi_i^0 = S_{1,2,3}$ denote the charged and neutral components of the inert scalar sector. The corresponding production cross sections are fixed by electroweak quantum numbers and scalar masses and are therefore largely model independent, enabling a direct reinterpretation of existing hadron and/or muon collider sensitivity studies for inert doublet-like scenarios.

The subsequent decay chains depend on the identity of the dark matter particle. In the fermionic DM scenario, the inert scalars decay into singlet fermions according to
\begin{equation}
\phi^\pm \to \ell^\pm \Psi_1,
\qquad
\phi_i^0 \to \nu \Psi_1,
\end{equation}
with the heavier fermions cascading into the lightest stable state $\psi_1$, which constitutes the dark matter. Among these signals, we have found that the dominant cross-section corresponds to the $\phi^\pm \to \ell^\pm \Psi_1$ process, so we have considered only this channel for our subsequent analysis. The dominant SM background relevant for this $\ell^+ \ell^- + \slashed{E}$ signature is via vector boson fusion (VBF) channels: $pp/ \mu^+ \mu^- \to WW/ZZ \to \ell^+ \ell^+ \nu \bar{\nu}$ where $\nu$ corresponds to all the SM active neutrino flavors combined. We have tabulated all the relevant cross-sections of this signal and SM background for $pp$ and $\mu^+ \mu^-$ colliders in table~\ref{tab:fermion_BP}. To simulate the signal and background processes in \texttt{MadGraph5\_aMC@NLO}~\cite{Alwall:2014hca}, we use the UFO output
of the \texttt{SARAH} model file. At the generation level, the default \texttt{MadGraph5\_aMC@NLO} cuts have been implemented to avoid the collinear and IR divergences. The subsequent analysis at parton-level has been performed using~\texttt{Delphes-3.5.0} \cite{deFavereau:2013fsa}. As benchmark scenarios, we have considered the DM masses $M_{\Psi_1}$ that satisfy the constraints from the current relic density and the latest direct detection bound (\texttt{LZ} (2025)), as discussed in previous sections, specifically we have chosen one benchmark value for a particular value of $M_{Z^\prime}$. All the other particle masses, couplings, and other parameters are set according to the top panel of figure~\ref{fig:DD1}.

\begin{table}[htb!]
    \centering
    \begin{tabular}{|c|c|c|}
    \hline
        Colliders & $\sigma_{\text{sig}}$ (fb) & $\sigma_{\text{bkg}}$ (fb)  \\
        \hline
        $pp$ & 75.39 ($M_{\Psi_1} = 130$ GeV, $M_{Z^\prime} = 360$ GeV) & 1929 \\
        & 0.051 ($M_{\Psi_1} = 716$ GeV, $M_{Z^\prime} = 1440$ GeV) & \\
        \hline
        $\mu^+\mu^-$ & 0.2744 ($M_{\Psi_1} = 130$ GeV, $M_{Z^\prime} = 360$ GeV) & 199 \\
        & 0.2702 ($M_{\Psi_1} = 716$ GeV, $M_{Z^\prime} = 1440$ GeV) & \\
        \hline
    \end{tabular}
    \caption{Cross-sections for signal and SM background for two benchmark values of fermionic DM $M_{\Psi_1}$ which satisfy relic density and DD bounds, for $pp$ and $\mu^+\mu^-$ colliders.}
    \label{tab:fermion_BP}
\end{table}

\begin{figure}[htb!]
    \centering
\includegraphics[width=0.49\textwidth]{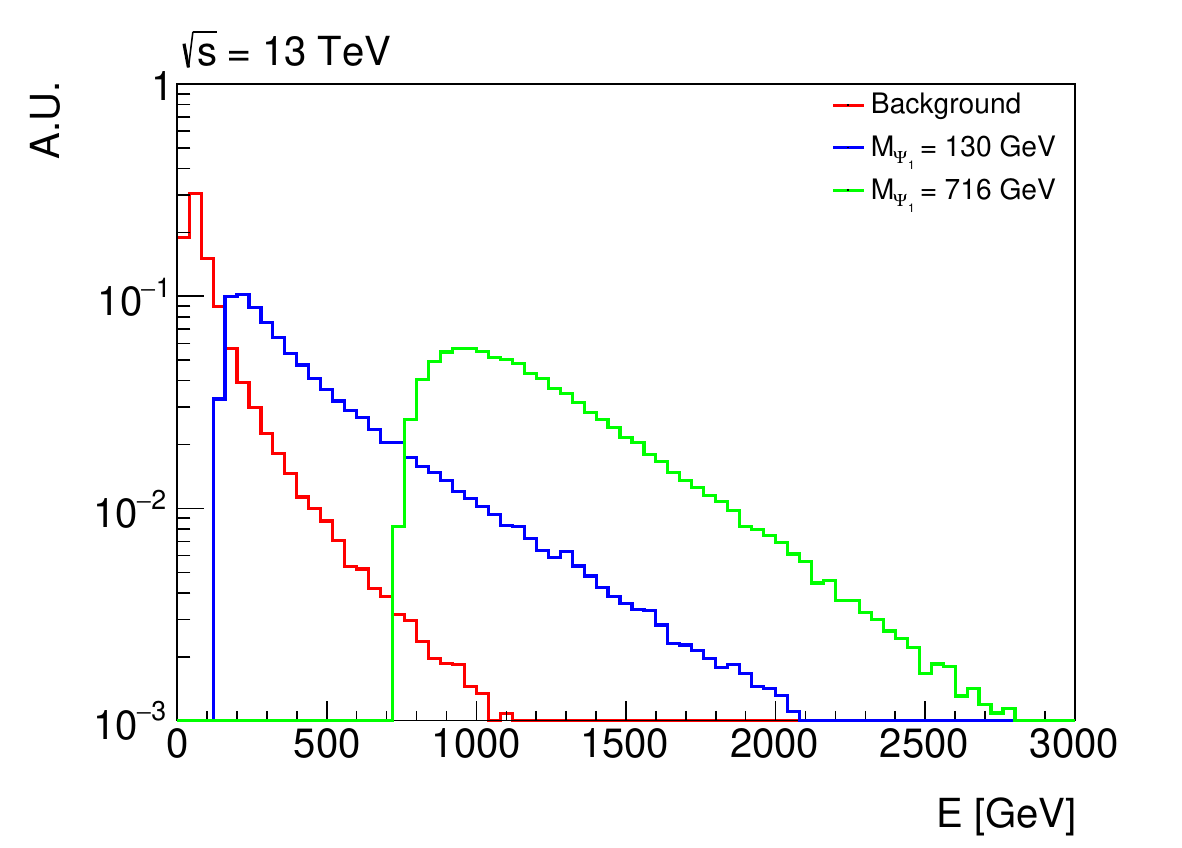}
\includegraphics[width=0.49\textwidth]{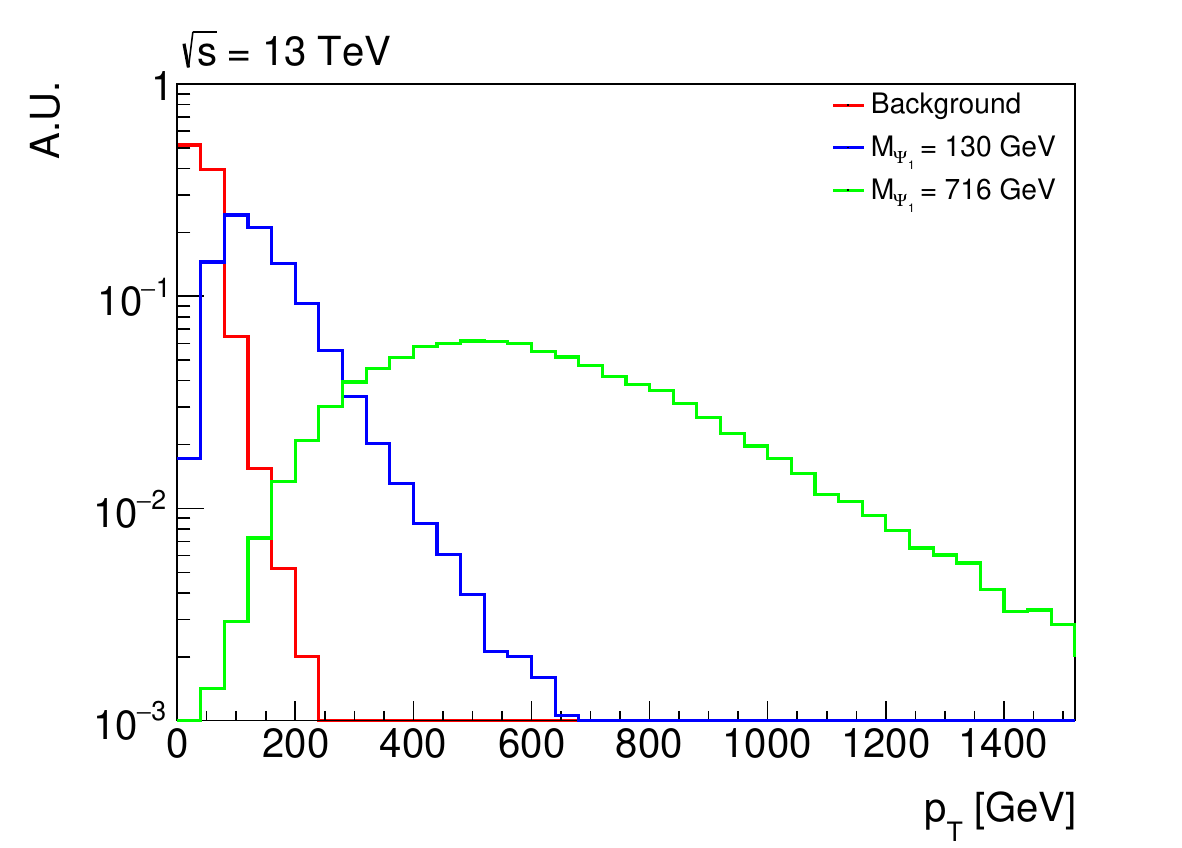}
\includegraphics[width=0.49\textwidth]{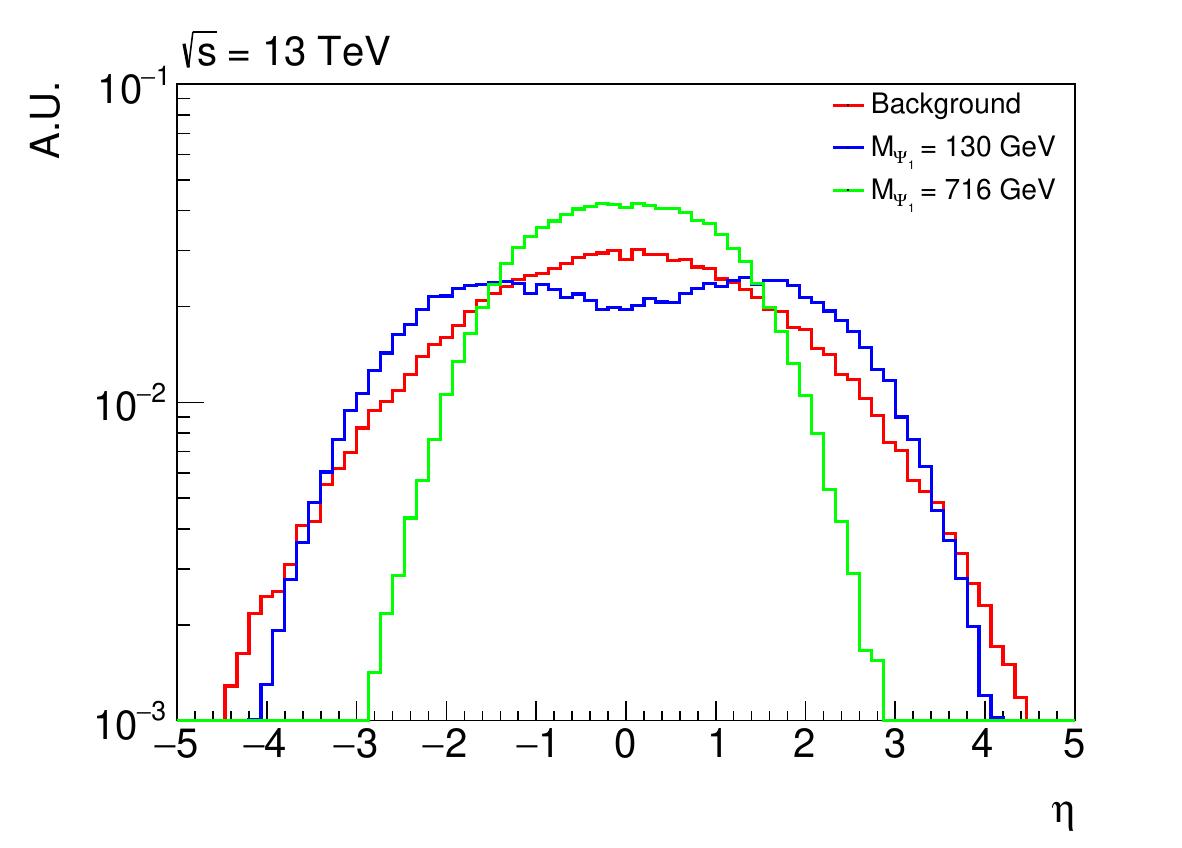}
\includegraphics[width=0.49\textwidth]{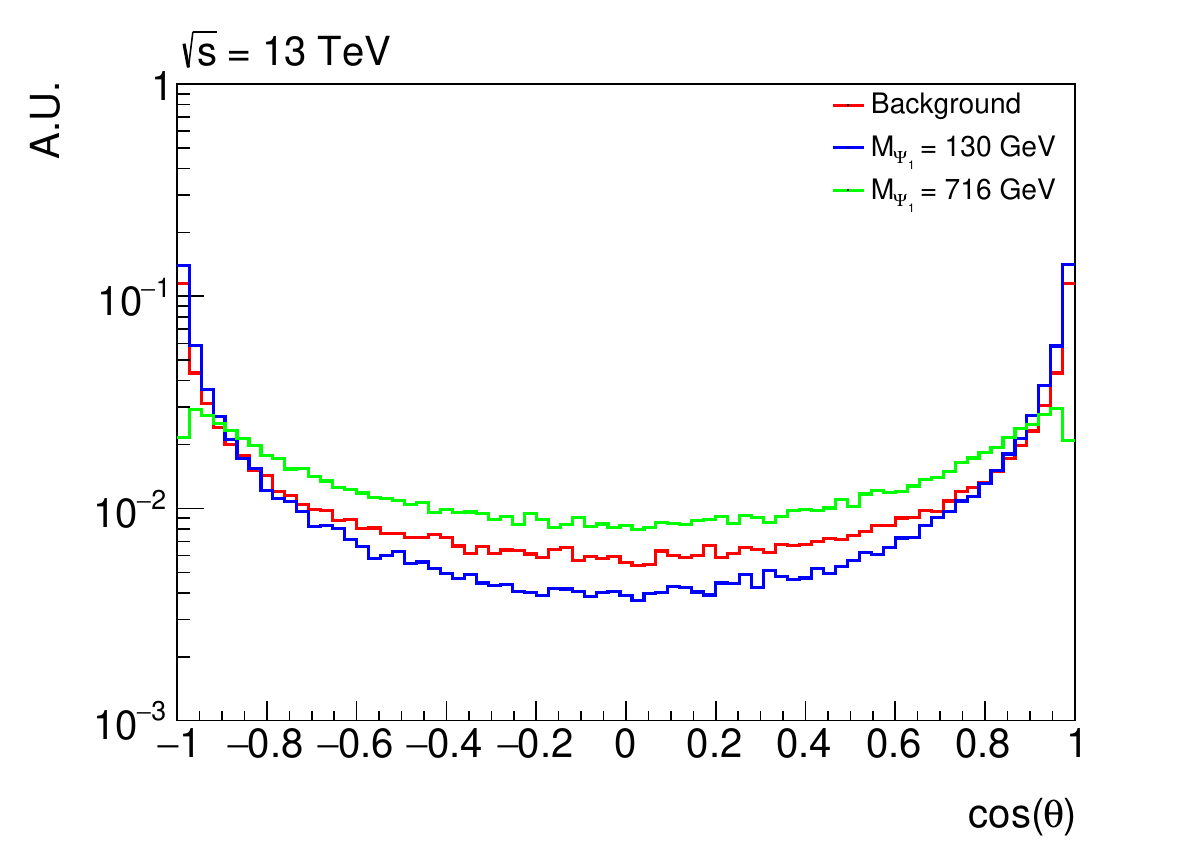}
\caption{Relevant kinematic distributions in context of $pp$ collider ($\sqrt{s} = 13$ TeV @ $\mathcal{L} = 140$ fb$^{-1}$) for fermionic DM $\Psi_1$ signal ($pp \to \phi^+ \phi^- \to \ell^\pm \ell^\mp \Psi_1 \bar{\Psi}_1$) with (\textit{blue}  : $M_{\Psi_1} = 130$ GeV, \textit{green} : $M_{\Psi_1} = 716$ GeV) and SM neutrinos for background (shown in \textit{red}). The distributions are normalised to unit area; the y-axis is therefore shown in arbitrary units (A.U.), reflecting the relative shape of each distribution.}
\label{fig:pp_fermion}
\end{figure}

\begin{figure}[htb!]
    \centering
\includegraphics[width=0.49\textwidth]{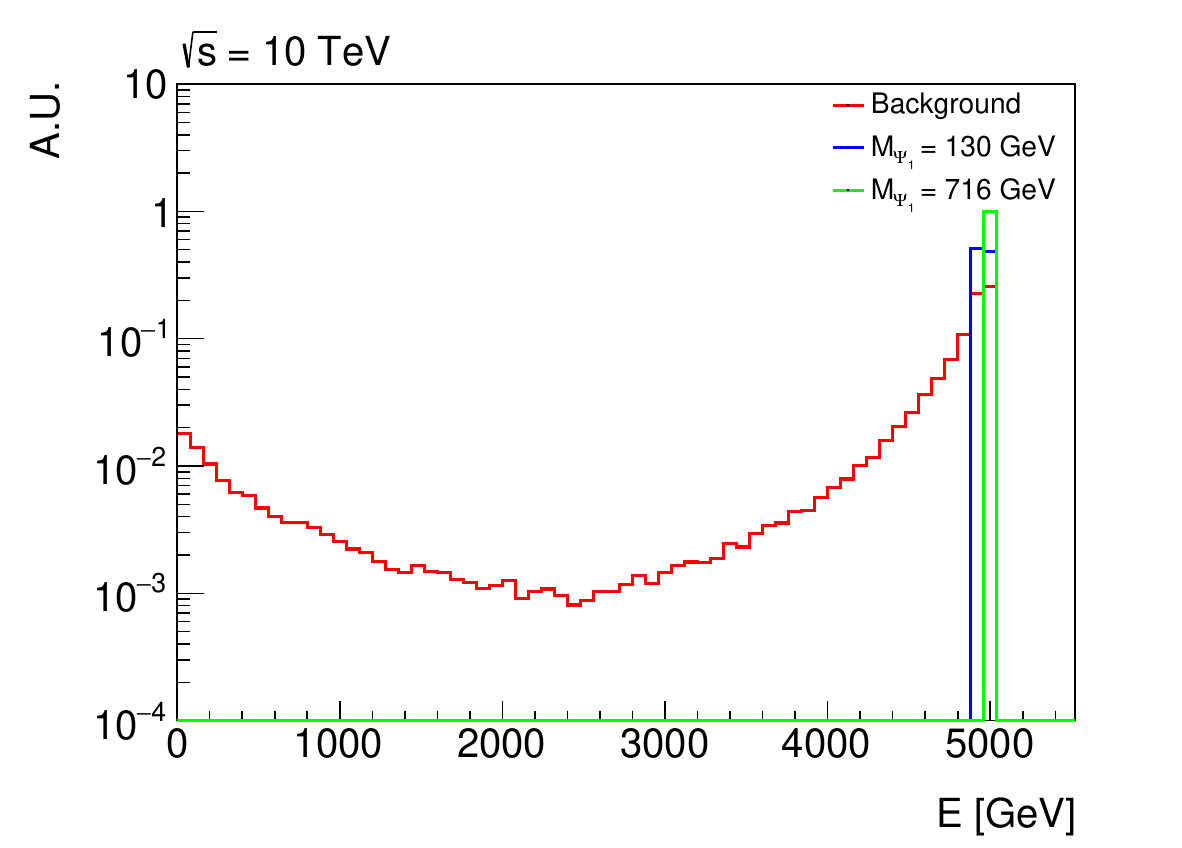}
\includegraphics[width=0.49\textwidth]{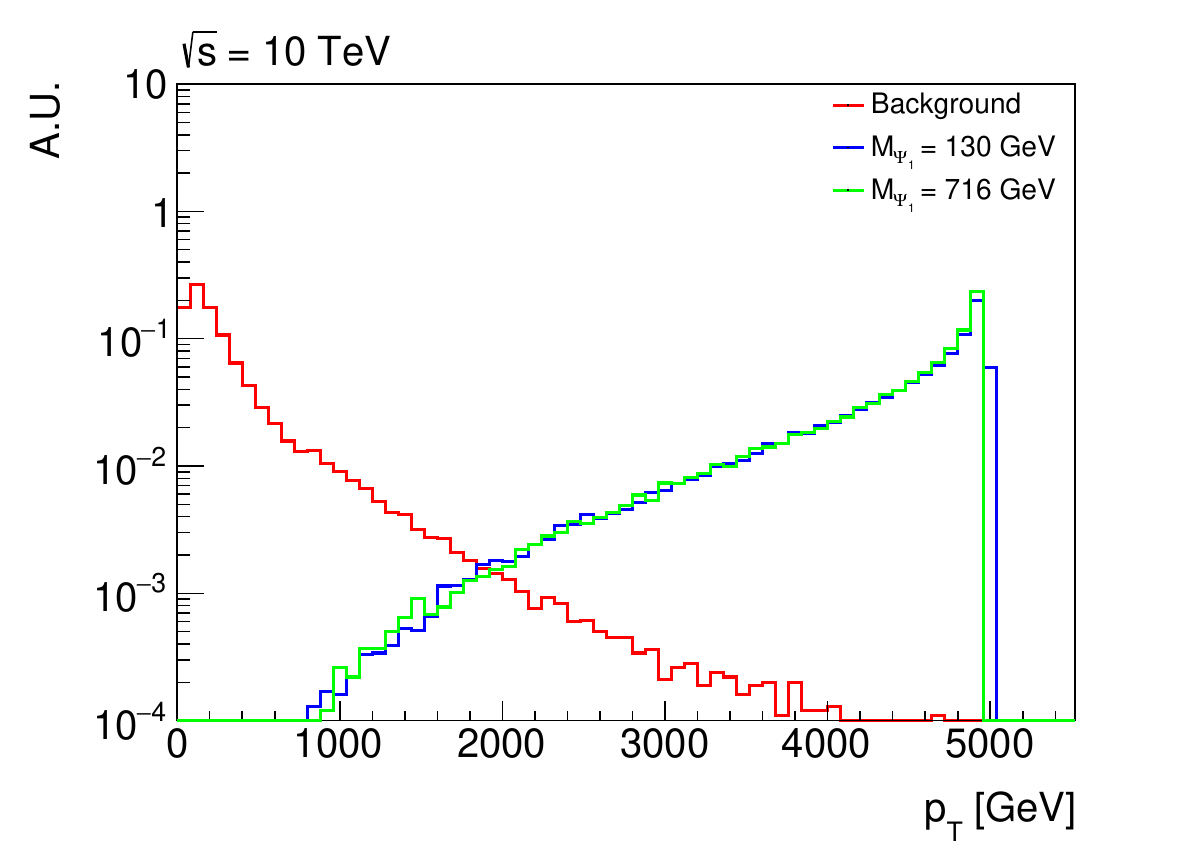}
\includegraphics[width=0.49\textwidth]{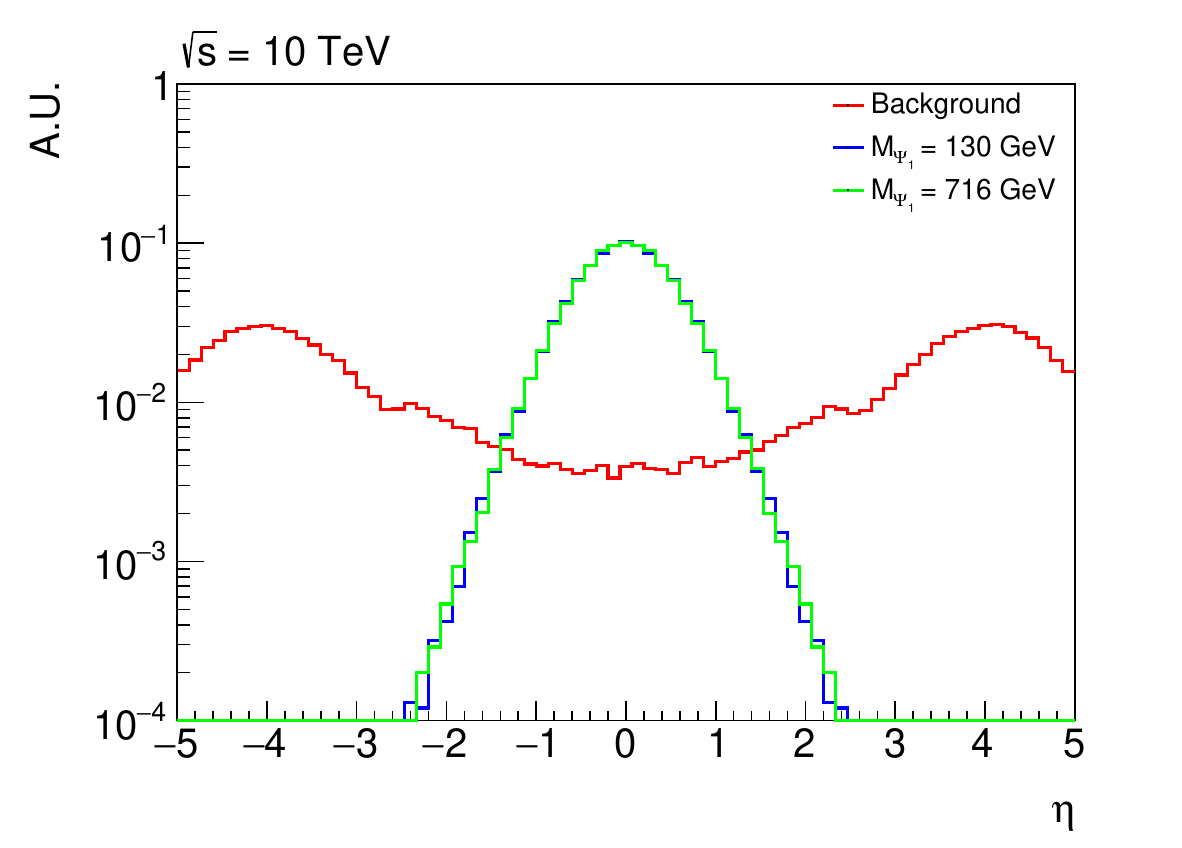}
\includegraphics[width=0.49\textwidth]{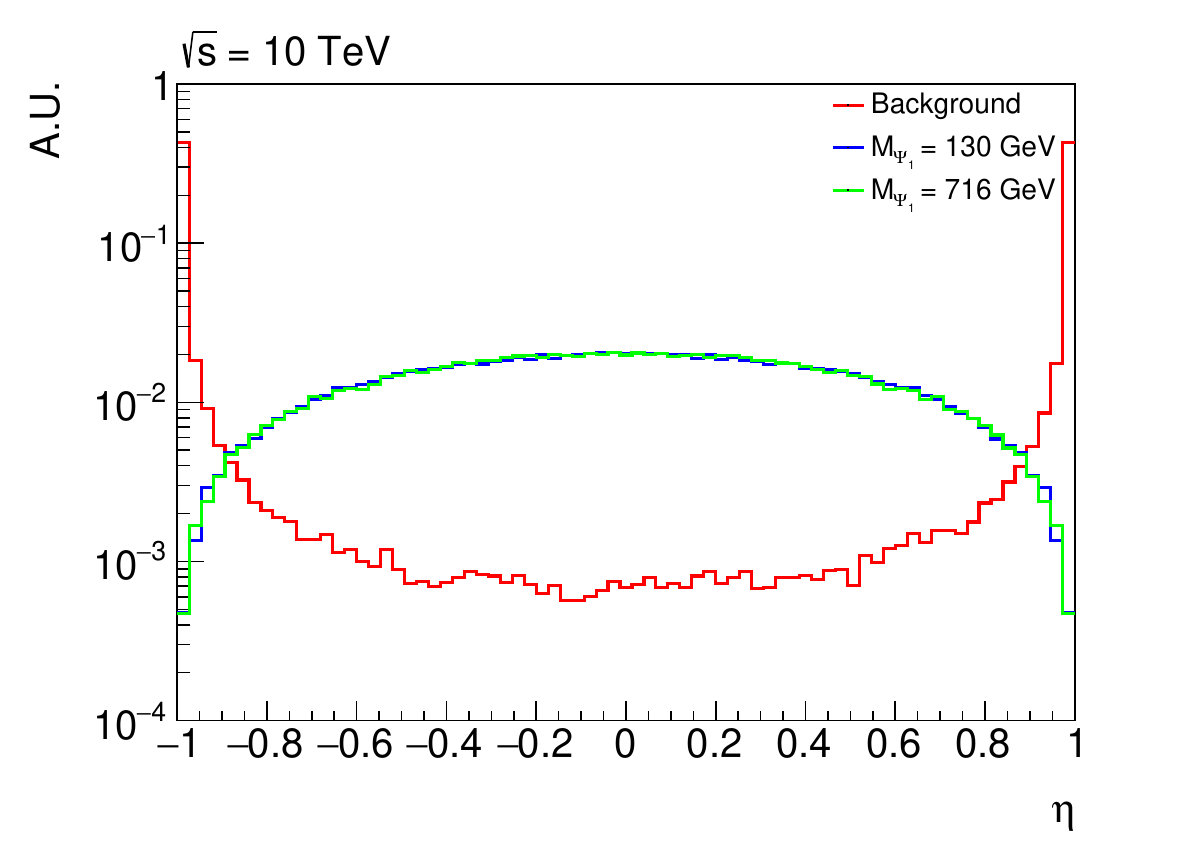}
\caption{Relevant kinematic distributions in context of $\mu^+\mu^-$ collider ($\sqrt{s} = 10$ TeV @ $\mathcal{L} = 10$ ab$^{-1}$) for fermionic DM $\Psi_1$ signal ($\mu^+ \mu^- \to \phi^+ \phi^- \to \ell^\pm \ell^\mp \Psi_1 \bar{\Psi}_1$) with (\textit{blue}  : $M_{\Psi_1} = 130$ GeV, \textit{green} : $M_{\Psi_1} = 716$ GeV) and SM neutrinos for background (shown in \textit{red}). Similar to figure~\ref{fig:pp_fermion}, in this scenario, the distributions are normalised such that the area under each curve remains unity, and we have denoted the y-axis by arbitrary units (A.U.).}
    \label{fig:mumu_fermion}
\end{figure}

We have shown the relevant kinematic distributions for of the outgoing DM particle (signal) or neutrinos (SM background) such as energy ($E$), transverse momentum ($p_T$), pseudo-rapidity ($\eta$) and the cosine of the angle ($\cos(\theta)$) of the particle, measured from the beam direction in figures~\ref{fig:pp_fermion} ($pp$ collider) and \ref{fig:mumu_fermion} ($\mu^+ \mu^-$ collider). The histograms for outgoing DM particles have been shown in \textit{blue} ($M_{\Psi_1} = 130$ GeV) and \textit{green} ($M_{\Psi_1} = 716$ GeV), while the distributions for outgoing neutrinos in the context of the SM background are shown in \textit{red}. All these distributions are normalised such that the area under the entire curve becomes unity. This normalization ensures no dependence on the absolute normalization and allows for a direct comparison of the shapes of the distributions. Consequently, the $y$-axis in each figure is presented in arbitrary units (A.U.), since only the relative variation of the distributions is physically relevant.

From these distributions, it is quite clear that we can implement several kinematic cuts so that the significance $\mathcal{S} = S/\sqrt{S+B}$ can be enhanced, where $S (B)$ corresponds to the effective number of signal (background) events i.e., $\sigma_{\text{sig}} (\sigma_{\text{bkg}}) \times \mathcal{L} \times \epsilon$, where efficiency $\epsilon$ corresponds to the ratio of total number of surviving events after implementing the cuts to the total number of events generated. For hadron collider, we have implemented the acceptance cuts, for $m_{\Psi_1} = 130$ GeV : 
\begin{equation}
    E > 100~\text{GeV},~~~ p_T > 80~\text{GeV},~~~ |\cos(\theta)| > 0.75,~~~ 3.5 > |\eta| > 1.4
    \label{eq:cut_130GeV}
\end{equation}
and for $m_{\Psi_1} = 716$ GeV :
\begin{equation}
    E > 700~\text{GeV},~~~ p_T > 150~\text{GeV},~~~ |\cos(\theta)| < 0.9,~~~ |\eta| < 1.5
    \label{eq:cut_716GeV}
\end{equation}
while for muon colliders, the acceptance cuts we have considered for maximising the signal significance :
\begin{equation}
    E > 4.8~\text{TeV},~~~ p_T > 1.9~\text{TeV},~~~ |\cos(\theta)| < 0.9,~~~ |\eta| < 1.5
\end{equation}
Owing to the relative shape of the signal distributions for two different DM masses with respect to the SM background for hadron collider, we have implemented two different kinematical cuts (as shown in Eqs.~\eqref{eq:cut_130GeV} and~\eqref{eq:cut_716GeV}) to maximise the signal significances. Corresponding significances before and after implementing the cuts are shown in table~\ref{tab:significance_fermion}. The significance for current hadron collider is impressively large even before the cuts are applied, for $m_{\Psi_1} = 130$ GeV benchmark point, but we can further increase it significantly after imposing the mentioned cuts. However, the signal significance, even after implementing the designated cuts, for $m_{\Psi_1} = 716$ GeV is not significant enough to expect to be observed in current hadron collider. In this context, we can consider studying this particular scenario at the proposed High-Luminosity LHC (HL-LHC), which is expected to achieve an integrated luminosity of about 3000 fb$^{-1}$~\cite{Barletta:2013ooa} over roughly twelve years following its planned upgrade. For HL-LHC, the proposed $\sqrt{s} = 14$ TeV, almost similar to the current LHC run with $\sqrt{s} = 13$ TeV with $\mathcal{L} = 140$ fb$^{-1}$, we will not further analyze the scenario in greater details. However, using the similar information of our current LHC run analysis, the expected signal sensitivity, in post-cut level, for $m_{\Psi_1} = 716$ GeV in HL-LHC will be around 6.06, which eventually increases the probability of observing such signal in this future hadron collider. 

\begin{figure}[htb!]
    \centering
    \includegraphics[width=0.7\textwidth]{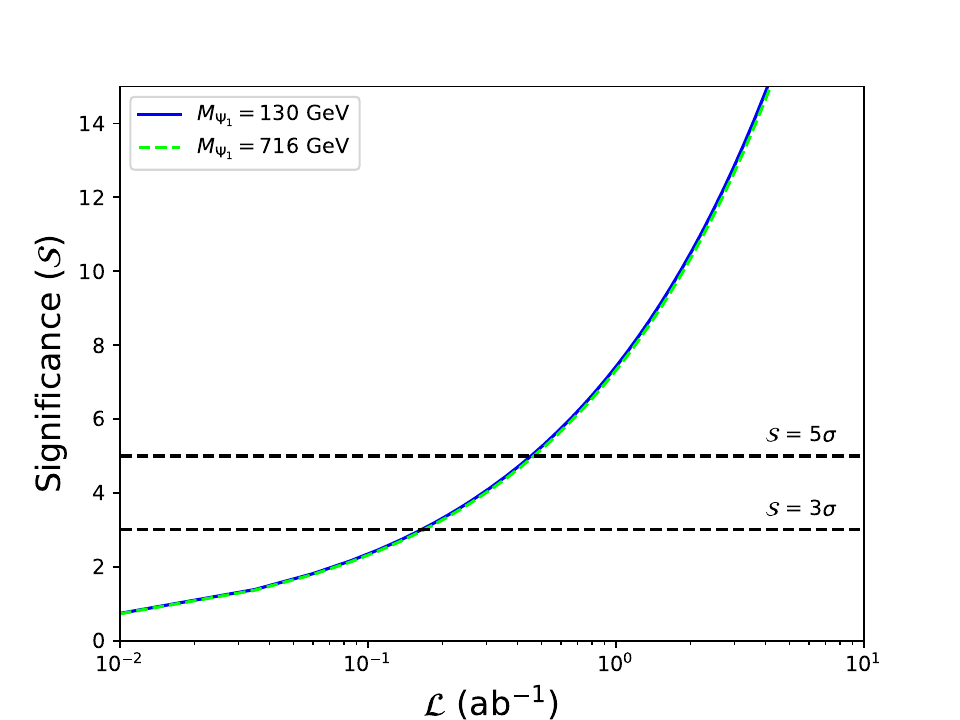}
    \caption{$\mu^+ \mu^-$ collider: Signal significances with respect to luminosity for fermionic DM. It is evident that one can achieve $3\sigma$ and/or $5\sigma$ signal significance for proposed muon collider for detecting such DM signal even with smaller integrated luminosity as compared to the proposed one.}
    \label{fig:lumi}
\end{figure}

For muon colliders, the pre-cut significances fall below the desired $3\sigma$ level, and they can be further enhanced by imposing cuts owing to distinctive distribution characteristics compared to SM background distributions. We have also plotted the significances after implementing the kinematic cuts discussed with respect to luminosity in context of muon collider in figure~\ref{fig:lumi}, here one can infer that $3\sigma$ and $5\sigma$ significances (shown in \textit{black} dashed lines) can be achieved for even smaller values of integrated luminosities as: 163.57 fb$^{-1}$ (168.01 fb$^{-1}$) and 454.35 fb$^{-1}$ (466.7 fb$^{-1}$) for $M_{\Psi_1} = 130~ (716)$ GeV as shown by \textit{blue solid} (\textit{green dashed}) contours, respectively.    
\begin{table}[htb!]
\small
    \centering
    \begin{tabular}{|c|c|c|c|}
    \hline
       Colliders & $M_{\psi_1}$ &$\mathcal{S}$ & $\mathcal{S}$  \\
       & (GeV) & (pre-cut) & (post-cut)\\
       \hline
        $pp ~@ 140$ fb$^{-1}$ & 130  & 19.925 & 37.642 \\
         & 716 & 0.014 & 1.308\\ 
        \hline
        $\mu^+\mu^- ~@ 10$ ab$^{-1}$ & 130 & 1.944 & 23.457 \\
         & 716 & 1.914 & 23.145\\ 
        \hline
    \end{tabular}
    \caption{Significances before and after kinematic cuts implemented for fermionic DM.}
    \label{tab:significance_fermion}
\end{table}

On the other hand, in the scalar DM scenario, the heavier neutral scalars decay dominantly via $pp/\mu^+\mu^- \to S_1 S_1$, where $S_1$ is the scalar dark matter particle. In both realizations, all decay chains terminate in invisible states, resulting in sizable missing energy. The corresponding cross-sections are too small to be realized in the respective $pp$ and $\mu^+\mu^-$ colliders with current and future energy and luminosity reach. So, we need either very high-energy and/or high-luminosity colliders to detect this scalar DM candidate. Henceforth, we will not discuss the detection prospects of our scalar DM candidate further.

\section{Conclusion}
\label{sec:conclusion} 

In this work, we have investigated an anomaly-free \(U(1)_{B-L}\) extension of the SM that simultaneously addresses the origin of tiny neutrino masses and the nature of dark matter within a unified framework. The model includes three generations of right-handed neutrinos, vector-like leptons, and an extended scalar sector. Dirac neutrino masses are generated radiatively at the one-loop level, and we have shown that the observed neutrino mass scale can be reproduced for TeV-scale new particles with Yukawa couplings of \(\mathcal{O}(10^{-6})\). The presence of additional scalars and fermions, in particular the vector-like fermion $\Psi$ and the charged scalar $\phi^+$, induces new contributions to cLFV processes such as $\mu\to e\gamma$, $\mu \to 3e$, and coherent $\mu$--$e$ conversion in nuclei. A correlated analysis of these observables demonstrates that the framework remains consistent with current experimental bounds, while future improvements in experimental sensitivity are expected to further constrain the allowed parameter space.

A distinctive feature of the model is the presence of stable DM candidates, whose stability is guaranteed by a residual \(\mathcal{Z}_6\) symmetry originating from the spontaneous breaking of \(U(1)_{B-L}\). Depending on the mass hierarchy, either the lightest vector-like fermion or the lightest scalar mass eigenstate can serve as a viable DM candidate. In the fermionic scenario, the lightest vector-like fermion $\Psi_1$, with $M_{\Psi_1} < M_{\Psi_{2}},\, M_{S_{1,2,3}}$ (where $S_{1,2,3}$ denote the physical BSM scalars), can account for the observed relic abundance, satisfying the \texttt{Planck} $2\sigma$ bound. Interestingly, the co-annihilation between the lightest and next-to-lightest fermionic states will affect adversely here due to the reduced available kinematic phase space or weaker effective couplings after the mass diagonalization, while the co-annihilation with the lightest scalar state will be effective for smaller values of DM mass. The fermionic DM scenario is sensitive to constraints from direct detection experiments, reaching current exclusion limits from \texttt{XENONnT} (2025), \texttt{PandaX} (2025), and \texttt{LZ} (2025) for specific parameter choices. Alternatively, the lightest scalar mass eigenstate $S_1$ coming from either the scalar $\eta_1$ or from $\phi$ can separately serve as a viable DM candidate. Similar to the fermionic case, the adverse effect of co-annihilation between the lightest and next-to-lightest scalar states on the relic density is also studied here. However, in contrast to the fermionic scenario, scalar DM remains compatible with relic density and direct detection constraints over a much broader mass range. For both the scenarios, the parameter points satisfying the relic density and recent direct detection bounds can simultaneously saturate other phenomenological constraints such as neutrino masses, charged lepton flavor violation and lepton mixing.

We have further explored the collider phenomenology of the model, focusing on the characteristic \(\ell^+\ell^- + \slashed{E}\) final states arising from dark sector production. Our parton-level analysis indicates that fermionic DM can be efficiently probed at current hadron colliders, and even more sensitively at future muon colliders, which can achieve a signal significance with $3\sigma$ and/or $5\sigma$ C.L. with relatively low integrated luminosity as compared to the proposed one. In contrast, the discovery of scalar dark matter signatures generally requires higher energies and/or luminosities. In summary, the \(U(1)_{B-L}\) extended SM studied here provides a coherent and phenomenologically rich framework that links neutrino mass generation, cLFV, dark matter, and collider signatures. The model is highly testable and offers multiple complementary avenues for experimental verification at current and forthcoming facilities.

\begin{acknowledgments}
CM and SS would like to thank Sudip Jana for the collaboration during an initial stage of this work. UP acknowledges SINP, Kolkata, as his current source of research funding. SP would like to acknowledge the funding support from SERB, Government of India, under the MATRICS project with grant no. MTR/2023/000687.
\end{acknowledgments}

\newpage
\noindent 
\begin{LARGE}\textbf{Appendix}\end{LARGE}

\appendix
\label{APP:app}

\section{Relevant Feynman Diagrams for processes setting 
\texorpdfstring{$\Psi_1$}{Psi1} relic abundance}
\label{app:A}
Here, we display the relevant Feynman diagrams contributing
to fermionic dark matter~$(\Psi_1)$ annihilation and co-annihilation processes into the SM final states. The diagrams include $s$- and $t$-channel processes mediated by model scalars, gauge bosons, and fermions. These channels provide the dominant contributions to the fermionic DM relic abundance.
\begin{figure}[htbp]
    \centering
    \begin{subfigure}[b]{0.48\textwidth}
        \centering
        \begin{tikzpicture}[line width=0.5 pt, scale=0.85]
            \draw[fermion] (-3.0,1.0)--(-1.5,0.0);
            \draw[antifermion] (-3.0,-1.0)--(-1.5,0.0);
            \draw[snake] (-1.5,0.0)--(0.8,0.0);
            \draw[fermion] (0.8,0.0)--(2.5,1.0);
            \draw[antifermion] (0.8,0.0)--(2.5,-1.0);
            \node at (-3.3,1.0) {${\Psi_i}$};
            \node at (-3.3,-1.0) {$\overline{\Psi}_i$};
            \node [above] at (0.0,0.05) {$Z/Z'$};
            \node at (3.15,1.0) {$f/W^+$};
            \node at (3.18,-1.0) {$\overline{f}/W^-$};
        \end{tikzpicture}
        \caption{}
        \label{fig:DM1FD1}
    \end{subfigure}
    \hfill
    \begin{subfigure}[b]{0.48\textwidth}
        \centering
        \begin{tikzpicture}[line width=0.5 pt, scale=1.35]
            \draw[fermion] (-3.5,1.0)--(-1.6,1.0);
            \draw[antifermion] (-3.5,-0.5)--(-1.6,-0.5);
            \draw[fermionTB](-1.6,1.0)--(-1.6,-0.5);
            \draw[snake] (-1.6,1.0)--(0.0,1.0);
            \draw[snake] (-1.6,-0.5)--(0.0,-0.5);
            \node at (-3.7,1.0) {${\Psi_i}$};
            \node at (-3.7,-0.5) {$\overline{\Psi}_i$};
            \node [right] at (-1.60,0.25) {$\Psi_j$};
            \node at (0.28,1.0) {$Z,Z'$};
            \node at (0.28,-0.5) {$Z,Z'$};
        \end{tikzpicture}
        \caption{}
        \label{fig:DM1FD2}
    \end{subfigure}

    \vspace{0.3cm} 
    \begin{subfigure}[b]{0.48\textwidth}
        \centering
        \begin{tikzpicture}[line width=0.5 pt, scale=1.35]
            \draw[fermion] (-3.5,1.0)--(-1.6,1.0);
            \draw[antifermion] (-3.5,-0.5)--(-1.6,-0.5);
            \draw[fermionTB](-1.6,1.0)--(-1.6,-0.5);
            \draw[dashed] (-1.6,1.0)--(0.0,1.0);
            \draw[dashed] (-1.6,-0.5)--(0.0,-0.5);
            \node at (-3.7,1.0) {${\Psi_i}$};
            \node at (-3.7,-0.5) {$\overline{\Psi}_i$};
            \node [right] at (-1.60,0.25) {$\nu_l$};
            \node at (0.28,1.0) {$S_k$};
            \node at (0.28,-0.5) {$S_k$};
        \end{tikzpicture}
        \caption{}
        \label{fig:DM1FD3}
    \end{subfigure}
    \hfill
        \begin{subfigure}[b]{0.48\textwidth}
        \centering
        \begin{tikzpicture}[line width=0.5 pt, scale=0.85]
            \draw[fermion] (-3.0,1.0)--(-1.5,0.0);
            \draw[antifermion] (-3.0,-1.0)--(-1.5,0.0);
            \draw[snake] (-1.5,0.0)--(0.8,0.0);
            \draw[dashed] (0.8,0.0)--(2.8,1.0);
            \draw[dashed] (0.8,0.0)--(2.8,-1.0);
            \node at (-3.3,1.0) {${\Psi_i}$};
            \node at (-3.3,-1.0) {$\overline{\Psi}_i$};
            \node [above] at (0.0,0.05) {$Z/Z'$};
            \node at (3.15,1.0) {$\phi^+$};
            \node at (3.18,-1.0) {$\phi^-$};
        \end{tikzpicture}
        \caption{}
        \label{fig:DM1FD4}
    \end{subfigure}

    \vspace{0.3cm}
    \begin{subfigure}[b]{0.48\textwidth}
        \centering
        \begin{tikzpicture}[line width=0.5 pt, scale=1.35]
            \draw[fermion] (1.2,1.0)--(3.2,1.0);
            \draw[antifermion] (1.2,-0.5)--(3.2,-0.5);
            \draw[dashed](3.2,1.0)--(3.2,-0.5);
            \draw[fermion] (3.2,1.0)--(4.9,1.0);
            \draw[antifermion] (3.2,-0.5)--(4.9,-0.5);
            \node at (1.05,1.0) {${\Psi_i}$};
            \node at (1.05,-0.5) {$\overline{\Psi}_j$};
            \node [right] at (3.15,0.25) {${\phi}$};
            \node at (5.05,1.0) {$e_k$};
            \node at (5.05,-0.5) {$\overline{e}_k$};
        \end{tikzpicture}
        \caption{}
        \label{fig:DM1FD5}
    \end{subfigure}
    \hfill
    \begin{subfigure}[b]{0.48\textwidth}
        \centering
        \begin{tikzpicture}[line width=0.5 pt, scale=1.35]
            \draw[fermion] (-3.5,1.0)--(-2.0,1.0);
            \draw[fermion] (-3.5,-0.5)--(-2.0,-0.5);
            \draw[dashed](-1.8,1.0)--(-1.8,-0.5);
            \draw[antifermion] (-2.0,1.0)--(0.0,1.0);
            \draw[antifermion] (-2.0,-0.5)--(0.0,-0.5);
            \node at (-3.7,1.0) {${\Psi_i}$};
            \node at (-3.7,-0.5) {$\overline{\Psi}_j$};
            \node [right] at (-1.86,0.25) {$S_k$};
            \node at (0.28,1.0) {$\nu_k$};
            \node at (0.28,-0.5) {$\overline{\nu}_k$};
        \end{tikzpicture}
        \caption{}
        \label{fig:DM1FD6}
    \end{subfigure}

    \vspace{0.3cm} 
    \begin{subfigure}[b]{0.48\textwidth}
        \centering
        \begin{tikzpicture}[line width=0.5 pt, scale=1.35]
            \draw[fermion] (-3.5,1.0)--(-1.6,1.0);
            \draw[dashed] (-3.5,-0.5)--(-1.6,-0.5);
            \draw[fermionTB](-1.6,1.0)--(-1.6,-0.5);
            \draw[fermion] (-1.6,1.0)--(0.0,1.0);
            \draw[snake] (-1.6,-0.5)--(0.0,-0.5);
            \node at (-3.7,1.0) {${\Psi_i}$};
            \node at (-3.7,-0.5) {$\overline{S}_l$};
            \node [right] at (-1.65,0.25) {$\nu_k$};
            \node at (0.28,1.0) {$\nu_k$};
            \node at (0.28,-0.5) {$Z,Z'$};
        \end{tikzpicture}
        \caption{}
        \label{fig:DM1FD7}
    \end{subfigure}
    \hfill
        \begin{subfigure}[b]{0.48\textwidth}
        \centering
        \begin{tikzpicture}[line width=0.5 pt, scale=0.85]
            \draw[fermion] (-3.0,1.0)--(-1.5,0.0);
            \draw[dashed] (-3.0,-1.0)--(-1.5,0.0);
            \draw[fermion] (-1.5,0.0)--(0.8,0.0);
            \draw[fermion] (0.8,0.0)--(2.8,1.0);
            \draw[snake] (0.8,0.0)--(2.8,-1.0);
            \node at (-3.3,1.0) {${\Psi_i}$};
            \node at (-3.3,-1.0) {$\overline{S}_l$};
            \node [above] at (-0.2,0.05) {$\Psi_i/S_k$};
            \node at (3.15,1.0) {$\nu_m$};
            \node at (3.25,-1.0) {$Z,Z'$};
        \end{tikzpicture}
        \caption{}
        \label{fig:DM1FD8}
    \end{subfigure}
    \caption{Feynman diagrams for annihilation and co-annihilation channels of fermionic DM, $\Psi_1$, into final SM states, mediated by scalars, gauge bosons, and fermions in the $s$- and $t$-channels. The indices $i,j$ take values $1$, $2$ while indices $k,l,m$ run from $1$ to $3$.}
    \label{fig:FD_DM1}
\end{figure}
\section{Relevant Feynman Diagrams for processes setting \texorpdfstring{$S_1$}{S1} relic abundance}
\label{app:B}
Here, we display the relevant Feynman diagrams contributing
to scalar dark matter~$(S_1)$ annihilation and co-annihilation processes into SM final states. The diagrams include $s$- and $t$-channel processes mediated by model scalars, gauge bosons, and fermions. These channels provide the dominant contributions to the scalar DM relic abundance.
\begin{figure}[htbp]
    \centering
    \begin{subfigure}[b]{0.48\textwidth}
        \centering
        \begin{tikzpicture}[line width=0.5 pt, scale=0.85]
            \draw[dashed] (-3.0,1.0)--(-1.5,0.0);
            \draw[dashed] (-3.0,-1.0)--(-1.5,0.0);
            \draw[dashed] (-1.5,0.0)--(0.8,0.0);
            \draw[fermion] (0.8,0.0)--(2.7,1.0);
            \draw[antifermion] (0.8,0.0)--(2.7,-1.0);
            \node at (-3.3,1.0) {${S_i}$};
            \node at (-3.3,-1.0) {$\overline{S}_i$};
            \node [above] at (0.0,0.05) {$h/H$};
            \node at (3.00,1.0) {$f$};
            \node at (3.00,-1.0) {$\overline{f}$};
        \end{tikzpicture}
        \caption{}
        \label{fig:DM2FD1}
    \end{subfigure}
    \hfill
    \begin{subfigure}[b]{0.48\textwidth}
        \centering
        \begin{tikzpicture}[line width=0.5 pt, scale=1.35]
            \draw[dashed] (-3.5,1.0)--(-1.6,1.0);
            \draw[dashed] (-3.5,-0.5)--(-1.6,-0.5);
            \draw[dashed](-1.6,1.0)--(-1.6,-0.5);
            \draw[snake] (-1.6,1.0)--(0.0,1.0);
            \draw[snake] (-1.6,-0.5)--(0.0,-0.5);
            \node at (-3.7,1.0) {${S_i}$};
            \node at (-3.7,-0.5) {$\overline{S}_i$};
            \node [right] at (-1.65,0.25) {$S_j$};
            \node at (0.28,1.0) {$Z,Z'$};
            \node at (0.28,-0.5) {$Z,Z'$};
        \end{tikzpicture}
        \caption{}
        \label{fig:DM2FD2}
    \end{subfigure}

    \vspace{0.3cm} 
    \begin{subfigure}[b]{0.48\textwidth}
        \centering
        \begin{tikzpicture}[line width=0.5 pt, scale=1.35]
            \draw[dashed] (-3.5,1.0)--(-1.6,1.0);
            \draw[dashed] (-3.5,-0.5)--(-1.6,-0.5);
            \draw[dashed](-1.6,1.0)--(-1.6,-0.5);
            \draw[dashed] (-1.6,1.0)--(0.0,1.0);
            \draw[dashed] (-1.6,-0.5)--(0.0,-0.5);
            \node at (-3.7,1.0) {${S_i}$};
            \node at (-3.7,-0.5) {$\overline{S}_i$};
            \node [right] at (-1.65,0.25) {$h/H$};
            \node at (0.28,1.0) {$S_j$};
            \node at (0.28,-0.5) {$S_j$};
        \end{tikzpicture}
        \caption{}
        \label{fig:DM2FD3}
    \end{subfigure}
    \hfill
        \begin{subfigure}[b]{0.48\textwidth}
        \centering
        \begin{tikzpicture}[line width=0.5 pt, scale=0.85]
            \draw[dashed] (-3.0,1.0)--(-1.5,0.0);
            \draw[dashed] (-3.0,-1.0)--(-1.5,0.0);
            \draw[dashed] (-1.5,0.0)--(0.8,0.0);
            \draw[dashed] (0.8,0.0)--(2.8,1.0);
            \draw[dashed] (0.8,0.0)--(2.8,-1.0);
            \node at (-3.3,1.0) {${S_i}$};
            \node at (-3.3,-1.0) {$\overline{S}_i$};
            \node [above] at (0.0,0.05) {$h/H$};
            \node at (3.15,1.0) {$\phi^+$};
            \node at (3.18,-1.0) {$\phi^-$};
        \end{tikzpicture}
        \caption{}
        \label{fig:DM2FD4}
    \end{subfigure}

    \vspace{0.3cm}
    \begin{subfigure}[b]{0.48\textwidth}
        \centering
        \begin{tikzpicture}[line width=0.5 pt, scale=1.35]
            \draw[dashed] (1.2,1.0)--(3.2,1.0);
            \draw[dashed] (1.2,-0.5)--(3.2,-0.5);
            \draw[dashed](3.2,1.0)--(3.2,-0.5);
            \draw[dashed] (3.2,1.0)--(4.9,1.0);
            \draw[dashed] (3.2,-0.5)--(4.9,-0.5);
            \node at (1.05,1.0) {${S_i}$};
            \node at (1.05,-0.5) {$\overline{S}_j$};
            \node [right] at (3.15,0.25) {$S_k/h/H$};
            \node at (5.15,1.0) {$h,H$};
            \node at (5.15,-0.5) {$h,H$};
        \end{tikzpicture}
        \caption{}
        \label{fig:DM2FD5}
    \end{subfigure}
    \hfill
    \begin{subfigure}[b]{0.48\textwidth}
        \centering
        \begin{tikzpicture}[line width=0.5 pt, scale=1.35]
            \draw[dashed] (-3.5,1.0)--(-2.0,1.0);
            \draw[dashed] (-3.5,-0.5)--(-2.0,-0.5);
            \draw[fermionBT](-1.8,1.0)--(-1.8,-0.5);
            \draw[fermion] (-2.0,1.0)--(0.0,1.0);
            \draw[antifermion] (-2.0,-0.5)--(0.0,-0.5);
            \node at (-3.7,1.0) {${S_i}$};
            \node at (-3.7,-0.5) {$\overline{S}_j$};
            \node [right] at (-1.80,0.25) {$\Psi_n$};
            \node at (0.28,1.0) {$\nu_l$};
            \node at (0.28,-0.5) {$\overline{\nu}_m$};
        \end{tikzpicture}
        \caption{}
        \label{fig:DM2FD6}
    \end{subfigure}

    \vspace{0.3cm} 
    \begin{subfigure}[b]{0.48\textwidth}
        \centering
        \begin{tikzpicture}[line width=0.5 pt, scale=1.35]
            \draw[dashed] (-3.5,1.0)--(-1.6,1.0);
            \draw[antifermion] (-3.5,-0.5)--(-1.6,-0.5);
            \draw[fermionBT](-1.6,1.0)--(-1.6,-0.5);
            \draw[fermion] (-1.6,1.0)--(0.0,1.0);
            \draw[snake] (-1.6,-0.5)--(0.0,-0.5);
            \node at (-3.7,1.0) {$S_i$};
            \node at (-3.7,-0.5) {$\overline{\Psi_n}$};
            \node [right] at (-1.60,0.25) {$\nu_k$};
            \node at (0.28,1.0) {$\nu_k$};
            \node at (0.28,-0.5) {$Z,Z'$};
        \end{tikzpicture}
        \caption{}
        \label{fig:DM2FD7}
    \end{subfigure}
    \hfill
        \begin{subfigure}[b]{0.48\textwidth}
        \centering
        \begin{tikzpicture}[line width=0.5 pt, scale=0.85]
            \draw[dashed] (-3.0,1.0)--(-1.5,0.0);
            \draw[antifermion] (-3.0,-1.0)--(-1.5,0.0);
            \draw[fermion] (-1.5,0.0)--(0.8,0.0);
            \draw[fermion] (0.8,0.0)--(2.8,1.0);
            \draw[snake] (0.8,0.0)--(2.8,-1.0);
            \node at (-3.3,1.0) {${S_i}$};
            \node at (-3.3,-1.0) {$\overline{\Psi}_n$};
            \node [above] at (-0.2,0.05) {$\Psi_n/S_k$};
            \node at (3.15,1.0) {$\nu_l$};
            \node at (3.25,-1.0) {$Z,Z'$};
        \end{tikzpicture}
        \caption{}
        \label{fig:DM2FD8}
    \end{subfigure}
    \caption{Feynman diagrams for annihilation and co-annihilation channels of scalar DM, $S_1$, into final SM states, mediated by scalars, gauge bosons, and fermions in the $s$- and $t$-channels. The indices $i,j,k,l,m$ run from $1$ to $3$, while index $n$ take values as $1$, $2$.}
    \label{fig:FD_DM2}
\end{figure}
\bibliographystyle{JHEP}
\bibliography{Diracnu_DM}
\end{document}